\documentclass[showpacs,twocolumn,aps,prx,superscriptaddress,10pt]{revtex4-1}
\usepackage{amsmath,amssymb,amsfonts,bm,graphicx,mathtools,braket}
\usepackage{hyperref}
\hypersetup{%
	pdftitle={charge pump}, %
	pdfauthor={Zhiyuan Sun},%
	pdfpagemode={UseNone},%
	pdfstartview={FitH},%
	breaklinks=true,%
	citecolor=blue,%
	colorlinks=true,%
	linkcolor=blue,%
	urlcolor=blue
}

\usepackage[T1]{fontenc}
\usepackage{fourier}
\usepackage{baskervald}

\newcommand{\unit}[1]{\,\mathrm{#1}} 
\newcommand{\equa}[1]{Eq.~\eqref{#1}} 
\newcommand{\fig}[1]{Fig.~\ref{#1}}

\newcommand{\rom}[1]{\uppercase\expandafter{\romannumeral #1\relax}}

\graphicspath{{./}}

\begin{document}

\title{Topological charge pumping in excitonic insulators}

\author{Zhiyuan Sun}
\affiliation{Department of Physics, Columbia University,
	538 West 120th Street, New York, New York 10027}

\author{Andrew J. Millis}
\affiliation{Department of Physics, Columbia University,
	538 West 120th Street, New York, New York 10027}
\affiliation{Center for Computational Quantum Physics, Flatiron Institute, 162 5th Avenue, New York, NY 10010}
\date{\today}

\begin{abstract} 
We show that in excitonic insulators with $s$-wave electron-hole pairing, an applied electric field (either pulsed or static) can induce a $p$-wave component to the order parameter, and further drive it to rotate in the $s+ip$  plane, realizing a Thouless charge pump. In one dimension, each cycle of rotation pumps exactly two electrons across the sample. Higher dimensional systems can be viewed as a stack of one dimensional chains in momentum space in which each chain crossing the fermi surface contributes a channel of charge pumping. 
Physics beyond the adiabatic limit, including in particular dissipative effects is  discussed. 
\end{abstract}

\maketitle

Controlling many-body systems, and in particular using appropriately applied external fields to `steer' order parameters of symmetry broken phases, has emerged as a central theme in current physics \cite{Kirilyuk10,Mentik15,Byrnes14,Zhang2014a,Basov17,GoleZ2016, Claassen2019, Sun2019metastable}.  The excitonic insulator (EI) is state of matter first proposed in the 1960s  \cite{Mott1961,Kozlov1965,Jerome1967,Portengen1996} with an order parameter defined as a  condensate of bound electron hole pairs that activates a hybridization between two otherwise (in the simplest case) decoupled bands and opens a gap in the electronic spectrum.  Several candidate materials including electron-hole bilayers \cite{Fogler2014a,Li2017, Du.2017}, Ta$_2$NiSe$_5$ \cite{Lu2017, Werdehausen2018, Wakisaka.2009,Kaneko2013, Sugimoto2018, Mazza2020}, $1T$-TiSe$_2$ \cite{Kogar2017,Cercellier2007, Kaneko2018, Chen2018} and monolayer WTe$_2$ \cite{jia.2020} are objects of current intensive study; recent work \cite{Du.2017,Hu2018,Wang2019,Hu2019,Varsano.2020,Perfetto2020, Hou.2019} has pointed out their possible topological aspects.  While the early theories of EI considered a one component  order parameter, typically of inversion symmetric $s$-wave type, realistic interactions also allow for electron-hole pairing in sub-dominant channels including $p$-wave (inversion-odd) ones. In equilibrium,  the $s$-wave ground state is favored, with the potential for $p$-wave order revealed by its fluctuations accompanied by dipole moment oscillations: the `Bardasis-Schrieffer' collective mode~\cite{sun2020BS}.  

\begin{figure}
	\includegraphics[width=\linewidth]{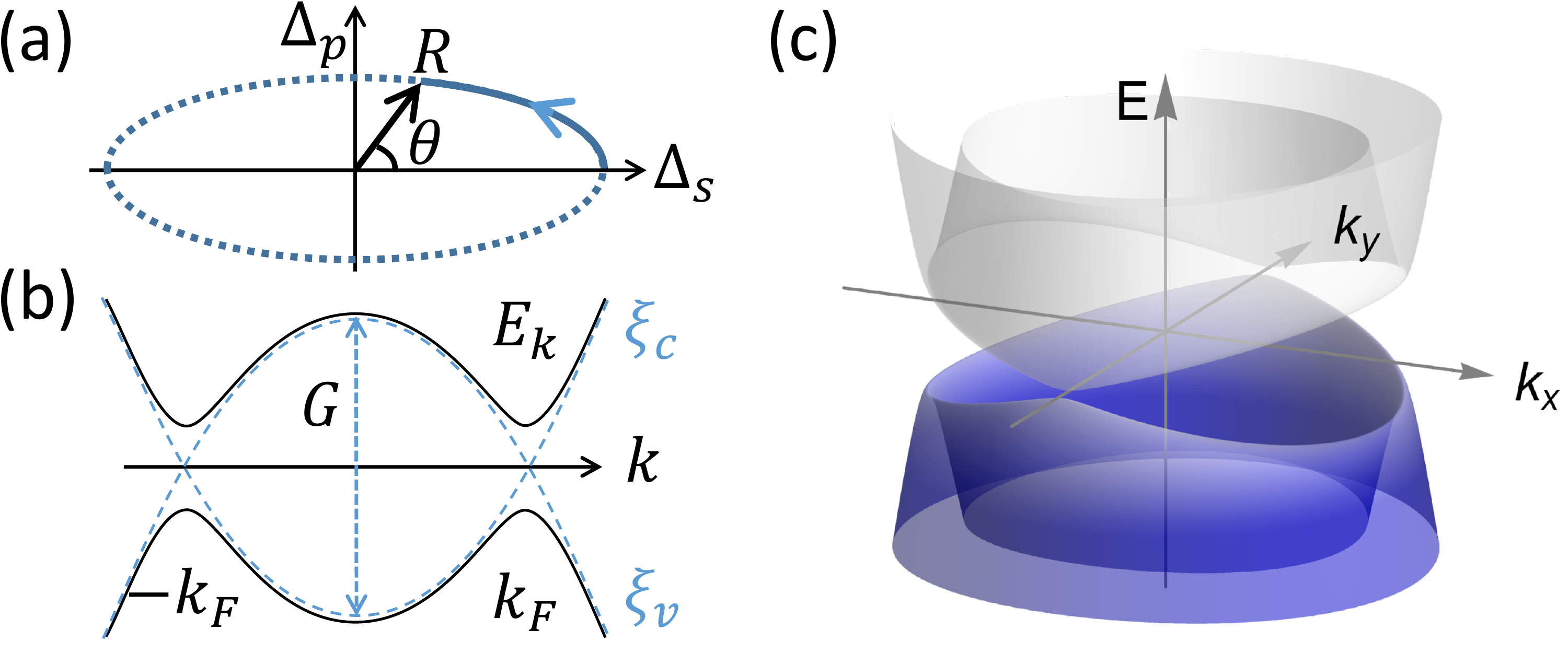}
	\caption{(a) The $s+ip$ plane for the excitonic order parameter, with electric field-driven evolution shown as dashed line. (b) The quasiparticle dispersion in a one dimensional EI (solid lines) along with bands in metallic phase (dashed lines). (c) The band dispersion of a two dimensional EI with an $s+ip$ order parameter and $\Delta_s \ll\Delta_p$.}
	\label{fig:band}
\end{figure}

In this paper we show that applied electric fields can steer the order parameter to rotate in the  space of s and p symmetry components, as shown in Fig.~\ref{fig:band}(a), leading to a realization of the `Thouless charge pump' \cite{Thouless1983,Rice1982,King-Smith1993,Zhang.2020}, providing quantized charge transport across an insulating sample.

The minimal model of an EI involves two electron bands shown in Fig.~\ref{fig:band}(b): a valence band with energy $\xi_{v,k}$ that disperses downwards from a high symmetry point (taken to have zero momentum) and a conduction band ($\xi_{c,k}$) that disperses upwards. For simplicity we assume that their energies are equal and opposite ($\xi_{c}=-\xi_{v}=\xi$). Relaxing this assumption does not change our results in an essential way. Defining the overlap $G=2 \xi_{v,0}$,  we distinguish the `BCS' case  $G>0$ where  the two bands cross at a fermi wavevector $k_F$ with fermi velocity $v_F$ as shown by the dashed lines, leading to electron and hole pockets, and the `BEC' case where $G<0$  and the bands do not cross.  Excitonic order corresponds to the spontaneous formation of a hybridization between the two bands due to the electron-electron interaction $V$, leading to an order parameter $\Delta(k)=\sum_{k^\prime}V_{k-k^\prime}\left<\psi^\dagger_{c,k^\prime} \psi_{v,k^\prime}\right>+ c.c.$ where $\psi_{c/v, k}$ is the electron annihilation operator at momentum $k$ of the conduction/valence band and $V_q$ is the Fourier transform of the density-density interaction potential $V(r)$. 
The $s$-wave order parameter $\Delta_s(k)$ is invariant under crystal symmetry operations while  $p$-wave order parameters are odd under inversion: $\Delta_p(k)=-\Delta_p(-k)$, and often transform as a multi-dimensional representation of the crystal symmetry group. For simplicity  we  neglect the $k$-dependence of $\Delta_s$, and define $\Delta_p(k)=\Delta_p f_k$ where the  pairing function $f_k$ carries the momentum dependence and satisfies $\text{max}(|f_{k_F}|)=1$. We focus on the $p_x$ pairing channel which is induced by the x-direction electric fields we consider here.
While the qualitative conclusions hold generically for all spatial dimensions, we will indicate the dimensionality if a specific d-dimensional system is discussed where the momentum $k$ means a d-dimensional vector.

Writing the partition function $Z$ as a path integral  over fermion fields $\psi=(\psi_c , \, \psi_v)$,  performing a  Hubbard-Stratonovich transformation of the interaction term in the excitonic pairing channel and subsuming the  intraband  interaction into $\xi$ one obtains the action (see \cite{SI} Sec.~\rom{1})
\begin{align}
S = \int d\tau dr 
\Bigg\{
&
\psi^\dagger 
\left( \partial_\tau +
H_m
\right)
\psi
+ \frac{1}{g_s}|\Delta_s|^2  + \frac{1}{g_p}|\Delta_p|^2 
\Bigg\}
\label{eqn:HS_action}
\end{align}
as an integral over space-time $(r, \tau)$ and the partition function is $Z=\int  D[\bar{\psi},\psi] D[\bar{\Delta},\Delta]e^{-S}$.
For physically reasonable interactions such as the screened Coulomb interaction, the $s$-wave pairing interaction $g_s$ is typically the strongest while
$g_p$ is the  leading subdominant one. 
We may write the mean field Hamiltonian as $\int dr \psi^\dagger H_m \psi=\sum_k \psi_k^\dagger H_m^k \psi_k$ with
\begin{equation}
H_m^k[\Delta_s,\Delta_p]=\xi_{k} \sigma_3 + \Delta_s\sigma_1+ \Delta_p f_k \sigma_2
\label{eqn:mean_field_H}
\end{equation}
where $\sigma_i$ are the Pauli matrices acting in the c/v band space. 
The  vector potential $A$ enters  \equa{eqn:mean_field_H} through the minimal coupling $k\rightarrow k-A$ required by local gauge invariance (electric field is $E=-\partial_t A$)  and we set electron charge $e$, speed of light and the Planck constant $\hbar$ to be one. 
Interband dipolar couplings could also occur \cite{GoleZ2016, Murakami.2020} but do not affect our results.
Since the global phase is not important, we choose the $s$-wave order parameter to be real. As we will show, the system develops an electrical polarization as a $p$-wave component $\pi/2$ out of phase with the equilibrium $\Delta_s$ is introduced. Due to an emergent `particle hole' symmetry in the BCS weak coupling case defined as $|\Delta_s|,|\Delta_p| \ll G$ which we focus on, applied electric fields create $\Delta_p$ primarily in this channel (see Sec.~\rom{6}~A of \cite{SI} for a rigorous proof), so we write $p$-wave pairing in the $\sigma_2$ channel \cite{sun2020BS}.
The quasiparticle spectrum is $E_k=\pm\sqrt{\xi_k^2+\Delta_s^2+\Delta_p^2f_k^2}$. As shown by Fig.~\ref{fig:band}(c), the spectrum will have gapless points (nodes) at $(k_x,k_y)=(0,\pm k_F)$ in the pure $p$-wave state ($\Delta_s=0$) in two dimensions (2D).

\begin{figure}
	\includegraphics[width= \linewidth]{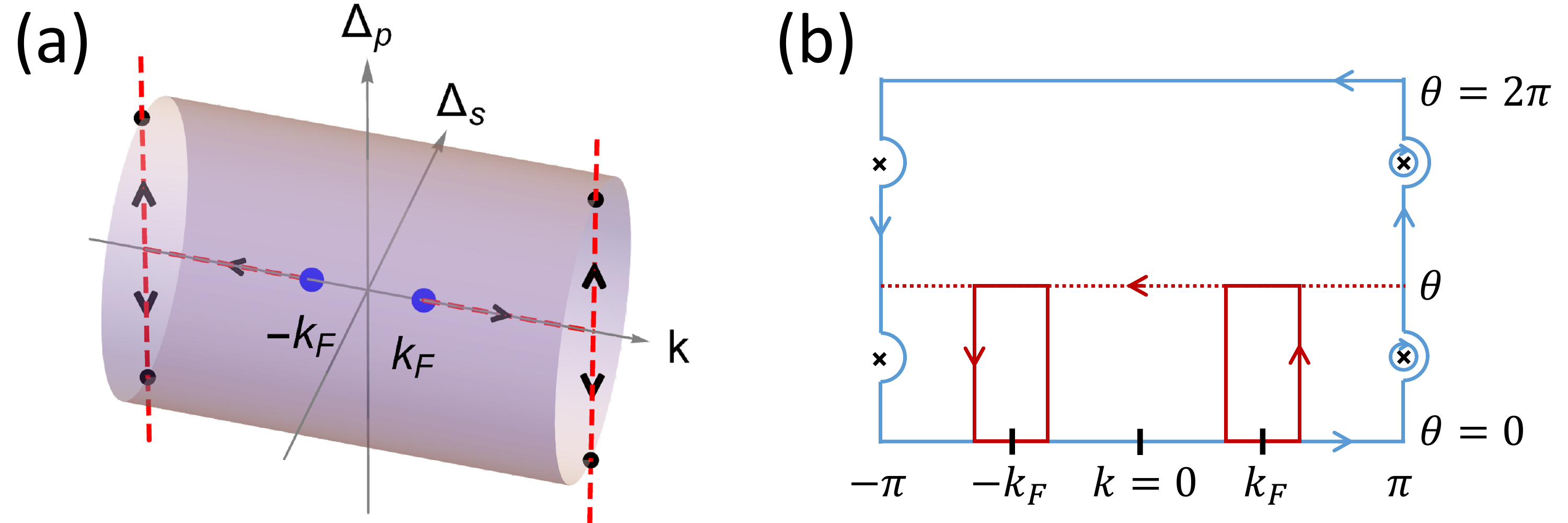}
	\caption{(a) The surface $S$  in the $(k,\Delta_s,\Delta_p)$ space used to calculate the flux of the Berry curvature for a 1D EI for which the order parameter evolution completes a full cycle in the $s+ip$ plane. The left and right ends of the cylinder are identified so that $S$ is a 2-torus. Blue dots are Berry curvature monopoles and red dashed lines are `Dirac strings' with direction shown by black arrows. (b) The surface of the  torus shown in (a)  parametrized by $k$ and $\theta$  with $k=\pm \pi$ and $\theta=0,\, 2\pi$ identified.  The blue contours yield the charge pumped during a full cycle. The red rectangles are used to compute the flux for a partial cycle in the  BCS limit.}
	\label{fig:path}
\end{figure}

\emph{Charge pump---}Spatially uniform changes in $\Delta_{s,p}$ produce uniform currents $J=\langle \sum_k \partial_k H_m^k \rangle$ (see \cite{SI} Sec.~\rom{2}), whose time integral from the initial $(\Delta_s,\Delta_p)=(\Delta,0)$ to the final point then gives the pumped charge $P$ (difference of polarization between the final state and the initial state). 
In a one dimensional (1D) system, $P$ has a  geometrical meaning \cite{King-Smith1993,Resta1994} in the limit of slow order parameter dynamics. It is the flux of the Berry curvature 2-form $B$ through the 2D surface $S$ spanned by the occupied 1D crystal momentum  $k$ and the time varying trajectory of $\Delta_{s,p}$, or alternatively by the line integral of the Berry connection $\mathcal{A}_\mu=i\langle \psi|\partial_\mu |\psi \rangle$ around its boundary:
\begin{align}
P= \frac{1}{2\pi} \int_S dS \cdot B
=\frac{1}{2\pi} \oint dl \cdot \mathcal{A}
\,
\label{eqn:berry_phase_formula}
\end{align}
where $\mu=(k,\Delta_s,\Delta_p)$
(see Fig.~\ref{fig:path}). 

The Berry curvature $B$ from the valence band of \equa{eqn:mean_field_H} is sourced by monopoles at the points $\xi_k=\Delta_s=\Delta_p=0$, i.e., the points $(k,\Delta_s,\Delta_p)=(\pm k_F,0,0)$ each of which has monopole charge $1$.   If the order parameter evolution completes a full cycle on the $s+ip$ plane, $S$ becomes the  surface of the 2-torus shown in Fig.~\ref{fig:path}(a) and the net charge pumped is the total  flux from the enclosed monopoles which is an integer $N=2$, the Chern number of the process. This quantized change in the polarization is known as the Thouless pump~\cite{Thouless1983}, a topological phenomenon immune to disorder.  Note that  the monopoles exist only for the `BCS'  ($G>0$, band inversion) case where the excitons strongly overlap such that charge can jump between them. In the `BEC' case $\xi_k\neq0$ for all $k$ and there are no monopoles enclosed in $S$ (see \cite{SI} Sec.~\rom{2} C).  

To compute the polarization for the case the order parameter does not complete a full cycle,  we use the line integral approach. For notational simplicity, we suppress the  subscripts `$k$' without causing ambiguities. An explicit expression for the valence band wave function from  \eqref{eqn:mean_field_H} at $(k,\Delta_s,\Delta_p)$ is
\begin{align}
|\psi \rangle=(-v^\ast,\,u^\ast)= \frac{1}{\sqrt{2E(E-\xi)}} \left(\xi-E ,\, \Delta^\ast \right)
\label{eqn:berry_connection}
\end{align} 
where $\Delta=\Delta_s + i\Delta_p f_k \equiv |\Delta| e^{i\phi}$ and $|u|^2(|v^2|)=\frac{1}{2}\left(1\pm\frac{\xi}{E}\right)$. 
The Berry connection  $\mathcal{A}_\mu= |u|^2 \partial_\mu \phi$  has singularities  associated with the Dirac strings, the intersections of which with $S$ (marked by crosses in Fig.~\ref{fig:path}(b)) must be correctly treated in the evaluation of the line integral. 
Noting that $|u|^2 \rightarrow 0$ when $\xi \ll -|\Delta|$ and $|u|^2 \rightarrow 1$ when $\xi \gg |\Delta|$, we see that in the weak coupling BCS limit the contour can be collapsed to the red rectangles in Fig.~\ref{fig:path}(b).  
Parameterizing $S$ using $k$ and the angle $\theta$ defined by $\Delta_s+i\Delta_p=Re^{i\theta}$ in Fig.~\ref{fig:band}(a), 
one observes that the polarization of an state on the $s+ip$ plane depends only on the angle $\theta$. Specifically, we found 
\begin{align}
P=\theta/\pi
\label{eqn:P_1D}
\end{align}
for a 1D  EI (see \cite{SI} Sec.~\rom{2}).
This may be understood by noting that the low energy physics around $\pm k_F$ is of two massive Dirac models,  each of which realizes a Goldstone-Wilczek \cite{Goldstone1981} mechanism of charge pumping. 

Higher dimensional systems can be viewed as 1D chains along $x$ direction stacked in momentum space. For a 2D circular fermi surface one finds 
\begin{align}
P(\theta)=
\left\{
\begin{array}{lc}
\frac{k_F}{2\pi} \tan \frac{\theta}{2}
&  
\,(0<\theta < \pi/2)
\\
\frac{k_F}{2\pi} \left(2-\cot  \frac{\theta}{2} \right)
&  (\pi/2<\theta < \pi)
\\
\frac{k_F}{\pi} +P(\theta-\pi)
&  (\pi<\theta < 2\pi)
\end{array}
\right. \,.
\label{eqn:P_2D}
\end{align}
A full cycle pumps exactly two electrons along each 1D momentum chain that crosses the fermi surface, giving
\begin{align}
P_{1D} = 2 \,,\quad 
P_{2D} = \frac{2k_F}{\pi} \,,\quad  P_{3D} = \frac{ k_F^2}{2\pi}
\label{eqn:p_2d_3d}
\end{align}
for 1D, 2D and three dimensional (3D) isotropic systems respectively.


\begin{figure}
	\includegraphics[width=\linewidth]{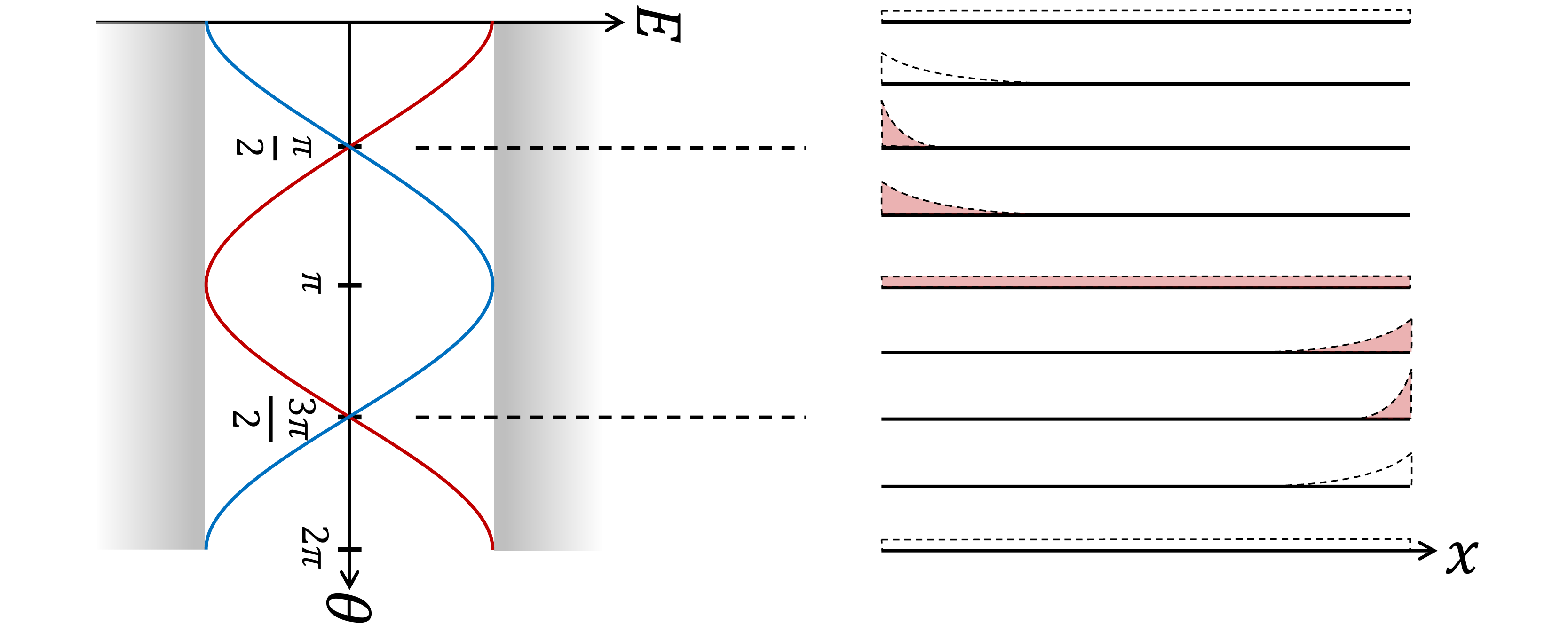}
	\caption{Left is the evolution of the edge state energies as a function of $\theta$. Right is the spatial profile of one of the two edge states (labeled by red line on the left) neglecting its quick oscillating detail. Being filled means the state is occupied. The blue edge state is not shown but is the mirror image of the red one. }
	\label{fig:edge_state}
\end{figure}

Although the charge pump is a bulk property carried by all valence band electrons, it is also revealed by the evolution of edge states \cite{Yang.2020} as  $\Delta_s$ and $\Delta_p$ are varied, as shown in Fig.~\ref{fig:edge_state} for a 1D wire connected with reservoirs.  
In the BCS limit, with open boundary conditions $\psi(0)=\psi(L)=0$, there are two edge states 
\begin{align}
\psi_\pm = \frac{1}{C_\pm} \left(1, \pm1 \right) \sin (k_F x) e^{\mp x \Delta_p/v_F} 
\,,\quad E= \pm \Delta_s
\label{eqn:edge_state}
\end{align}
where $C_\pm$ is a normalization constant. We suppose $\Delta_s+i\Delta_p=R e^{i\theta}$ and follow the evolution of $\psi_+$ as $\theta$ is varied (see Fig.\ref{fig:edge_state}). At $\theta=0$ the state is delocalized and unoccupied with energy $R$.  As $\theta$ is increased the state becomes localized near $x=0$ and decreases in energy. When $\theta$ passes through $\pi/2$, the state becomes maximally localized and becomes occupied by an electron from the left reservoir since its energy crosses the chemical potential. As $\theta$ further increases the state becomes delocalized and then localized at the right edge, delivering its electron to the right reservoir when $\theta$ crosses $3\pi/2$. Considering the $\psi_-$ state during the same cycle,  two electrons in total are pumped.
In higher dimensions, each 1D $k_x$ chain crossing the fermi surface has a similar edge state evolution (see \cite{SI} Sec.~\rom{3}).

\begin{figure}
	\includegraphics[width=0.7 \linewidth]{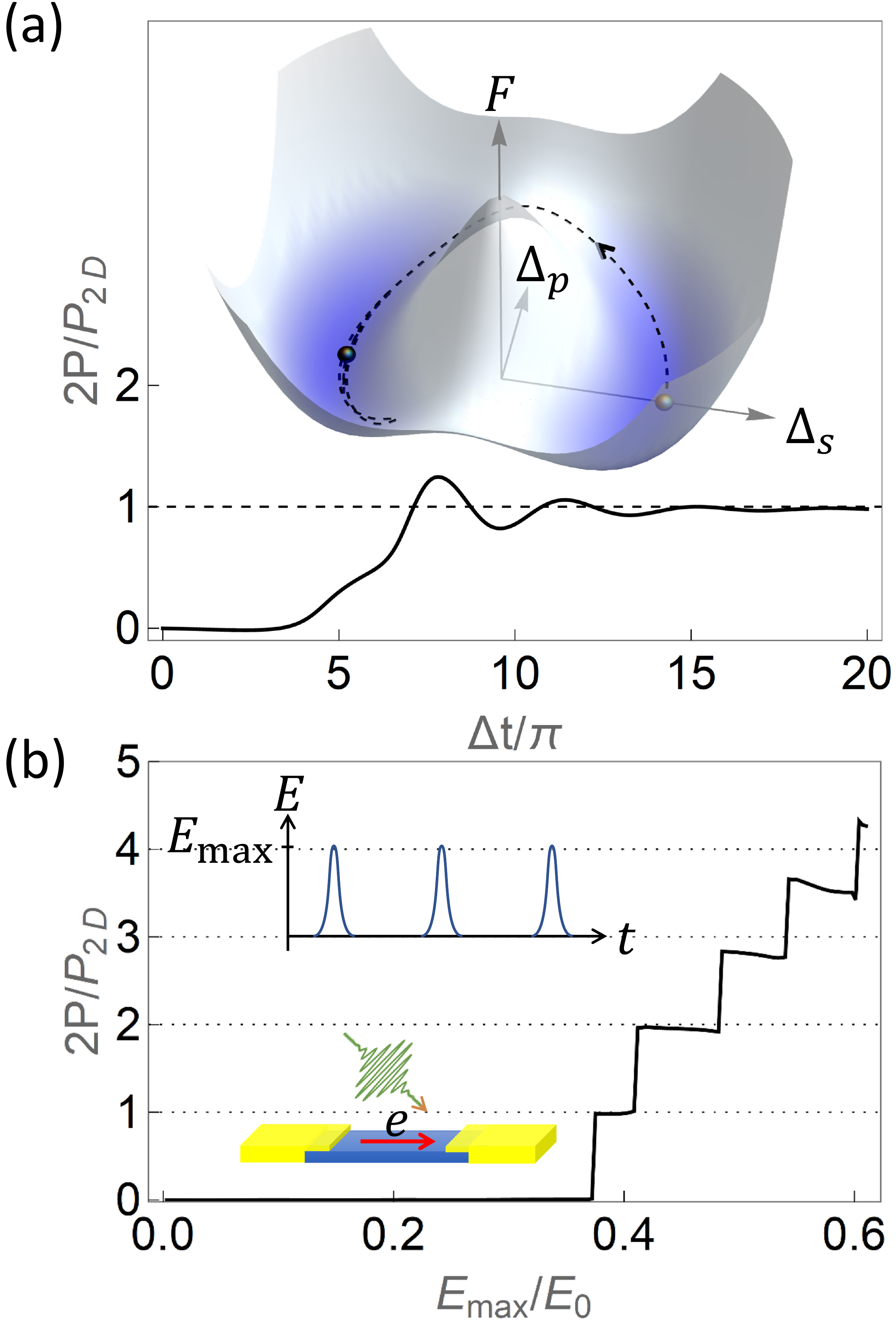}
	\caption{Electric field pulse induced  dynamics of a 2D isotropic EI. (a)  The polarization as a function of time during the dynamics with $P_{\text{dis}}$ being small. Inset is the free energy landscape $F(\Delta_s,\Delta_p)$ plotted on the $s+ip$ plane. Lower energy appears bluer. The black dashed order parameter trajectory is caused by a pulse $E(t)=E_{\text{max}}\tanh^\prime\left((t-t_0)/w\right)$ with maximum electric field  $E_{\text{max}}=0.39 E_0$.
	(b) The pumped charge by a single pulse as a function of $E_{\text{max}}$. 
	The units are $P_{2D}=2k_F/\pi$ and $E_0=\Delta^2/v_F$. Top inset is a schematic of a train of well separated pulses which can induce a `steady' current. Bottom inset is a  schematic of the device with EI shown in blue and the contacts in gold. The parameters are $g_s \nu=0.3$, $g_p \nu=0.58$, $\Delta=2\Lambda e^{-1/(g_s \nu)}$, $\gamma=0.07 \Delta$ and $w=2\pi/(2\Delta)$.}
	\label{fig:landscape}
\end{figure}
\emph{Dynamics---}The coupled dynamics of electrons and the order parameters in the presence of an applied electric field is described by the action \equa{eqn:HS_action}.
To understand the qualitative dynamics, we use a low energy effective Ginzburg-Landau Lagrangian
\begin{align}
L(\Delta_s,\Delta_p;E) = F - K +L_{\text{drive}}
\label{eqn:GL_lagrangian}
\end{align}
for fields $\Delta_s,\,\Delta_p$ obtained by integrating out the Fermions (see \cite{SI} Sec.~\rom{4}). The dynamics is  given by the standard Euler-Lagrange equation $\frac{d}{dt} \frac{\delta L}{\delta \dot{\Delta}_i} = \frac{\delta L}{\delta \Delta_i}$ and is that of a point particle moving in the landscape defined by $F$, with kinetic energy $K$ and driven by an electric field through $L_{\text{drive}}$. We find 
\begin{equation}
L_{\text{drive}}= -P(\theta) E -s(\Delta_s,\Delta_p) E^2 + O(E^3)
\label{eqn:drive}
\end{equation}
where $P$ is the adiabatic polarization in Eqs.~\eqref{eqn:P_1D} or \eqref{eqn:P_2D}, $s=\lim_{\omega\rightarrow 0} \sigma(\omega)/(2i\omega)$ and $\sigma(\omega)$ is the optical conductivity from virtual interband excitations (see \cite{SI} Sec.~\rom{4}). It is natural that electric field couples linearly to the polarization and therefore provides a `force' $E \partial_\theta P$ to rotate the order parameter in the $\Delta_s,\Delta_p$ plane.

$F(\Delta_s,\Delta_p)$ gives the potential landscape in which the dynamics takes place; it has the anisotropic `Mexican hat' form shown in Fig.~\ref{fig:landscape}(a). 
For (quasi) 1D systems in the weak coupling BCS limit \cite{Altland.2010,Sun2020a}:
\begin{align}
F = -\nu \left(\Delta_s^2 + \Delta_p^2 \right) \ln \frac{2\Lambda}{\sqrt{\Delta_s^2 + \Delta_p^2}} + \frac{1}{g_s} \Delta_s^2 + \frac{1}{g_p} \Delta_p^2\label{eqn:F_1D}
\end{align}
where $\nu$ is density of states in the metallic phase and $\Lambda \gg \sqrt{\Delta_s^2+\Delta_p^2}$ is a high energy cutoff \cite{Kozlov1965}. The first term becomes
$
-\nu \int \frac{d\theta_k}{2\pi} \left(\Delta_s^2 + \Delta_p^2 \cos^2 \theta_k \right) \ln \left(2\Lambda/\sqrt{\Delta_s^2 + \Delta_p^2 \cos^2 \theta_k}\right)
$
for a 2D isotropic Fermi surface
and $\frac{d\theta_k}{2\pi} \rightarrow \frac{\sin \theta_k d\theta_k d\phi_k}{4\pi}$ for 3D where $\theta_k$ and $\phi_k$ are angular variables on the Fermi surface. The landscape has  a local maximum at $R=0$ surrounded by a trough at $R(\theta)$ of lower values of $F$. The ground state minima are at $(\pm \Delta, 0)$ and the pure p-wave phases  at $(0, \pm \Delta_{p0})$ are saddle points with energy higher by $F_b=\nu(\Delta^2-c \Delta_{p0}^2)/2$ where $c$ is a constant depending on the space dimension. 

We may estimate the minimal electric field required to drive the system from the minimum through the $p$-wave saddle point by equating the potential energy barrier $F_b$ to the work $EP(\theta=\pi/2)+\mathcal{O}(E^2)$ done by the electric field, obtaining
\begin{align}
E_c \approx \kappa E_0
,\quad E_0=\Delta/\xi^0=\Delta^2/v_F
\label{eqn:Ec}
\end{align}
where $\xi^0= v_F/\Delta$ is the coherence length (exciton size), $\kappa=\frac{1}{\pi}(1-\Delta_{p0}^2/\Delta^2)$ in 1D and $\kappa= \frac{1}{2}-\frac{1}{4}\frac{\Delta_{p0}^2}{\Delta^2} $ in 2D, and $E_0$ is at the order of the dielectric breakdown field.      For $v_F=10^6 \unit{m/s}$, $\Delta=10 \unit{meV}$ and $\Delta_{p0} \ll \Delta$, such as the case of electron hole bilayers, the threshold field is $E_c \sim 10^3 \unit{V/cm}$ which can be easily achieved by modern optical technique. For a $100 \unit{meV}$ gap such as that in Ta$_2$NiSe$_5$ \cite{Lu2017,Werdehausen2018} (assuming it is in the BCS regime), the threshold field is about $10^5 \unit{V/cm}$. At such large field, $O(E^2)$ terms in the Lagrangian 
will be important, which pushes the order parameter closer to zero but does not destroy the qualitative dynamics in the transient regime (See \cite{SI} Sec.~\rom{4} D).

The dynamical term  $K$ has a relatively simple form if the gap never closes on the Fermi surface and the order parameter variation timescale is long compared to the inverse of the gap.  For example for (quasi) 1D 
\begin{equation}
K\approx \nu \left(\dot{R}^2/R^2+3\dot{\theta}^2\right)/12
\label{eqn:kinetic}
\end{equation}
to lowest order in time derivatives. For higher dimensions with closed Fermi surfaces, there are $O(1)$ changes to the coefficients and, crucially,  dissipation and time non-locality arises from quasiparticle excitations near the nodes of the p-wave gap when $\Delta_s$ passes zero. This dissipation also brings a correction to the pumped charge: $P\rightarrow P_{2D}+P_{\text{dis}}$. To estimate  $P_{\text{dis}}$, we observe that
as the order parameter passes this gapless regime with a velocity $\dot{\Delta}_s$, the probability for exciting a particle-hole pair at $k$ is given by the Landau-Zener formula \cite{Wittig.2005}: $P_k= e^{-2\pi \delta_k^2/|\partial_t 2 \Delta_s|}$ where $\delta_k=\sqrt{\xi_k^2+\Delta_p^2 f_k^2}$ is half of its minimal excitation energy during the dynamics. In 2D, summing over momenta, one obtains the number of excited quasi particles $N= \frac{k_F}{2\pi^2 v_F}
|\frac{\dot{\Delta}_s}{\Delta_{p}}|$ and the non-adiabatic correction to the pumped charge 
\begin{align}
P_{\text{dis}} = -P_{2D} \frac{1}{8\pi^2} 
\frac{|\dot{\Delta}_s|}{\Delta_p^2}
\label{eqn_P_dis_maintext}
\,
\end{align}
valid if $\sqrt{|\dot{\Delta}_s|} \ll |\Delta_p|$
(see \cite{SI} Sec.~\rom{6} C). Therefore, if the sub-dominant $p$-wave coupling constant is too small such that $\Delta_{p0}\sim \Lambda e^{-\frac{1}{g_p \nu}} <\sqrt{|\dot{\Delta}_s|/(8\pi^2)} $, this dissipative correction will dominate over the adiabatic charge.

\emph{Numerics and Experiment---}We numerically solved the mean field dynamics implied by \equa{eqn:HS_action} for a BCS weak coupling EI in 2D, driven by a train of widely separated electric field pulses (Fig.~\ref{fig:landscape}(b)).
Mean field dynamics \cite{Barankov.2004} means that each momentum state evolves in the time dependent mean field $(\Delta_s,\, \Delta_p f_{k-A},\, \xi_{k-A})$ with $\Delta_{s,p}$ determined self consistently by the gap equation, neglecting any spatial fluctuations. We include a weak phenomenological damping $\gamma$ to represent energy loss caused by, e.g., a phonon bath (see \cite{SI} Sec.~\rom{6}). Each pulse drives the order parameter along the trajectory shown as  the black dashed line  in Fig.~\ref{fig:landscape}(a), advancing it by $\theta=\pi$ to stabilize the system in the other $s$-wave ground state.  The total duration of the evolution from one minimum to the next is  $T_s\approx 20\pi/\Delta$ and the amount of charge pumped is $W P_{2D}/2$ where $W$ is the width of the sample, as shown by Fig.~\ref{fig:landscape}(a).
In a  train of pulses with inter pulse separation $T_0 \gg T_s$, each pulse advances the order parameter from one minimum to the next and allows it to stabilize before next pulse arrives, leading to a time-averaged current  $I_0=e W k_F/(\pi T_0)$.
For a $10 \unit{\mu m}$ wide sample with normal state carrier density of $10^{12} \unit{cm^{-2}}$ and inter pulse time $T_0=1 \unit{ns}$, the current is $I_0=255 \unit{nA}$ considering spin degeneracy.  

A minimum field strength $\sim E_c$ (\ref{eqn:Ec}) is required: as the maximum electric field $E_{\text{max}}$ of the pulse is increased beyond the threshold, the charge pumping (DC current) will onset sharply, as shown in Fig.~\ref{fig:landscape}(b). As $E_{\text{max}}$ further increases, each pulse induces a rotation of more cycles which pumps more charge, giving rise to the step structure.
Deviations from perfect quantization arise from fast order parameter dynamics caused by the short duration pulse. A precisely engineered long duration pulse can substantially reduce these deviations; see \cite{SI} Sec.~\rom{5}.

A static electric field in the DC transport regime could also drive such an order parameter rotation and charge pumping. However, unlike the case of well separated pulses, there is no time break to dump the generated heat into the environment which might destroy the system.

\emph{Discussion---}We have shown theoretically that a Thouless charge pump may be realized as a collective many-body effect arising from order parameter steering in BCS type excitonic insulators. Similar dynamics and charge
pumping can occur in general when the ground state order parameter and the sub dominant one have different parities under inversion.
Its observation would provide both a verification of order parameter steering and a  probe  of the excitonic insulating state, in particular, distinguishing BCS and BEC states. It is interesting to study the dynamics in the vicinity of the BCS-BEC crossover and effects beyond mean field.

\begin{acknowledgements}
	We acknowledge support from the
	Energy Frontier Research Center on Programmable Quantum Materials funded
	by the US Department of Energy (DOE), Office of Science, Basic Energy
	Sciences (BES), under award No. DE-SC0019443. We thank W. Yang, D. Golez and T. Kaneko for helpful discussions.
\end{acknowledgements}
\bibliographystyle{apsrev4-1}
\bibliography{./Excitonic_Insulator}

\begin{thebibliography}{46}%
\makeatletter
\providecommand \@ifxundefined [1]{%
 \@ifx{#1\undefined}
}%
\providecommand \@ifnum [1]{%
 \ifnum #1\expandafter \@firstoftwo
 \else \expandafter \@secondoftwo
 \fi
}%
\providecommand \@ifx [1]{%
 \ifx #1\expandafter \@firstoftwo
 \else \expandafter \@secondoftwo
 \fi
}%
\providecommand \natexlab [1]{#1}%
\providecommand \enquote  [1]{``#1''}%
\providecommand \bibnamefont  [1]{#1}%
\providecommand \bibfnamefont [1]{#1}%
\providecommand \citenamefont [1]{#1}%
\providecommand \href@noop [0]{\@secondoftwo}%
\providecommand \href [0]{\begingroup \@sanitize@url \@href}%
\providecommand \@href[1]{\@@startlink{#1}\@@href}%
\providecommand \@@href[1]{\endgroup#1\@@endlink}%
\providecommand \@sanitize@url [0]{\catcode `\\12\catcode `\$12\catcode
  `\&12\catcode `\#12\catcode `\^12\catcode `\_12\catcode `\%12\relax}%
\providecommand \@@startlink[1]{}%
\providecommand \@@endlink[0]{}%
\providecommand \url  [0]{\begingroup\@sanitize@url \@url }%
\providecommand \@url [1]{\endgroup\@href {#1}{\urlprefix }}%
\providecommand \urlprefix  [0]{URL }%
\providecommand \Eprint [0]{\href }%
\providecommand \doibase [0]{http://dx.doi.org/}%
\providecommand \selectlanguage [0]{\@gobble}%
\providecommand \bibinfo  [0]{\@secondoftwo}%
\providecommand \bibfield  [0]{\@secondoftwo}%
\providecommand \translation [1]{[#1]}%
\providecommand \BibitemOpen [0]{}%
\providecommand \bibitemStop [0]{}%
\providecommand \bibitemNoStop [0]{.\EOS\space}%
\providecommand \EOS [0]{\spacefactor3000\relax}%
\providecommand \BibitemShut  [1]{\csname bibitem#1\endcsname}%
\let\auto@bib@innerbib\@empty
\bibitem [{\citenamefont {Kirilyuk}\ \emph {et~al.}(2010)\citenamefont
  {Kirilyuk}, \citenamefont {Kimel},\ and\ \citenamefont
  {Rasing}}]{Kirilyuk10}%
  \BibitemOpen
  \bibfield  {author} {\bibinfo {author} {\bibfnamefont {A.}~\bibnamefont
  {Kirilyuk}}, \bibinfo {author} {\bibfnamefont {A.~V.}\ \bibnamefont {Kimel}},
  \ and\ \bibinfo {author} {\bibfnamefont {T.}~\bibnamefont {Rasing}},\ }\href
  {\doibase 10.1103/RevModPhys.82.2731} {\bibfield  {journal} {\bibinfo
  {journal} {Rev. Mod. Phys.}\ }\textbf {\bibinfo {volume} {82}},\ \bibinfo
  {pages} {2731} (\bibinfo {year} {2010})}\BibitemShut {NoStop}%
\bibitem [{\citenamefont {Mentink}\ \emph {et~al.}(2015)\citenamefont
  {Mentink}, \citenamefont {Balzer},\ and\ \citenamefont
  {Eckstein}}]{Mentik15}%
  \BibitemOpen
  \bibfield  {author} {\bibinfo {author} {\bibfnamefont {J.~H.}\ \bibnamefont
  {Mentink}}, \bibinfo {author} {\bibfnamefont {K.}~\bibnamefont {Balzer}}, \
  and\ \bibinfo {author} {\bibfnamefont {M.}~\bibnamefont {Eckstein}},\ }\href
  {https://doi.org/10.1038/ncomms7708} {\bibfield  {journal} {\bibinfo
  {journal} {Nature Communications}\ }\textbf {\bibinfo {volume} {6}},\
  \bibinfo {pages} {6708} (\bibinfo {year} {2015})}\BibitemShut {NoStop}%
\bibitem [{\citenamefont {Byrnes}\ \emph {et~al.}(2014)\citenamefont {Byrnes},
  \citenamefont {Kim},\ and\ \citenamefont {Yamamoto}}]{Byrnes14}%
  \BibitemOpen
  \bibfield  {author} {\bibinfo {author} {\bibfnamefont {T.}~\bibnamefont
  {Byrnes}}, \bibinfo {author} {\bibfnamefont {N.~Y.}\ \bibnamefont {Kim}}, \
  and\ \bibinfo {author} {\bibfnamefont {Y.}~\bibnamefont {Yamamoto}},\ }\href
  {https://doi.org/10.1038/nphys3143} {\bibfield  {journal} {\bibinfo
  {journal} {Nature Physics}\ }\textbf {\bibinfo {volume} {10}},\ \bibinfo
  {pages} {803} (\bibinfo {year} {2014})}\BibitemShut {NoStop}%
\bibitem [{\citenamefont {Zhang}\ and\ \citenamefont
  {Averitt}(2014)}]{Zhang2014a}%
  \BibitemOpen
  \bibfield  {author} {\bibinfo {author} {\bibfnamefont {J.}~\bibnamefont
  {Zhang}}\ and\ \bibinfo {author} {\bibfnamefont {R.}~\bibnamefont
  {Averitt}},\ }\href {\doibase 10.1146/annurev-matsci-070813-113258}
  {\bibfield  {journal} {\bibinfo  {journal} {Annu. Rev. Mater. Res.}\ }\textbf
  {\bibinfo {volume} {44}},\ \bibinfo {pages} {19} (\bibinfo {year}
  {2014})}\BibitemShut {NoStop}%
\bibitem [{\citenamefont {Basov}\ \emph {et~al.}(2017)\citenamefont {Basov},
  \citenamefont {Averitt},\ and\ \citenamefont {Hsieh}}]{Basov17}%
  \BibitemOpen
  \bibfield  {author} {\bibinfo {author} {\bibfnamefont {D.~N.}\ \bibnamefont
  {Basov}}, \bibinfo {author} {\bibfnamefont {R.~D.}\ \bibnamefont {Averitt}},
  \ and\ \bibinfo {author} {\bibfnamefont {D.}~\bibnamefont {Hsieh}},\ }\href
  {https://doi.org/10.1038/nmat5017} {\bibfield  {journal} {\bibinfo  {journal}
  {Nature Materials}\ }\textbf {\bibinfo {volume} {16}},\ \bibinfo {pages}
  {1077} (\bibinfo {year} {2017})}\BibitemShut {NoStop}%
\bibitem [{\citenamefont {Gole\ifmmode~\check{z}\else \v{z}\fi{}}\ \emph
  {et~al.}(2016)\citenamefont {Gole\ifmmode~\check{z}\else \v{z}\fi{}},
  \citenamefont {Werner},\ and\ \citenamefont {Eckstein}}]{GoleZ2016}%
  \BibitemOpen
  \bibfield  {author} {\bibinfo {author} {\bibfnamefont {D.}~\bibnamefont
  {Gole\ifmmode~\check{z}\else \v{z}\fi{}}}, \bibinfo {author} {\bibfnamefont
  {P.}~\bibnamefont {Werner}}, \ and\ \bibinfo {author} {\bibfnamefont
  {M.}~\bibnamefont {Eckstein}},\ }\href {\doibase 10.1103/PhysRevB.94.035121}
  {\bibfield  {journal} {\bibinfo  {journal} {Phys. Rev. B}\ }\textbf {\bibinfo
  {volume} {94}},\ \bibinfo {pages} {035121} (\bibinfo {year}
  {2016})}\BibitemShut {NoStop}%
\bibitem [{\citenamefont {Claassen}\ \emph {et~al.}(2019)\citenamefont
  {Claassen}, \citenamefont {Kennes}, \citenamefont {Zingl}, \citenamefont
  {Sentef},\ and\ \citenamefont {Rubio}}]{Claassen2019}%
  \BibitemOpen
  \bibfield  {author} {\bibinfo {author} {\bibfnamefont {M.}~\bibnamefont
  {Claassen}}, \bibinfo {author} {\bibfnamefont {D.~M.}\ \bibnamefont
  {Kennes}}, \bibinfo {author} {\bibfnamefont {M.}~\bibnamefont {Zingl}},
  \bibinfo {author} {\bibfnamefont {M.~A.}\ \bibnamefont {Sentef}}, \ and\
  \bibinfo {author} {\bibfnamefont {A.}~\bibnamefont {Rubio}},\ }\href
  {https://doi.org/10.1038/s41567-019-0532-6} {\bibfield  {journal} {\bibinfo
  {journal} {Nature Physics}\ }\textbf {\bibinfo {volume} {15}},\ \bibinfo
  {pages} {766} (\bibinfo {year} {2019})}\BibitemShut {NoStop}%
\bibitem [{\citenamefont {Sun}\ and\ \citenamefont
  {Millis}(2020{\natexlab{a}})}]{Sun2019metastable}%
  \BibitemOpen
  \bibfield  {author} {\bibinfo {author} {\bibfnamefont {Z.}~\bibnamefont
  {Sun}}\ and\ \bibinfo {author} {\bibfnamefont {A.~J.}\ \bibnamefont
  {Millis}},\ }\href {\doibase 10.1103/PhysRevX.10.021028} {\bibfield
  {journal} {\bibinfo  {journal} {Phys. Rev. X}\ }\textbf {\bibinfo {volume}
  {10}},\ \bibinfo {pages} {021028} (\bibinfo {year}
  {2020}{\natexlab{a}})}\BibitemShut {NoStop}%
\bibitem [{\citenamefont {Mott}(1961)}]{Mott1961}%
  \BibitemOpen
  \bibfield  {author} {\bibinfo {author} {\bibfnamefont {N.~F.}\ \bibnamefont
  {Mott}},\ }\href {\doibase 10.1080/14786436108243318} {\bibfield  {journal}
  {\bibinfo  {journal} {Philos. Mag.}\ }\textbf {\bibinfo {volume} {6}},\
  \bibinfo {pages} {287} (\bibinfo {year} {1961})}\BibitemShut {NoStop}%
\bibitem [{\citenamefont {Kozlov}\ and\ \citenamefont
  {Maksimov}(1965)}]{Kozlov1965}%
  \BibitemOpen
  \bibfield  {author} {\bibinfo {author} {\bibfnamefont {A.}~\bibnamefont
  {Kozlov}}\ and\ \bibinfo {author} {\bibfnamefont {L.}~\bibnamefont
  {Maksimov}},\ }\href@noop {} {\bibfield  {journal} {\bibinfo  {journal} {Sov.
  J. Exp. Theor. Phys.}\ }\textbf {\bibinfo {volume} {21}},\ \bibinfo {pages}
  {790} (\bibinfo {year} {1965})}\BibitemShut {NoStop}%
\bibitem [{\citenamefont {J{\'{e}}rome}\ \emph {et~al.}(1967)\citenamefont
  {J{\'{e}}rome}, \citenamefont {Rice},\ and\ \citenamefont
  {Kohn}}]{Jerome1967}%
  \BibitemOpen
  \bibfield  {author} {\bibinfo {author} {\bibfnamefont {D.}~\bibnamefont
  {J{\'{e}}rome}}, \bibinfo {author} {\bibfnamefont {T.~M.}\ \bibnamefont
  {Rice}}, \ and\ \bibinfo {author} {\bibfnamefont {W.}~\bibnamefont {Kohn}},\
  }\href {\doibase 10.1103/PhysRev.158.462} {\bibfield  {journal} {\bibinfo
  {journal} {Phys. Rev.}\ }\textbf {\bibinfo {volume} {158}},\ \bibinfo {pages}
  {462} (\bibinfo {year} {1967})}\BibitemShut {NoStop}%
\bibitem [{\citenamefont {Portengen}\ \emph {et~al.}(1996)\citenamefont
  {Portengen}, \citenamefont {{\"{O}}streich},\ and\ \citenamefont
  {Sham}}]{Portengen1996}%
  \BibitemOpen
  \bibfield  {author} {\bibinfo {author} {\bibfnamefont {T.}~\bibnamefont
  {Portengen}}, \bibinfo {author} {\bibfnamefont {T.}~\bibnamefont
  {{\"{O}}streich}}, \ and\ \bibinfo {author} {\bibfnamefont {L.~J.}\
  \bibnamefont {Sham}},\ }\href {\doibase 10.1103/PhysRevB.54.17452} {\bibfield
   {journal} {\bibinfo  {journal} {Phys. Rev. B}\ }\textbf {\bibinfo {volume}
  {54}},\ \bibinfo {pages} {17452} (\bibinfo {year} {1996})}\BibitemShut
  {NoStop}%
\bibitem [{\citenamefont {Fogler}\ \emph {et~al.}(2014)\citenamefont {Fogler},
  \citenamefont {Butov},\ and\ \citenamefont {Novoselov}}]{Fogler2014a}%
  \BibitemOpen
  \bibfield  {author} {\bibinfo {author} {\bibfnamefont {M.~M.}\ \bibnamefont
  {Fogler}}, \bibinfo {author} {\bibfnamefont {L.~V.}\ \bibnamefont {Butov}}, \
  and\ \bibinfo {author} {\bibfnamefont {K.~S.}\ \bibnamefont {Novoselov}},\
  }\href {\doibase 10.1038/ncomms5555} {\bibfield  {journal} {\bibinfo
  {journal} {Nat. Commun.}\ }\textbf {\bibinfo {volume} {5}},\ \bibinfo {pages}
  {4555} (\bibinfo {year} {2014})}\BibitemShut {NoStop}%
\bibitem [{\citenamefont {Li}\ \emph {et~al.}(2017)\citenamefont {Li},
  \citenamefont {Taniguchi}, \citenamefont {Watanabe}, \citenamefont {Hone},\
  and\ \citenamefont {Dean}}]{Li2017}%
  \BibitemOpen
  \bibfield  {author} {\bibinfo {author} {\bibfnamefont {J.~I.~A.}\
  \bibnamefont {Li}}, \bibinfo {author} {\bibfnamefont {T.}~\bibnamefont
  {Taniguchi}}, \bibinfo {author} {\bibfnamefont {K.}~\bibnamefont {Watanabe}},
  \bibinfo {author} {\bibfnamefont {J.}~\bibnamefont {Hone}}, \ and\ \bibinfo
  {author} {\bibfnamefont {C.~R.}\ \bibnamefont {Dean}},\ }\href
  {https://doi.org/10.1038/nphys4140} {\bibfield  {journal} {\bibinfo
  {journal} {Nature Physics}\ }\textbf {\bibinfo {volume} {13}},\ \bibinfo
  {pages} {751} (\bibinfo {year} {2017})}\BibitemShut {NoStop}%
\bibitem [{\citenamefont {Du}\ \emph {et~al.}(2017)\citenamefont {Du},
  \citenamefont {Li}, \citenamefont {Lou}, \citenamefont {Sullivan},
  \citenamefont {Chang}, \citenamefont {Kono},\ and\ \citenamefont
  {Du}}]{Du.2017}%
  \BibitemOpen
  \bibfield  {author} {\bibinfo {author} {\bibfnamefont {L.}~\bibnamefont
  {Du}}, \bibinfo {author} {\bibfnamefont {X.}~\bibnamefont {Li}}, \bibinfo
  {author} {\bibfnamefont {W.}~\bibnamefont {Lou}}, \bibinfo {author}
  {\bibfnamefont {G.}~\bibnamefont {Sullivan}}, \bibinfo {author}
  {\bibfnamefont {K.}~\bibnamefont {Chang}}, \bibinfo {author} {\bibfnamefont
  {J.}~\bibnamefont {Kono}}, \ and\ \bibinfo {author} {\bibfnamefont {R.-R.}\
  \bibnamefont {Du}},\ }\href {https://doi.org/10.1038/s41467-017-01988-1}
  {\bibfield  {journal} {\bibinfo  {journal} {Nature Communications}\ }\textbf
  {\bibinfo {volume} {8}},\ \bibinfo {pages} {1971} (\bibinfo {year}
  {2017})}\BibitemShut {NoStop}%
\bibitem [{\citenamefont {Lu}\ \emph {et~al.}(2017)\citenamefont {Lu},
  \citenamefont {Kono}, \citenamefont {Larkin}, \citenamefont {Rost},
  \citenamefont {Takayama}, \citenamefont {Boris}, \citenamefont {Keimer},\
  and\ \citenamefont {Takagi}}]{Lu2017}%
  \BibitemOpen
  \bibfield  {author} {\bibinfo {author} {\bibfnamefont {Y.~F.}\ \bibnamefont
  {Lu}}, \bibinfo {author} {\bibfnamefont {H.}~\bibnamefont {Kono}}, \bibinfo
  {author} {\bibfnamefont {T.~I.}\ \bibnamefont {Larkin}}, \bibinfo {author}
  {\bibfnamefont {A.~W.}\ \bibnamefont {Rost}}, \bibinfo {author}
  {\bibfnamefont {T.}~\bibnamefont {Takayama}}, \bibinfo {author}
  {\bibfnamefont {A.~V.}\ \bibnamefont {Boris}}, \bibinfo {author}
  {\bibfnamefont {B.}~\bibnamefont {Keimer}}, \ and\ \bibinfo {author}
  {\bibfnamefont {H.}~\bibnamefont {Takagi}},\ }\href {\doibase
  10.1038/ncomms14408} {\bibfield  {journal} {\bibinfo  {journal} {Nat.
  Commun.}\ }\textbf {\bibinfo {volume} {8}},\ \bibinfo {pages} {1} (\bibinfo
  {year} {2017})}\BibitemShut {NoStop}%
\bibitem [{\citenamefont {Werdehausen}\ \emph {et~al.}(2018)\citenamefont
  {Werdehausen}, \citenamefont {Takayama}, \citenamefont {H{\"o}ppner},
  \citenamefont {Albrecht}, \citenamefont {Rost}, \citenamefont {Lu},
  \citenamefont {Manske}, \citenamefont {Takagi},\ and\ \citenamefont
  {Kaiser}}]{Werdehausen2018}%
  \BibitemOpen
  \bibfield  {author} {\bibinfo {author} {\bibfnamefont {D.}~\bibnamefont
  {Werdehausen}}, \bibinfo {author} {\bibfnamefont {T.}~\bibnamefont
  {Takayama}}, \bibinfo {author} {\bibfnamefont {M.}~\bibnamefont
  {H{\"o}ppner}}, \bibinfo {author} {\bibfnamefont {G.}~\bibnamefont
  {Albrecht}}, \bibinfo {author} {\bibfnamefont {A.~W.}\ \bibnamefont {Rost}},
  \bibinfo {author} {\bibfnamefont {Y.}~\bibnamefont {Lu}}, \bibinfo {author}
  {\bibfnamefont {D.}~\bibnamefont {Manske}}, \bibinfo {author} {\bibfnamefont
  {H.}~\bibnamefont {Takagi}}, \ and\ \bibinfo {author} {\bibfnamefont
  {S.}~\bibnamefont {Kaiser}},\ }\href {\doibase 10.1126/sciadv.aap8652}
  {\bibfield  {journal} {\bibinfo  {journal} {Science Advances}\ }\textbf
  {\bibinfo {volume} {4}},\ \bibinfo {pages} {eaap8652} (\bibinfo {year}
  {2018})}\BibitemShut {NoStop}%
\bibitem [{\citenamefont {Wakisaka}\ \emph {et~al.}(2009)\citenamefont
  {Wakisaka}, \citenamefont {Sudayama}, \citenamefont {Takubo}, \citenamefont
  {Mizokawa}, \citenamefont {Arita}, \citenamefont {Namatame}, \citenamefont
  {Taniguchi}, \citenamefont {Katayama}, \citenamefont {Nohara},\ and\
  \citenamefont {Takagi}}]{Wakisaka.2009}%
  \BibitemOpen
  \bibfield  {author} {\bibinfo {author} {\bibfnamefont {Y.}~\bibnamefont
  {Wakisaka}}, \bibinfo {author} {\bibfnamefont {T.}~\bibnamefont {Sudayama}},
  \bibinfo {author} {\bibfnamefont {K.}~\bibnamefont {Takubo}}, \bibinfo
  {author} {\bibfnamefont {T.}~\bibnamefont {Mizokawa}}, \bibinfo {author}
  {\bibfnamefont {M.}~\bibnamefont {Arita}}, \bibinfo {author} {\bibfnamefont
  {H.}~\bibnamefont {Namatame}}, \bibinfo {author} {\bibfnamefont
  {M.}~\bibnamefont {Taniguchi}}, \bibinfo {author} {\bibfnamefont
  {N.}~\bibnamefont {Katayama}}, \bibinfo {author} {\bibfnamefont
  {M.}~\bibnamefont {Nohara}}, \ and\ \bibinfo {author} {\bibfnamefont
  {H.}~\bibnamefont {Takagi}},\ }\href {\doibase
  10.1103/PhysRevLett.103.026402} {\bibfield  {journal} {\bibinfo  {journal}
  {Phys. Rev. Lett.}\ }\textbf {\bibinfo {volume} {103}},\ \bibinfo {pages}
  {026402} (\bibinfo {year} {2009})}\BibitemShut {NoStop}%
\bibitem [{\citenamefont {Kaneko}\ \emph {et~al.}(2013)\citenamefont {Kaneko},
  \citenamefont {Toriyama}, \citenamefont {Konishi},\ and\ \citenamefont
  {Ohta}}]{Kaneko2013}%
  \BibitemOpen
  \bibfield  {author} {\bibinfo {author} {\bibfnamefont {T.}~\bibnamefont
  {Kaneko}}, \bibinfo {author} {\bibfnamefont {T.}~\bibnamefont {Toriyama}},
  \bibinfo {author} {\bibfnamefont {T.}~\bibnamefont {Konishi}}, \ and\
  \bibinfo {author} {\bibfnamefont {Y.}~\bibnamefont {Ohta}},\ }\href {\doibase
  10.1103/PhysRevB.87.035121} {\bibfield  {journal} {\bibinfo  {journal} {Phys.
  Rev. B}\ }\textbf {\bibinfo {volume} {87}},\ \bibinfo {pages} {035121}
  (\bibinfo {year} {2013})}\BibitemShut {NoStop}%
\bibitem [{\citenamefont {Sugimoto}\ \emph {et~al.}(2018)\citenamefont
  {Sugimoto}, \citenamefont {Nishimoto}, \citenamefont {Kaneko},\ and\
  \citenamefont {Ohta}}]{Sugimoto2018}%
  \BibitemOpen
  \bibfield  {author} {\bibinfo {author} {\bibfnamefont {K.}~\bibnamefont
  {Sugimoto}}, \bibinfo {author} {\bibfnamefont {S.}~\bibnamefont {Nishimoto}},
  \bibinfo {author} {\bibfnamefont {T.}~\bibnamefont {Kaneko}}, \ and\ \bibinfo
  {author} {\bibfnamefont {Y.}~\bibnamefont {Ohta}},\ }\href {\doibase
  10.1103/PhysRevLett.120.247602} {\bibfield  {journal} {\bibinfo  {journal}
  {Phys. Rev. Lett.}\ }\textbf {\bibinfo {volume} {120}},\ \bibinfo {pages}
  {247602} (\bibinfo {year} {2018})}\BibitemShut {NoStop}%
\bibitem [{\citenamefont {Mazza}\ \emph {et~al.}(2020)\citenamefont {Mazza},
  \citenamefont {R\"osner}, \citenamefont {Windg\"atter}, \citenamefont
  {Latini}, \citenamefont {H\"ubener}, \citenamefont {Millis}, \citenamefont
  {Rubio},\ and\ \citenamefont {Georges}}]{Mazza2020}%
  \BibitemOpen
  \bibfield  {author} {\bibinfo {author} {\bibfnamefont {G.}~\bibnamefont
  {Mazza}}, \bibinfo {author} {\bibfnamefont {M.}~\bibnamefont {R\"osner}},
  \bibinfo {author} {\bibfnamefont {L.}~\bibnamefont {Windg\"atter}}, \bibinfo
  {author} {\bibfnamefont {S.}~\bibnamefont {Latini}}, \bibinfo {author}
  {\bibfnamefont {H.}~\bibnamefont {H\"ubener}}, \bibinfo {author}
  {\bibfnamefont {A.~J.}\ \bibnamefont {Millis}}, \bibinfo {author}
  {\bibfnamefont {A.}~\bibnamefont {Rubio}}, \ and\ \bibinfo {author}
  {\bibfnamefont {A.}~\bibnamefont {Georges}},\ }\href {\doibase
  10.1103/PhysRevLett.124.197601} {\bibfield  {journal} {\bibinfo  {journal}
  {Phys. Rev. Lett.}\ }\textbf {\bibinfo {volume} {124}},\ \bibinfo {pages}
  {197601} (\bibinfo {year} {2020})}\BibitemShut {NoStop}%
\bibitem [{\citenamefont {Kogar}\ \emph {et~al.}(2017)\citenamefont {Kogar},
  \citenamefont {Rak}, \citenamefont {Vig}, \citenamefont {Husain},
  \citenamefont {Flicker}, \citenamefont {Joe}, \citenamefont {Venema},
  \citenamefont {MacDougall}, \citenamefont {Chiang}, \citenamefont {Fradkin},
  \citenamefont {{Van Wezel}},\ and\ \citenamefont {Abbamonte}}]{Kogar2017}%
  \BibitemOpen
  \bibfield  {author} {\bibinfo {author} {\bibfnamefont {A.}~\bibnamefont
  {Kogar}}, \bibinfo {author} {\bibfnamefont {M.~S.}\ \bibnamefont {Rak}},
  \bibinfo {author} {\bibfnamefont {S.}~\bibnamefont {Vig}}, \bibinfo {author}
  {\bibfnamefont {A.~A.}\ \bibnamefont {Husain}}, \bibinfo {author}
  {\bibfnamefont {F.}~\bibnamefont {Flicker}}, \bibinfo {author} {\bibfnamefont
  {Y.~I.}\ \bibnamefont {Joe}}, \bibinfo {author} {\bibfnamefont
  {L.}~\bibnamefont {Venema}}, \bibinfo {author} {\bibfnamefont {G.~J.}\
  \bibnamefont {MacDougall}}, \bibinfo {author} {\bibfnamefont {T.~C.}\
  \bibnamefont {Chiang}}, \bibinfo {author} {\bibfnamefont {E.}~\bibnamefont
  {Fradkin}}, \bibinfo {author} {\bibfnamefont {J.}~\bibnamefont {{Van
  Wezel}}}, \ and\ \bibinfo {author} {\bibfnamefont {P.}~\bibnamefont
  {Abbamonte}},\ }\href {\doibase 10.1126/science.aam6432} {\bibfield
  {journal} {\bibinfo  {journal} {Science}\ }\textbf {\bibinfo {volume}
  {358}},\ \bibinfo {pages} {1314} (\bibinfo {year} {2017})}\BibitemShut
  {NoStop}%
\bibitem [{\citenamefont {Cercellier}\ \emph {et~al.}(2007)\citenamefont
  {Cercellier}, \citenamefont {Monney}, \citenamefont {Clerc}, \citenamefont
  {Battaglia}, \citenamefont {Despont}, \citenamefont {Garnier}, \citenamefont
  {Beck}, \citenamefont {Aebi}, \citenamefont {Patthey}, \citenamefont
  {Berger},\ and\ \citenamefont {Forr\'o}}]{Cercellier2007}%
  \BibitemOpen
  \bibfield  {author} {\bibinfo {author} {\bibfnamefont {H.}~\bibnamefont
  {Cercellier}}, \bibinfo {author} {\bibfnamefont {C.}~\bibnamefont {Monney}},
  \bibinfo {author} {\bibfnamefont {F.}~\bibnamefont {Clerc}}, \bibinfo
  {author} {\bibfnamefont {C.}~\bibnamefont {Battaglia}}, \bibinfo {author}
  {\bibfnamefont {L.}~\bibnamefont {Despont}}, \bibinfo {author} {\bibfnamefont
  {M.~G.}\ \bibnamefont {Garnier}}, \bibinfo {author} {\bibfnamefont
  {H.}~\bibnamefont {Beck}}, \bibinfo {author} {\bibfnamefont {P.}~\bibnamefont
  {Aebi}}, \bibinfo {author} {\bibfnamefont {L.}~\bibnamefont {Patthey}},
  \bibinfo {author} {\bibfnamefont {H.}~\bibnamefont {Berger}}, \ and\ \bibinfo
  {author} {\bibfnamefont {L.}~\bibnamefont {Forr\'o}},\ }\href {\doibase
  10.1103/PhysRevLett.99.146403} {\bibfield  {journal} {\bibinfo  {journal}
  {Phys. Rev. Lett.}\ }\textbf {\bibinfo {volume} {99}},\ \bibinfo {pages}
  {146403} (\bibinfo {year} {2007})}\BibitemShut {NoStop}%
\bibitem [{\citenamefont {Kaneko}\ \emph {et~al.}(2018)\citenamefont {Kaneko},
  \citenamefont {Ohta},\ and\ \citenamefont {Yunoki}}]{Kaneko2018}%
  \BibitemOpen
  \bibfield  {author} {\bibinfo {author} {\bibfnamefont {T.}~\bibnamefont
  {Kaneko}}, \bibinfo {author} {\bibfnamefont {Y.}~\bibnamefont {Ohta}}, \ and\
  \bibinfo {author} {\bibfnamefont {S.}~\bibnamefont {Yunoki}},\ }\href
  {\doibase 10.1103/PhysRevB.97.155131} {\bibfield  {journal} {\bibinfo
  {journal} {Phys. Rev. B}\ }\textbf {\bibinfo {volume} {97}},\ \bibinfo
  {pages} {155131} (\bibinfo {year} {2018})}\BibitemShut {NoStop}%
\bibitem [{\citenamefont {Chen}\ \emph {et~al.}(2018)\citenamefont {Chen},
  \citenamefont {Singh}, \citenamefont {Lin},\ and\ \citenamefont
  {Pereira}}]{Chen2018}%
  \BibitemOpen
  \bibfield  {author} {\bibinfo {author} {\bibfnamefont {C.}~\bibnamefont
  {Chen}}, \bibinfo {author} {\bibfnamefont {B.}~\bibnamefont {Singh}},
  \bibinfo {author} {\bibfnamefont {H.}~\bibnamefont {Lin}}, \ and\ \bibinfo
  {author} {\bibfnamefont {V.~M.}\ \bibnamefont {Pereira}},\ }\href {\doibase
  10.1103/PhysRevLett.121.226602} {\bibfield  {journal} {\bibinfo  {journal}
  {Phys. Rev. Lett.}\ }\textbf {\bibinfo {volume} {121}},\ \bibinfo {pages}
  {226602} (\bibinfo {year} {2018})}\BibitemShut {NoStop}%
\bibitem [{\citenamefont {Jia}\ \emph {et~al.}(2020)\citenamefont {Jia},
  \citenamefont {Wang}, \citenamefont {Chiu}, \citenamefont {Song},
  \citenamefont {Yu}, \citenamefont {J.}, \citenamefont {Lei}, \citenamefont
  {Klemenz}, \citenamefont {Cevallos}, \citenamefont {Onyszczak}, \citenamefont
  {Fishchenko}, \citenamefont {Liu}, \citenamefont {Farahi}, \citenamefont
  {Xie}, \citenamefont {Xu}, \citenamefont {Watanabe}, \citenamefont
  {Taniguchi}, \citenamefont {Bernevig}, \citenamefont {Cava}, \citenamefont
  {Schoop}, \citenamefont {Yazdani},\ and\ \citenamefont {Wu}}]{jia.2020}%
  \BibitemOpen
  \bibfield  {author} {\bibinfo {author} {\bibfnamefont {Y.}~\bibnamefont
  {Jia}}, \bibinfo {author} {\bibfnamefont {P.}~\bibnamefont {Wang}}, \bibinfo
  {author} {\bibfnamefont {C.-L.}\ \bibnamefont {Chiu}}, \bibinfo {author}
  {\bibfnamefont {Z.}~\bibnamefont {Song}}, \bibinfo {author} {\bibfnamefont
  {G.}~\bibnamefont {Yu}}, \bibinfo {author} {\bibfnamefont {B.}~\bibnamefont
  {J.}}, \bibinfo {author} {\bibfnamefont {S.}~\bibnamefont {Lei}}, \bibinfo
  {author} {\bibfnamefont {S.}~\bibnamefont {Klemenz}}, \bibinfo {author}
  {\bibfnamefont {F.~A.}\ \bibnamefont {Cevallos}}, \bibinfo {author}
  {\bibfnamefont {M.}~\bibnamefont {Onyszczak}}, \bibinfo {author}
  {\bibfnamefont {N.}~\bibnamefont {Fishchenko}}, \bibinfo {author}
  {\bibfnamefont {X.}~\bibnamefont {Liu}}, \bibinfo {author} {\bibfnamefont
  {G.}~\bibnamefont {Farahi}}, \bibinfo {author} {\bibfnamefont
  {F.}~\bibnamefont {Xie}}, \bibinfo {author} {\bibfnamefont {Y.}~\bibnamefont
  {Xu}}, \bibinfo {author} {\bibfnamefont {K.}~\bibnamefont {Watanabe}},
  \bibinfo {author} {\bibfnamefont {T.}~\bibnamefont {Taniguchi}}, \bibinfo
  {author} {\bibfnamefont {B.~A.}\ \bibnamefont {Bernevig}}, \bibinfo {author}
  {\bibfnamefont {R.~J.}\ \bibnamefont {Cava}}, \bibinfo {author}
  {\bibfnamefont {L.~M.}\ \bibnamefont {Schoop}}, \bibinfo {author}
  {\bibfnamefont {A.}~\bibnamefont {Yazdani}}, \ and\ \bibinfo {author}
  {\bibfnamefont {S.}~\bibnamefont {Wu}},\ }\href@noop {} {\enquote {\bibinfo
  {title} {Evidence for a monolayer excitonic insulator},}\ } (\bibinfo {year}
  {2020}),\ \Eprint {http://arxiv.org/abs/2010.05390} {arXiv:2010.05390
  [cond-mat.mes-hall]} \BibitemShut {NoStop}%
\bibitem [{\citenamefont {Hu}\ \emph {et~al.}(2018)\citenamefont {Hu},
  \citenamefont {Venderbos},\ and\ \citenamefont {Kane}}]{Hu2018}%
  \BibitemOpen
  \bibfield  {author} {\bibinfo {author} {\bibfnamefont {Y.}~\bibnamefont
  {Hu}}, \bibinfo {author} {\bibfnamefont {J.~W.~F.}\ \bibnamefont
  {Venderbos}}, \ and\ \bibinfo {author} {\bibfnamefont {C.~L.}\ \bibnamefont
  {Kane}},\ }\href {\doibase 10.1103/PhysRevLett.121.126601} {\bibfield
  {journal} {\bibinfo  {journal} {Phys. Rev. Lett.}\ }\textbf {\bibinfo
  {volume} {121}},\ \bibinfo {pages} {126601} (\bibinfo {year}
  {2018})}\BibitemShut {NoStop}%
\bibitem [{\citenamefont {Wang}\ \emph {et~al.}(2019)\citenamefont {Wang},
  \citenamefont {Erten}, \citenamefont {Wang},\ and\ \citenamefont
  {Xing}}]{Wang2019}%
  \BibitemOpen
  \bibfield  {author} {\bibinfo {author} {\bibfnamefont {R.}~\bibnamefont
  {Wang}}, \bibinfo {author} {\bibfnamefont {O.}~\bibnamefont {Erten}},
  \bibinfo {author} {\bibfnamefont {B.}~\bibnamefont {Wang}}, \ and\ \bibinfo
  {author} {\bibfnamefont {D.~Y.}\ \bibnamefont {Xing}},\ }\href {\doibase
  10.1038/s41467-018-08203-9} {\bibfield  {journal} {\bibinfo  {journal} {Nat.
  Commun.}\ }\textbf {\bibinfo {volume} {10}},\ \bibinfo {pages} {1} (\bibinfo
  {year} {2019})}\BibitemShut {NoStop}%
\bibitem [{\citenamefont {Hu}\ \emph {et~al.}(2020)\citenamefont {Hu},
  \citenamefont {Zhang}, \citenamefont {Zhang},\ and\ \citenamefont
  {Wu}}]{Hu2019}%
  \BibitemOpen
  \bibfield  {author} {\bibinfo {author} {\bibfnamefont {L.-H.}\ \bibnamefont
  {Hu}}, \bibinfo {author} {\bibfnamefont {R.-X.}\ \bibnamefont {Zhang}},
  \bibinfo {author} {\bibfnamefont {F.-C.}\ \bibnamefont {Zhang}}, \ and\
  \bibinfo {author} {\bibfnamefont {C.}~\bibnamefont {Wu}},\ }\href {\doibase
  10.1103/PhysRevB.102.235115} {\bibfield  {journal} {\bibinfo  {journal}
  {Phys. Rev. B}\ }\textbf {\bibinfo {volume} {102}},\ \bibinfo {pages}
  {235115} (\bibinfo {year} {2020})}\BibitemShut {NoStop}%
\bibitem [{\citenamefont {Varsano}\ \emph {et~al.}(2020)\citenamefont
  {Varsano}, \citenamefont {Palummo}, \citenamefont {Molinari},\ and\
  \citenamefont {Rontani}}]{Varsano.2020}%
  \BibitemOpen
  \bibfield  {author} {\bibinfo {author} {\bibfnamefont {D.}~\bibnamefont
  {Varsano}}, \bibinfo {author} {\bibfnamefont {M.}~\bibnamefont {Palummo}},
  \bibinfo {author} {\bibfnamefont {E.}~\bibnamefont {Molinari}}, \ and\
  \bibinfo {author} {\bibfnamefont {M.}~\bibnamefont {Rontani}},\ }\href
  {https://doi.org/10.1038/s41565-020-0650-4} {\bibfield  {journal} {\bibinfo
  {journal} {Nature Nanotechnology}\ }\textbf {\bibinfo {volume} {15}},\
  \bibinfo {pages} {367} (\bibinfo {year} {2020})}\BibitemShut {NoStop}%
\bibitem [{\citenamefont {Perfetto}\ and\ \citenamefont
  {Stefanucci}(2020)}]{Perfetto2020}%
  \BibitemOpen
  \bibfield  {author} {\bibinfo {author} {\bibfnamefont {E.}~\bibnamefont
  {Perfetto}}\ and\ \bibinfo {author} {\bibfnamefont {G.}~\bibnamefont
  {Stefanucci}},\ }\href {\doibase 10.1103/PhysRevLett.125.106401} {\bibfield
  {journal} {\bibinfo  {journal} {Phys. Rev. Lett.}\ }\textbf {\bibinfo
  {volume} {125}},\ \bibinfo {pages} {106401} (\bibinfo {year}
  {2020})}\BibitemShut {NoStop}%
\bibitem [{\citenamefont {Hou}\ \emph {et~al.}(2019)\citenamefont {Hou},
  \citenamefont {Wang}, \citenamefont {Xiao}, \citenamefont {McClintock},
  \citenamefont {Clark~Travaglini}, \citenamefont {Paulus~Francia},
  \citenamefont {Fetsch}, \citenamefont {Erten}, \citenamefont {Savrasov},
  \citenamefont {Wang}, \citenamefont {Rossi}, \citenamefont {Vishik},
  \citenamefont {Rotenberg},\ and\ \citenamefont {Yu}}]{Hou.2019}%
  \BibitemOpen
  \bibfield  {author} {\bibinfo {author} {\bibfnamefont {Y.}~\bibnamefont
  {Hou}}, \bibinfo {author} {\bibfnamefont {R.}~\bibnamefont {Wang}}, \bibinfo
  {author} {\bibfnamefont {R.}~\bibnamefont {Xiao}}, \bibinfo {author}
  {\bibfnamefont {L.}~\bibnamefont {McClintock}}, \bibinfo {author}
  {\bibfnamefont {H.}~\bibnamefont {Clark~Travaglini}}, \bibinfo {author}
  {\bibfnamefont {J.}~\bibnamefont {Paulus~Francia}}, \bibinfo {author}
  {\bibfnamefont {H.}~\bibnamefont {Fetsch}}, \bibinfo {author} {\bibfnamefont
  {O.}~\bibnamefont {Erten}}, \bibinfo {author} {\bibfnamefont {S.~Y.}\
  \bibnamefont {Savrasov}}, \bibinfo {author} {\bibfnamefont {B.}~\bibnamefont
  {Wang}}, \bibinfo {author} {\bibfnamefont {A.}~\bibnamefont {Rossi}},
  \bibinfo {author} {\bibfnamefont {I.}~\bibnamefont {Vishik}}, \bibinfo
  {author} {\bibfnamefont {E.}~\bibnamefont {Rotenberg}}, \ and\ \bibinfo
  {author} {\bibfnamefont {D.}~\bibnamefont {Yu}},\ }\href {\doibase
  10.1038/s41467-019-13711-3} {\bibfield  {journal} {\bibinfo  {journal}
  {Nature Communications}\ }\textbf {\bibinfo {volume} {10}},\ \bibinfo {pages}
  {5723} (\bibinfo {year} {2019})}\BibitemShut {NoStop}%
\bibitem [{\citenamefont {Sun}\ and\ \citenamefont
  {Millis}(2020{\natexlab{b}})}]{sun2020BS}%
  \BibitemOpen
  \bibfield  {author} {\bibinfo {author} {\bibfnamefont {Z.}~\bibnamefont
  {Sun}}\ and\ \bibinfo {author} {\bibfnamefont {A.~J.}\ \bibnamefont
  {Millis}},\ }\href {\doibase 10.1103/PhysRevB.102.041110} {\bibfield
  {journal} {\bibinfo  {journal} {Phys. Rev. B}\ }\textbf {\bibinfo {volume}
  {102}},\ \bibinfo {pages} {041110(R)} (\bibinfo {year}
  {2020}{\natexlab{b}})}\BibitemShut {NoStop}%
\bibitem [{\citenamefont {Thouless}(1983)}]{Thouless1983}%
  \BibitemOpen
  \bibfield  {author} {\bibinfo {author} {\bibfnamefont {D.~J.}\ \bibnamefont
  {Thouless}},\ }\href {\doibase 10.1103/PhysRevB.27.6083} {\bibfield
  {journal} {\bibinfo  {journal} {Phys. Rev. B}\ }\textbf {\bibinfo {volume}
  {27}},\ \bibinfo {pages} {6083} (\bibinfo {year} {1983})}\BibitemShut
  {NoStop}%
\bibitem [{\citenamefont {Rice}\ and\ \citenamefont {Mele}(1982)}]{Rice1982}%
  \BibitemOpen
  \bibfield  {author} {\bibinfo {author} {\bibfnamefont {M.~J.}\ \bibnamefont
  {Rice}}\ and\ \bibinfo {author} {\bibfnamefont {E.~J.}\ \bibnamefont
  {Mele}},\ }\href {\doibase 10.1103/PhysRevLett.49.1455} {\bibfield  {journal}
  {\bibinfo  {journal} {Phys. Rev. Lett.}\ }\textbf {\bibinfo {volume} {49}},\
  \bibinfo {pages} {1455} (\bibinfo {year} {1982})}\BibitemShut {NoStop}%
\bibitem [{\citenamefont {King-Smith}\ and\ \citenamefont
  {Vanderbilt}(1993)}]{King-Smith1993}%
  \BibitemOpen
  \bibfield  {author} {\bibinfo {author} {\bibfnamefont {R.~D.}\ \bibnamefont
  {King-Smith}}\ and\ \bibinfo {author} {\bibfnamefont {D.}~\bibnamefont
  {Vanderbilt}},\ }\href {\doibase 10.1103/PhysRevB.47.1651} {\bibfield
  {journal} {\bibinfo  {journal} {Phys. Rev. B}\ }\textbf {\bibinfo {volume}
  {47}},\ \bibinfo {pages} {1651(R)} (\bibinfo {year} {1993})}\BibitemShut
  {NoStop}%
\bibitem [{\citenamefont {Zhang}\ \emph {et~al.}(2020)\citenamefont {Zhang},
  \citenamefont {Gao},\ and\ \citenamefont {Xiao}}]{Zhang.2020}%
  \BibitemOpen
  \bibfield  {author} {\bibinfo {author} {\bibfnamefont {Y.}~\bibnamefont
  {Zhang}}, \bibinfo {author} {\bibfnamefont {Y.}~\bibnamefont {Gao}}, \ and\
  \bibinfo {author} {\bibfnamefont {D.}~\bibnamefont {Xiao}},\ }\href {\doibase
  10.1103/PhysRevB.101.041410} {\bibfield  {journal} {\bibinfo  {journal}
  {Phys. Rev. B}\ }\textbf {\bibinfo {volume} {101}},\ \bibinfo {pages}
  {041410(R)} (\bibinfo {year} {2020})}\BibitemShut {NoStop}%
\bibitem [{SI()}]{SI}%
  \BibitemOpen
  \href@noop {} {}\bibinfo {note} {See Supplemental Material for details of
  derivation.}\BibitemShut {Stop}%
\bibitem [{\citenamefont {Murakami}\ \emph {et~al.}(2020)\citenamefont
  {Murakami}, \citenamefont {Gole\ifmmode~\check{z}\else \v{z}\fi{}},
  \citenamefont {Kaneko}, \citenamefont {Koga}, \citenamefont {Millis},\ and\
  \citenamefont {Werner}}]{Murakami.2020}%
  \BibitemOpen
  \bibfield  {author} {\bibinfo {author} {\bibfnamefont {Y.}~\bibnamefont
  {Murakami}}, \bibinfo {author} {\bibfnamefont {D.}~\bibnamefont
  {Gole\ifmmode~\check{z}\else \v{z}\fi{}}}, \bibinfo {author} {\bibfnamefont
  {T.}~\bibnamefont {Kaneko}}, \bibinfo {author} {\bibfnamefont
  {A.}~\bibnamefont {Koga}}, \bibinfo {author} {\bibfnamefont {A.~J.}\
  \bibnamefont {Millis}}, \ and\ \bibinfo {author} {\bibfnamefont
  {P.}~\bibnamefont {Werner}},\ }\href {\doibase 10.1103/PhysRevB.101.195118}
  {\bibfield  {journal} {\bibinfo  {journal} {Phys. Rev. B}\ }\textbf {\bibinfo
  {volume} {101}},\ \bibinfo {pages} {195118} (\bibinfo {year}
  {2020})}\BibitemShut {NoStop}%
\bibitem [{\citenamefont {Resta}(1994)}]{Resta1994}%
  \BibitemOpen
  \bibfield  {author} {\bibinfo {author} {\bibfnamefont {R.}~\bibnamefont
  {Resta}},\ }\href {\doibase 10.1103/RevModPhys.66.899} {\bibfield  {journal}
  {\bibinfo  {journal} {Rev. Mod. Phys.}\ }\textbf {\bibinfo {volume} {66}},\
  \bibinfo {pages} {899} (\bibinfo {year} {1994})}\BibitemShut {NoStop}%
\bibitem [{\citenamefont {Goldstone}\ and\ \citenamefont
  {Wilczek}(1981)}]{Goldstone1981}%
  \BibitemOpen
  \bibfield  {author} {\bibinfo {author} {\bibfnamefont {J.}~\bibnamefont
  {Goldstone}}\ and\ \bibinfo {author} {\bibfnamefont {F.}~\bibnamefont
  {Wilczek}},\ }\href {\doibase 10.1103/PhysRevLett.47.986} {\bibfield
  {journal} {\bibinfo  {journal} {Phys. Rev. Lett.}\ }\textbf {\bibinfo
  {volume} {47}},\ \bibinfo {pages} {986} (\bibinfo {year} {1981})}\BibitemShut
  {NoStop}%
\bibitem [{\citenamefont {Yang}\ \emph {et~al.}(2020)\citenamefont {Yang},
  \citenamefont {Xu},\ and\ \citenamefont {Wu}}]{Yang.2020}%
  \BibitemOpen
  \bibfield  {author} {\bibinfo {author} {\bibfnamefont {W.}~\bibnamefont
  {Yang}}, \bibinfo {author} {\bibfnamefont {C.}~\bibnamefont {Xu}}, \ and\
  \bibinfo {author} {\bibfnamefont {C.}~\bibnamefont {Wu}},\ }\href {\doibase
  10.1103/PhysRevResearch.2.042047} {\bibfield  {journal} {\bibinfo  {journal}
  {Phys. Rev. Research}\ }\textbf {\bibinfo {volume} {2}},\ \bibinfo {pages}
  {042047} (\bibinfo {year} {2020})}\BibitemShut {NoStop}%
\bibitem [{\citenamefont {Altland}\ and\ \citenamefont
  {Simons}(2010)}]{Altland.2010}%
  \BibitemOpen
  \bibfield  {author} {\bibinfo {author} {\bibfnamefont {A.}~\bibnamefont
  {Altland}}\ and\ \bibinfo {author} {\bibfnamefont {B.~D.}\ \bibnamefont
  {Simons}},\ }\href {\doibase 10.1017/CBO9780511789984} {\emph {\bibinfo
  {title} {{Condensed Matter Field Theory}}}}\ (\bibinfo  {publisher}
  {Cambridge University Press},\ \bibinfo {address} {Cambridge},\ \bibinfo
  {year} {2010})\BibitemShut {NoStop}%
\bibitem [{\citenamefont {Sun}\ \emph {et~al.}(2020)\citenamefont {Sun},
  \citenamefont {Fogler}, \citenamefont {Basov},\ and\ \citenamefont
  {Millis}}]{Sun2020a}%
  \BibitemOpen
  \bibfield  {author} {\bibinfo {author} {\bibfnamefont {Z.}~\bibnamefont
  {Sun}}, \bibinfo {author} {\bibfnamefont {M.~M.}\ \bibnamefont {Fogler}},
  \bibinfo {author} {\bibfnamefont {D.~N.}\ \bibnamefont {Basov}}, \ and\
  \bibinfo {author} {\bibfnamefont {A.~J.}\ \bibnamefont {Millis}},\ }\href
  {\doibase 10.1103/PhysRevResearch.2.023413} {\bibfield  {journal} {\bibinfo
  {journal} {Phys. Rev. Research}\ }\textbf {\bibinfo {volume} {2}},\ \bibinfo
  {pages} {023413} (\bibinfo {year} {2020})}\BibitemShut {NoStop}%
\bibitem [{\citenamefont {Wittig}(2005)}]{Wittig.2005}%
  \BibitemOpen
  \bibfield  {author} {\bibinfo {author} {\bibfnamefont {C.}~\bibnamefont
  {Wittig}},\ }\href {\doibase 10.1021/jp040627u} {\bibfield  {journal}
  {\bibinfo  {journal} {J. Phys. Chem. B}\ }\textbf {\bibinfo {volume} {109}},\
  \bibinfo {pages} {8428} (\bibinfo {year} {2005})}\BibitemShut {NoStop}%
\bibitem [{\citenamefont {Barankov}\ \emph {et~al.}(2004)\citenamefont
  {Barankov}, \citenamefont {Levitov},\ and\ \citenamefont
  {Spivak}}]{Barankov.2004}%
  \BibitemOpen
  \bibfield  {author} {\bibinfo {author} {\bibfnamefont {R.~A.}\ \bibnamefont
  {Barankov}}, \bibinfo {author} {\bibfnamefont {L.~S.}\ \bibnamefont
  {Levitov}}, \ and\ \bibinfo {author} {\bibfnamefont {B.~Z.}\ \bibnamefont
  {Spivak}},\ }\href {\doibase 10.1103/PhysRevLett.93.160401} {\bibfield
  {journal} {\bibinfo  {journal} {Phys. Rev. Lett.}\ }\textbf {\bibinfo
  {volume} {93}},\ \bibinfo {pages} {160401} (\bibinfo {year}
  {2004})}\BibitemShut {NoStop}%
\end{thebibliography}%

\pagebreak
\widetext
\begin{center}
	\textbf{\large Supplemental Material for `Topological charge pumping in excitonic insulators'}
\end{center}
\setcounter{equation}{0}
\setcounter{figure}{0}
\setcounter{table}{0}
\setcounter{page}{1}
\makeatletter
\renewcommand{\theequation}{S\arabic{equation}}
\renewcommand{\thefigure}{S\arabic{figure}}
\renewcommand{\bibnumfmt}[1]{[S#1]}

 \tableofcontents

\section{The Hamiltonian}
\label{app:Hamiltonian}
We base our discussion on  the two-band spinless Fermion Hamiltonian  that is a minimal model for excitonic insulators:
\begin{align}
	H =& \int {dr} \left[ \psi^\dagger
	\left(
	\sigma_3 \xi_{p-A}  + \phi_0 
	\right)
	\psi \right]
	+ 
	\int {dr dr^\prime} V(r-r^\prime) \rho(r) \rho(r^\prime)
	\label{eqn:hamiltonian}
\end{align}
where $\rho(r)=\psi^\dagger(r) \psi(r)$ is the local density at position $r$, $\psi^\dagger=(\psi_c^\dagger , \, \psi_v^\dagger)$ is the two component electron creation operator with c/v labeling the conduction/valence band, $\xi_p=\varepsilon_p-\mu$ is the kinetic energy shifted by the chemical potential, $p=-i\hbar\nabla$, $\sigma_i$ are the Pauli matrices in c-v space, $(\phi_0,\, A) \equiv A_\mu$ is the electromagnetic (EM) potential and we have set the electron charge, the speed of light and Planck constant $\hbar$ to be $1$. In the non interacting case, the overlap of the bands gives rise to an electron and a hole pocket, each with the Fermi momentum $k_F$, Fermi velocity $v_F$,  Fermi level density of states $\nu$ and carrier density $n/2$. We choose the gauge $\phi_0=0$ in this paper. 

The interaction is assumed to be density-density type, e.g., $V(r)=1/|r|$ for the Coulomb interaction. We denote its Fourier transform at momentum $q$ as $V_q$. The repulsive interaction is effectively attractive  between electrons and holes and can induce pairing in several  angular momentum channels, in formal analogy to Cooper pairing in superconductors. We write the model as a fermionic path integral   so the partition function is  $Z=\int  D[\bar{\psi},\psi] e^{\psi^\dagger\partial_\tau\psi-H[\psi]}$ and  decouple the interaction in the electron hole pairing channel: $Z
=\int  D[\bar{\psi},\psi] D[\bar{\Delta},\Delta] e^{-S[\psi,\Delta]}$. Note that `$D$' means functional field integral. The Hubbard-Stratonovich fields $\Delta$ then represent the order parameters. 
The interband interaction term in \equa{eqn:hamiltonian} can be decomposed into symmetry channels labeled by $l$ as $\hat{V}_{\text{inter}}=\sum_l g_l \hat{b}_l^\dagger \hat{b}_l$ where $\hat{b}_l =\sum_k f^l_k \psi^\dagger_{c,k} \psi_{v,k}$ and $f^l_k$ are the representation functions (pairing functions) of the point symmetry group.
Therefore, we can resolve  $\Delta$ into pairing functions as $\Delta_k=\sum_l\Delta_l f^l_k$. We assume for notational simplicity that the excitonic effects occur near a high symmetry point so that lattice effects are unimportant and that the interaction effects may be restricted to the Fermi surface. In this case $f^l$ become the usual d-dimensional rotational harmonics and the interaction $V_q$ is parameterized by the momentum transfer $q$ connecting two points on the Fermi surface.   We focus on $s$ pairing ($f^s_k$ has the full point symmetry of the lattice; we take $f^s_k=1$) and $p_x$ pairing $f^p_k=k_x/k_F$. We denote the latter as $f_k$ for notational simplicity.

The Hubbard Stratonovich transformed action is
\begin{align}
	S[\psi,\Delta_s,\Delta_p, A] = \int d\tau dr 
	\Bigg\{
	&
	\psi^\dagger 
	\left( \partial_\tau +
	H_m
	\right)
	\psi
	+ \frac{1}{g_s}|\Delta_s|^2  + \frac{1}{g_p}|\Delta_p|^2 
	\Bigg\}
	\,.
	\label{eqn:HS_action_SI}
\end{align}
Since the overall phase of the Hubbard Stratonovich field is not relevant for our considerations, we choose $\Delta_s$ to be real. As we will see in Sec.~\ref{app:exactmeanfield_pseudo_spin}, due to a `particle hole' symmetry in the BCS weak coupling case, the main effect of an electric field is to induce a $p$-wave field that is $\pi/2$ out of phase with $\Delta_s$. We restrict our attention to this case.
The mean field Hamiltonian is thus 
$
H_m^k=\xi_{k} \sigma_3 + \Delta_s\sigma_1+ \Delta_p f_k \sigma_2
$ 
with real $\Delta_s,\, \Delta_p$
and the EM vector potential enters as $k\rightarrow k-A$. 
Note that minimal coupling substitution is also applied to the $p$-wave decoupling term: $\Delta_p f_{k-A}$ although this term comes from the electron-electron interaction that contains no EM field. We discuss this choice here in terms of local gauge invariance.

In the full functional integral, the general gauge invariant  form of the decoupling term is $e^{-i\int_{r_1}^{r_2}dl A(l)}\Delta(r_1, r_2)\psi^\dagger_v(r_1)\psi_c(r_2)$, which preserves its form under the usual local gauge transformation  $U_g: \psi(r) \rightarrow \psi(r) e^{i \theta(r)},\,  A_\mu \rightarrow A_\mu + \partial_\mu \theta(r)$.  We write $\Delta(r_1,r_2)=|\Delta(r_1, r_2)|e^{i \varphi (r_1, r_2)}$; both the amplitude $|\Delta(r_1, r_2)|$ and the phase $\varphi (r_1, r_2)$ are dynamical variables.   The dependence on `center of mass' coordinate $r=(r_1+r_2)/2$ gives the spatial variation of the order parameter while the dependence on $r_1-r_2$ gives the internal structure of the electron hole pair (the momentum dependence of the pairing function). Writing the phase degrees of freedom as $\varphi (r_1, r_2) =  \varphi_0(r) +  (r_1-r_2)\alpha(r)$ in the slow varying limit,  the $\varphi_0$ is the usual order parameter phase and the combination $\alpha+A$ enters structure of pairing wave function as $f_{k-A-\alpha}$. Although $\alpha=0$ in the initial ground state,  one should in principle track the dynamics of $\alpha$. In the weak coupling BCS limit, we may neglect the dynamics of $\alpha$ since its appearance is equivalent to a $\Delta_s$ in the $\sigma_2$ channel which vanishes as we will prove in Sec.~\ref{app:exactmeanfield}. Even in the general case when the appearance of $\alpha$ must be considered at intermediate stages of the dynamics,  the amount of charge pumping during a full cycle is still the quantized value given that the system finally returns to its initial state, as shown in general by Thouless \cite{Thouless1983}.

If the two bands are formed by atomic orbitals having different parities, e.g., p and d orbitals, an interband dipolar moment $D_0$ adds the extra term $D_0 E \psi^\dagger \sigma_1 \psi$ to the Hamiltonian. This term also contributes to the EM response due to change of inter orbital hybridization. However, for a full cycle of order parameter dynamics, the amount of pumped charge  won't be affected if the initial and final states are the same such that they have the same inter orbital hybridization. The dynamics itself won't be qualitatively affected if the interband dipole $D_0$ is not large compared to the dipole formed between $s$ and $p$-symmetry electron and hole bound states that produce the order parameters. This is true in the BCS case since the former is proportional to the size of atomic orbitals while the latter is the size $\xi^0$ of the extended electron hole bound state.

\subsection{The pairing interactions}
For isotropic systems in $d$ dimensions, the pairing interaction in channel $l$ is 
\begin{align}
	g_l= \frac{1}{2|f^l|^2} \int d\Omega_1 d\Omega_2 f^{l \ast}_{k(\Omega_1)} V_{ k(\Omega_1) -k(\Omega_2) } f^l_{k(\Omega_2)}
	\label{eqn:gl}
\end{align}
where $\Omega$ is the angular variable on the $d-1$ dimensional Fermi surface,  $k(\Omega)$ is the electron momentum at angle $\Omega$ and $|f^l|^2=\int d\Omega |f^l_{k(\Omega)}|^2$ is the normalization factor. Since the symmetry group is $O(d)$, the $d$-dimensional rotations and inversions, the pairing functions $f^l$ are $d$-dimensional spherical harmonics. Below we discuss 1D and 2D separately.

\emph{One dimension---}There are only two pairing functions: the inversion symmetric one $f^s_{k_F}=1,\, f^s_{-k_F}=1$ and the antisymmetric one $f^p_{k_F}=1,\, f^p_{-k_F}=-1$.  Thus \equa{eqn:gl} leads to $g_s= V_{q=0}+  V_{q=2k_F}$ and $g_p= V_{q=0} - V_{q=2k_F}$.
Since the Coulumb interaction in 1D is $V_q \sim \ln(aq)$ where $a$ is a short distance cutoff, $V_{q=0}$ diverges Logarithmically which results in $g_s=g_p$. If the 1D system is put parallel to a metallic gate at a distance $h$, the screening kills the divergence of $V_{q=0}$ and renders $g_p/g_s<1$. As $h$ decreases, the screening get stronger and $g_p/g_s$ decreases continuously until reaching zero at $h\ll 1/k_F$. Therefore, $g_p/g_s$ is experimentally tunable by $h$. 


\emph{Two dimensions---}Examples are electron hole bilayers and other 2D excitonic insulators such as monolayer WTe$_2$ \cite{jia.2020}. The natural pairing functions are $f_k^l=e^{il \theta_k}$ which plugged into \equa{eqn:gl} renders $g_l= \frac{1}{2\pi} \int d\theta_k \cos(l\theta_k) V_{2 k_F \sin(\theta_k/2) }$. Note that for $l=0$, the $1/2\pi$ factor should be changed to $1/4\pi$. Since $e^{il \theta_k}$ and $e^{-il \theta_k}$ have the same $g_l$, we will use the real pairing functions $f_k^l=\cos(l\theta_k), \sin(l\theta_k)$ for convenience. For Thomas-Fermi screened interaction  $V_q= \frac{2\pi}{\epsilon (q +q_{\text{TF}})}$ in 2D where $q_{\text{TF}}/(2k_F)=e^2/(\epsilon \hbar v_F) \equiv \alpha$ is the `fine structure constant' in this system and $\epsilon$ is the dielectric constant of the environment, the $s$-wave pairing strength is 
\begin{align}
	\nu g_s&=\nu \frac{1}{4\pi} \int d\theta_k \frac{2\pi}{2 k_F |\sin(\theta_k/2)| + q_{\text{TF}} } 
	=\frac{\alpha}{\sqrt{1-\alpha^2}}
	\frac{1}{\pi}
	\mathrm{Tanh}^{-1}\left( \sqrt{1-\alpha^2} \right)
	\label{eqn:g_s}
\end{align} 
and the $p$-wave one is
\begin{align}
	\nu g_p&=\nu \frac{1}{2\pi} \int d\theta_k \frac{2\pi \cos\theta_k}{2 k_F |\sin(\theta_k/2)| + q_{\text{TF}} } 
	\notag\\
	&= \alpha
	\Bigg[-\frac{4}{\pi} + 2 \alpha + \frac{4}{\pi}
	\frac{\left(1-2\alpha^2\right)}{\sqrt{1-\alpha^2}} 
	\Bigg( 
	\mathrm{Tanh}^{-1} \left(\sqrt{1-\alpha^2} \right)
	- \mathrm{Tanh}^{-1}\left(\frac{\sqrt{1-\alpha^2}}{1+\alpha} \right)
	\Bigg) \Bigg]
	\label{eqn:g_p}
	\,.
\end{align} 
These equations were previously given \cite{sun2020BS} and are reproduced here for convenience.

\begin{figure}
	\includegraphics[width= 0.7\linewidth]{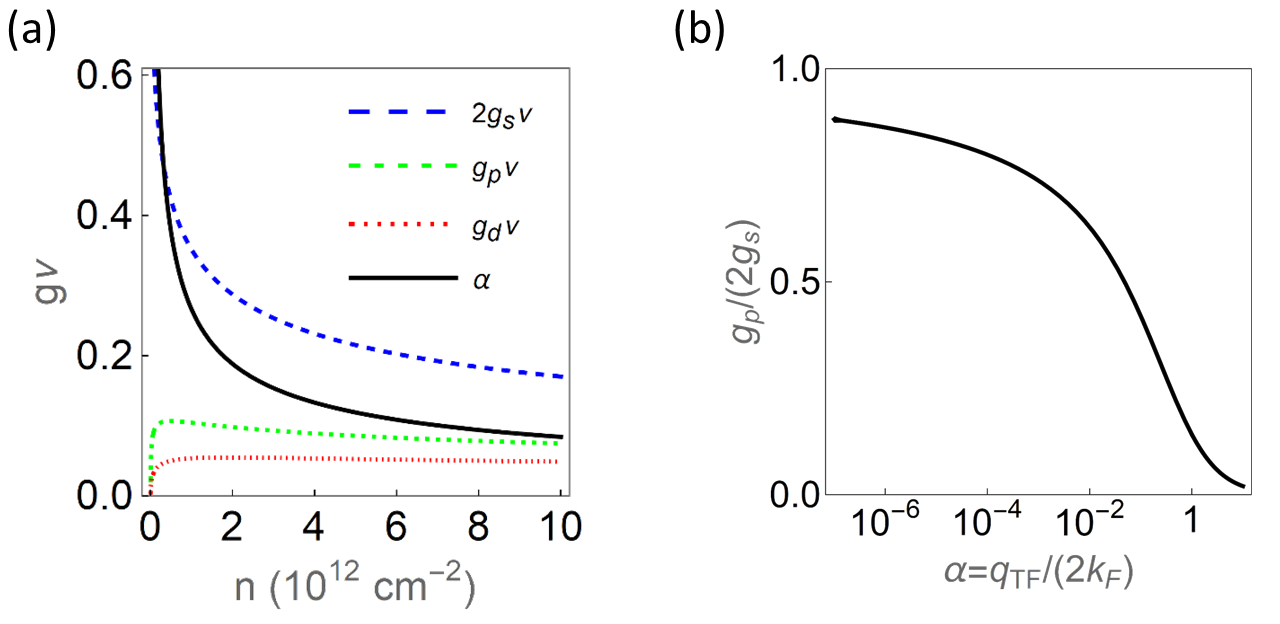}
	\caption{Pairing interaction plots reproduced from Ref.~\cite{sun2020BS}. (a) The $s$,$p$,$d$-wave components of the screened Coulomb interaction in 2D and the `fine structure constant' $\alpha=e^2/(\epsilon \hbar v_F) = q_{\text{TF}}/(2 k_F)$ as functions of electron density $n/2=m^2 v_F^2/(4\pi\hbar^2)$ computed from Eqs.~\eqref{eqn:g_s} and \eqref{eqn:g_p} using $m=0.05 m_e$ and $\epsilon=10$ where $m$ is the electron mass in the model and $m_e$ is the vacuum electron mass. (b) The ratio $g_p/(2g_s)$ as a function of $\alpha=q_{\text{TF}}/(2k_F)$.  For $\alpha \ll 1$, i.e., in the high density case, $g_p/(2g_s)$ becomes considerable and approaches one in the high density limit.  Spin degeneracy is neglected.}
	\label{fig:gs_gp}
\end{figure}

The pairing interactions are shown in Fig.~\ref{fig:gs_gp} for the screened Coulomb interaction in 2D (reproduced from the supplemental material of Ref.~\cite{sun2020BS}).
To obtain a substantial $g_p/(2 g_s)$, one needs the high density case where the fermi velocity is large so that the Thomas fermi wave vector is smaller than the fermi momentum: $q_{\text{TF}}/(2k_F)=\alpha=e^2/(\epsilon \hbar v_F) \ll 1$. Stronger dielectric screening of the environment can further reduce $\alpha$ and increase $g_p/(2 g_s)$. Moreover, a non-negligible interlayer distance $a$ changes the bare electron-hole Coulomb attraction into $V(r)=1/\sqrt{r^2+a^2}$, making it more nonlocal and thus can lead to a larger  $g_p/(2 g_s)$. Other types of interactions such as nearest neighbor Hubbard interaction (although originating from Coulomb) could give a very strong $g_p$, given that the band overlapping is suitable.

\section{Computation of the polarization }
\label{app:polarization}
In this section, we explicitly derive the  charge pumping in an 1D excitonic insulator by computing  the polarization $P$ (charge pumped) as a time integral of the  current $J$ induced by adiabatic changes to the order parameter over a time interval from $0$ to $t$ and comparing the result to the formula in terms of Berry curvature, consistent with previous results \cite{King-Smith1993}.

\subsection{Polarization}

For convenience we reproduce the mean field Hamiltonian $H_m^k$ and current operator $J$ here:
\begin{align}
	H_m^k=\xi_k\sigma_3+\Delta_s\sigma_1+\Delta_p f_k\sigma_2
	\,,\quad
	\hat{j}_k=\psi_k^\dagger j_k \psi_k
	=\psi_k^\dagger \left( \partial_k H_m^k \right) \psi_k
	=\psi_k^\dagger \left( v_k \sigma_3 +\Delta_p \partial_k f_k \sigma_2 \right) \psi_k
	\,,\quad
	J=\sum_k \hat{j}_k
	\label{eqn:current}
\end{align}
where $v_k=\partial_k\xi_k$ is the velocity and the energy eigenvalues are $ E_k=\pm\sqrt{\xi_k^2+\Delta_s^2+\Delta_p^2f^{2}_k}$.
The current $\delta J$ in response to a change in order parameter $\delta \Delta_k=\delta \Delta_s(t)\sigma_1+\delta \Delta_p(t) f_k \sigma_2$ is 
\begin{equation}
	\delta J(\Omega_n)=-T\sum_{\omega_n}\sum_k \mathrm{Tr}\left[\frac{j_k\left(i\omega_n+i\Omega_n-H_m^k\right)\delta \Delta_k\left(i\omega-H_m^k\right)}{\left(\left(\omega_n+\Omega_n\right)^2+E_k^2\right)\left(\omega_n^2+E_k^2\right)}\right]
	\label{eqn:current_response_SI}
\end{equation}
in frequency representation where $\omega_n= (2n+1) \pi T$, $\Omega_n= 2n \pi T$  are the Fermion and Boson Matsubara frequencies with $n\in \mathbb{Z}$ (not to be confused with the carrier density), $T$ is the temperature and we have set the Boltzmann constant to be $1$.
Carrying out the trace over band indices, performing the frequency summation at $T=0$ and analytically continuing $i\Omega_n$ to $\omega$, one obtains
\begin{equation}
	\delta J(\omega)=\frac{i \omega}{2}\left(\sum_k\frac{v_kf_k-\xi_k\partial_kf_k}{E_k\left(-\frac{\omega^2}{4}+E_k^2\right)}\Delta_p\delta\Delta_s(\omega) -
	\sum_k\frac{v_kf_k}{{E_k\left(-\frac{\omega^2}{4}+E_k^2\right)}}\Delta_s \delta\Delta_p(\omega)\right)
	\label{eqn:J2}
\end{equation}
Note that if the argument is $\omega$ in $\Delta_{s,p}(\omega)$, the latter means  $\Delta_{s,p}(t)$ Fourier transformed into frequency representation.
In the adiabatic limit  we may neglect the $\omega^2$ in the denominators; then transforming to the time domain we obtain
\begin{equation}
	J(t)=-\frac{1}{2}\left(\sum_k\frac{v_kf_k-\xi_k\partial_kf_k}{E_k^3}\Delta_p\frac{\partial\Delta_s}{\partial t}-\sum_k\frac{v_kf_k}{{E_k^3}}\Delta_s\frac{\partial \Delta_p}{\partial t}\right)
	\,
	\label{eqn:Jfinal1}
\end{equation}
where `$J$' now means the expectation value of the current operator.
Integrating in time gives the change in polarization:
\begin{equation}
	P= \int \left( -\frac{\Delta_p\left(v_kf_k-\xi_k\partial_kf_k\right)}{2E_k^3}\frac{d\Delta_sdk}{2\pi}+\frac{\Delta_sv_kf_k}{2E_k^3}\frac{d\Delta_pdk}{2\pi}
	\right)
	\,.
	\label{eqn:Jfinal2}
\end{equation}

\subsection{Berry Connection and Berry Curvature}
The Berry connection $\mathcal{A}_\mu$ is given in terms of the change in wave function under infinitesimal variation of the parameters $\mu=(k,\Delta_s,\Delta_p)$ as $\mathcal{A}_\mu=i\langle \psi|\partial_\mu |\psi\rangle$. We suppress the momentum subscripts whenever possible without causing ambiguity.  Defining $\Delta=\Delta_s + i\Delta_p f_k \equiv |\Delta| e^{i\phi}$  we may write the valence band wave function as
\begin{align}
	|\psi \rangle=(-v^\ast,\,u^\ast)= \frac{1}{\sqrt{2E(E-\xi)}} \left(\xi-E ,\, \Delta^\ast \right)
	\label{eqn:psi_SI}
\end{align} 
implying  $\mathcal{A}_\mu=|u|^2\partial_\mu\phi$ where $|u|^2=\frac{1}{2}\left(1+\frac{\xi}{E}\right)$. Explicitly,
\begin{eqnarray}
	\left(\mathcal{A}_{k}, \mathcal{A}_{\Delta_s}, \mathcal{A}_{\Delta_p} \right)
	=  \frac{|u|^2}{\Delta_s^2+f_k^2\Delta_p^2} \left(
	\Delta_s\Delta_p \partial_k f_k,\,
	-f_k\Delta_p
	,\,
	f_k\Delta_s
	\right)
	\label{eqn:A_SI}
	\,.
\end{eqnarray}
Note that $\mathcal{A}$ has singularities (``Dirac strings") along the line $\Delta_s=\Delta_p=0$ and also along the lines $\Delta_s=f_k=0$ unless $|u|^2$ vanishes  on parts of these lines. These are shown as dashed lines in Fig.~\ref{fig:monopoles_BEC}(a) for the BEC case ($G<0$).
The Berry curvature $B=dA$ is then
\begin{eqnarray}
	\left( B_{k}, B_{\Delta_s}, B_{\Delta_p} \right)
	=-\frac{1}{2E_k^3}  \left(  
	\xi_k f_k
	,\,
	\Delta_s v_k f_k
	,\, 
	\Delta_p (v_k f_k-\xi_k \partial_k f_k)
	\right)
	\,.
	\label{eqn:B_SI}
\end{eqnarray}
Considering now the flux of $B$ through a surface element of an oriented 2D manifold in $(k, \Delta_s,\Delta_p)$ space defined by a function $s(\Delta_s,\Delta_p)=\text{constant}$ and choosing the orientation to be pointing  `inside'  the cylinder in Fig.~\ref{fig:monopoles_BEC}(a) we see by comparison to \equa{eqn:Jfinal2} that the flux through the surface is just the polarization $P$. This conclusion is independent of the choice of coordinate.

\subsection{BCS-BEC crossover}
\label{app:BCS-BEC}
If the numbers of electrons and holes are separately conserved, the total number $n=\langle n_{electron}+n_{hole} \rangle=-\langle\sigma_3 \rangle+n_0$ is also conserved where $n_0$ is the particle number of a completely occupied band. $n$ is the analogy to the total charge in a superconductor, and gives the constraint that shifts $G$ from  positive to negative as interaction becomes stronger such that the system crossovers from a BCS to a BEC type condensate. This is the situation in electron hole bilayers with no interlayer tunneling. Moreover, $n$ can also be approximately fixed by gate voltage.
For natural crystals, $n$ is not fixed since there are always interband conversion mechanisms breaking this $U(1)$ symmetry. Hartree terms due to Coulomb repulsion between a/b orbitals will shift down $G$ and induce such a crossover in this case.

\begin{figure}
	\includegraphics[width=0.9 \linewidth]{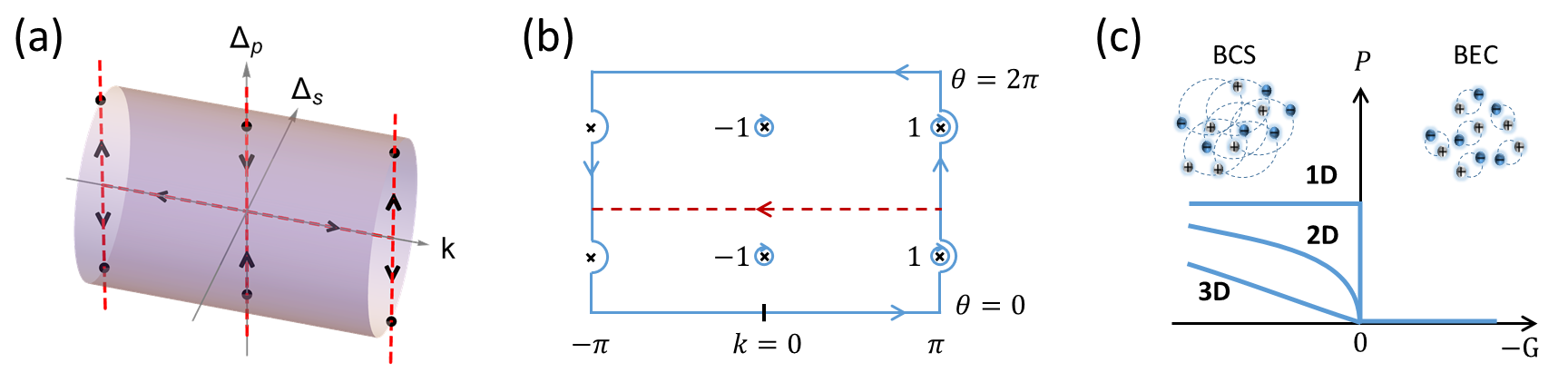}
	\caption{(a) The $(k,\theta)$ surface $S$ embedded in the $(k,\Delta_s,\Delta_p)$ space. Shown is the BEC case where the effective $G$ is negative in the mean field equation (band gap $-G$ is positive). In the Berry connection convention \equa{eqn:A_SI}, there are `Dirac strings' as shown by the red dashed lines, whose intersection with $S$ lead to the singular vortices in (b). The black arrows indicate the directions of the Dirac strings. (b) The torus parametrized with $(k,\theta)$. The integral of the Berry curvature over the torus is converted into the loop integral on its boundary and around the singular vortices. The vortices at $k=0$ contribute opposite values to those at $k=\pi/a$, rendering the net result to be $P=0$. (c) Schematic of the pumped charge in BCS and BEC regimes of excitonic insulators.}
	\label{fig:monopoles_BEC}
\end{figure}

In the BEC case ($G<0$, no band inversion), there are no monopoles and the Dirac string structure looks like that in Fig.~\ref{fig:monopoles_BEC}, rending zero pumped charge. Intuitively, the excitons in the BEC state are tightly bound electron hole pairs that don't overlap with other, and can be viewed as charge neutral point particles. Thus no charge transport can occur.

Therefore, there is a topological transition at $G=0$ during the BCS-BEC crossover, and the charge pumping $P$ can be viewed as an `order parameter' that separate these two regimes, as shown in Fig.~\ref{fig:monopoles_BEC}(c). However, we have focused on the dynamics in the BCS limit in this paper while it is interesting to investigate similar dynamics in the crossover regime.

\subsection{Pumped charge for arbitrary rotation angle} \label{app:Poftheta}

In this section we provide the details leading to Eqs.~(5) and (6) of the main text.
Parameterizing $S$ using $k$ and the angle $\theta \equiv \arg(\Delta_s, \Delta_p)$ defined in Fig.~1(a) of the main text,  the Berry connection and curvature can be projected onto the $(k,\theta)$ space. In other words, The wave function can now be viewed as a function of $(k,\theta)$ and $\Delta(k,\theta)=\Delta_s + i\Delta_p f_k=|\Delta(k,\theta)| e^{i\phi(k,\theta)}$ is the pairing field at $(k,\theta)$. Note that $f_k$ reverses sign for positive and negative $k$, and that $|u|^2=\frac{1}{2}\left(1+\frac{\xi}{E}\right)$ is nearly zero deep inside the fermi sea ($\xi \ll -|\Delta|$) and  $|u|^2 \rightarrow 1$ outside when $\xi \gg |\Delta|$.  

The flux can be converted to the line integral of the Berry connection over the edge of $S$ in Fig.~2(b) of the main text. The singularities (marked by crosses in Fig.~2(b)) are from the intersections with the Dirac strings, and must be correctly treated in the evaluation of the line integral, although they do not contain fluxes in $B$.  For a full cycle, the edge is the blue contour together with the two small circles surrounding the two vortices. The former gives zero net contribution due to periodicity in $k$ and $\theta$ while the latter contributes $N=2$, recovering the number of monopoles.

The polarization at arbitrary angle $\theta$ can be evaluated analytically in the BCS limit in which $|\Delta_s|,|\Delta_p|\ll G$ where $a$ is the lattice constant. The Berry curvature is concentrated in the region $|\xi_k |\lesssim |\Delta|$ so we may perform the integral in Eq.~(3) of the main text only on the red rectangles centered on $k=\pm k_F$ in Fig.~2(b) of the main text, chosen such that $u\approx 0$ on the vertical edges inside the fermi momentum $\pm k_F$ and thus $\mathcal{A}_\theta=0$ from \equa{eqn:A_SI}, while $|u| \approx 1$ on the vertical edges outside the fermi momentum  and $\mathcal{A}_\theta=\pm 1$. The top and bottom edges all contribute zero since $\partial_k f_k$ and thus $\mathcal{A}_k$ are negligible around $k=\pm k_F$. Therefore, the contour integral gives 
\begin{align}
	P=\theta/\pi
	\label{eqn:P_1D_SI}
\end{align}
for an 1D  excitonic insulator.
This result may also be understood by noting that the low energy physics around $\pm k_F$ is of two massive Dirac models,  each of which realizes a Goldstone-Wilczek \cite{Goldstone1981} mechanism of charge pumping.

In a 2D system one has two momenta, which we choose to be parallel ($k_x$) and vertical  $(k_y)$ to the direction defined by the antisymmetry of $f_k$.  The net charge pumped is then an integral over $k_y$ of the previously obtained formula. The only change is that now $\xi_{k_x} \rightarrow \xi_{k_x,k_y}$ and it may be that for some values of $k_y$ the sign of $\xi$ does not change as a function of $k_x$, meaning that for these $k_y$ the monopoles lie outside the torus of integration so no charge pumping occurs.  In the weak coupling limit the issue may be discussed in terms of the Fermi surface of the metallic ($\Delta=0$) phase. If the fermi surface (fermi crossings for each $k_y$  as $k_x$ is  varied) is open  the density of transferred charge is $2/a_y$ during a full cycle where $a_y$ is the lattice constant in $y$ direction; if the fermi surface is closed, then only the range of $k_y$ where crossings occur gives rise to a charge pumping; thus the net density of pumped charge during a full cycle is $2k_{Fy}/\pi$ where $k_{Fy}$ is the maximum extent of the fermi surface in the $y$ direction.

In an isotropic 2D system, for an incomplete cycle with arbitrary $\theta$,  note that each 1D momentum chain crossing the fermi surface at $\left(\pm\sqrt{k_F^2-k_y^2},k_y \right)=k_F\left(\pm \cos\theta_k,\sin\theta_k \right)$ contributes a charge pumping channel described by \equa{eqn:P_1D_SI}, with effective rotation angle $\varphi_{k_y}=\tan^{-1}\left(\cos\theta_k \tan\theta \right)$. Summing over all the chains, one obtains
\begin{align}
	P=\frac{1}{2\pi}\int_{-k_F}^{k_F} dk_y \frac{\varphi_{k_y}}{\pi}=\frac{k_F}{2\pi^2}\int_{-1}^{1} dt \tan^{-1}\left(\sqrt{1-t^2}\text{tan}\theta \right)=\frac{k_F}{2\pi} \tan \frac{\theta}{2}
	\label{eqn:P2D_SI}
\end{align}
for $0<\theta<\pi/2$. Extending the above integral to higher $\theta$, one obtains Eq.~(6) of the main text.

\subsection{Current response in time domain}
In this section, we try to expand
\equa{eqn:current_response_SI} to higher orders in frequency and show that this won't give corrections to the adiabatic result.
We focus on the nodes at $k=(0,\pm k_F)$ in 2D when the system is close to pure $p_x$-wave order. In 3D, the nodes become a nodal line and the result stays the same up to some $O(1)$ constants. 
Close to $(0,\Delta_p)$, the trajectory of motion is nearly along the $\Delta_s$ direction due to the saddle point structure on the free energy landscape in Fig.~4(a) of the main text. Thus it is enough to consider the current response to $\dot{\Delta}_s\equiv \partial_t\Delta_s$ in \equa{eqn:J2}: $J(\omega)=\chi_{j_x,\Delta_s}(\omega) \Delta_s(\omega)$.  We assume the order parameter passes the point $(0,\Delta_p)$ with nearly constant velocity $\dot{\Delta}_s$. In the BCS limit, the second term vanishes due to the $\xi$ factor and what remains is
\begin{align}
	\chi_{j_x,\Delta_s}(\omega,0)  = \sum_k
	v_k f_k \frac{\Delta_p}{E_k} \frac{-2i\omega}{\omega^2-4E_k^2} = C_0(\Delta_p,\Delta_s) (-i\omega) + C_1(\Delta_p,\Delta_s) (-i\omega)^2 + O(\omega^3)
	\label{eqn:chi_j_deltas}
\end{align}
where 
\begin{align}
	C_0=-\frac{1}{2}\Delta_p\sum_k
	v_k f_k \frac{1}{E_k^3}
	=-\Delta_p \nu \int d\theta_k \frac{1}{2\pi}
	v_F \cos^2\theta_k \frac{1}{\Delta_p^2 \cos^2\theta_k}
	=- \nu v_F \frac{1}{\Delta_p} 
	=-\frac{1}{2\pi} \frac{k_F}{\Delta_p} 
	\label{eqn:C0}
\end{align}
is the adiabatic current leading to \equa{eqn:P2D_SI} and $C_1$ is a dissipative term that arises from quasiparicles excitations. At exactly $(0,\Delta_p)$, it is
\begin{align}
	C_1  &= \frac{1}{-i\omega} \mathrm{Im}\left[\sum_k
	\frac{v_k f_k \Delta_p}{E_k^2}  \frac{2E_k}{\omega^2-4E_k^2} 
	\right]
	=\frac{\pi}{\omega} \Delta_p \sum_k
	\frac{v_k f_k }{E_k^2}  
	\left(
	\delta(\omega-2E_k) -\delta(\omega+2E_k)
	\right)
	\approx
	\frac{1}{8} \frac{k_F}{\Delta_p^2} 
	.
	\label{eqn:c1}
\end{align}
Another source for dissipative current is the quasiparticle contributed optical conductivity from the node:
\begin{align}
	\sigma_{xx} &= \frac{i}{\omega}\chi_{\Delta_p \partial_{k_x} f_k \sigma_2,\, \Delta_p \partial_{k_x} f_k \sigma_2}  =\frac{i}{\omega}\frac{\Delta_p^2}{k_F^2} \chi_{\sigma_2 \sigma_2}
	=\frac{i}{\omega} \frac{\Delta_p^2}{k_F^2} \sum_k \frac{4E_k}{\omega^2-4E_k^2} \frac{\xi_k^2}{E^2_k} 
	=  \frac{1}{8}\frac{\Delta_p}{k_F v_F} 
	+i\text{Im}[\sigma_{xx}]
	\,.
\end{align}
Its real part is suppressed by the small number $\Delta_p/\varepsilon_F=\Delta_p/(k_F v_F)$ and can thus be neglected.

It appears from \equa{eqn:c1} that there is a correction to the pumped charge as $\delta P=\int dt C_1 \partial_t^2 \Delta_s$. However, if one includes higher order terms in frequency, the current response from \equa{eqn:chi_j_deltas} can be written in time domain:
\begin{align}
	J(t) = \int_{-\infty}^{t} dt^\prime \chi(t,t^\prime) \partial_{t^\prime}\Delta_s 
\end{align}
where
\begin{align}
	\chi(t+t^\prime,t^\prime)  
	&= \sum_k v_{x,k} f_k \frac{\Delta_p}{E_k^2} \sin(2E_k t) 
	= \nu \frac{1}{2\pi}\int d\xi d\theta_k \Delta_p v_F \cos^2\theta_k \frac{ \sin(2E t)}{\xi^2+\Delta_p^2\cos^2\theta_k+\Delta_s^2}  \notag\\
	&= \text{node contribution} + \text{high energy state contribution}  
	\notag\\
	&\approx \nu  \frac{2}{2\pi} \Delta_p^{-2} v_F  \int_{-\Delta_p}^{\Delta_p} 
	d\xi dx x^2 \frac{ \sin(2E t)}{\xi^2+x^2+\Delta_s^2} 
	+
	\nu  \frac{2\pi}{2\pi} \Delta_p v_F \int_{\Delta_p}^{\infty} 
	d\xi \frac{ \sin(2E t)}{\xi^2+\Delta^2} 
	\notag\\
	&\approx \nu  \Delta_p^{-2} v_F  \int^{\Delta_p}_{0} 
	du u^3 \frac{ \sin(2E t)}{u^2+\Delta_s^2} 
	+ \text{high energy state contribution}  
	\notag\\
	&\approx \nu  \Delta_p^{-2} v_F  \int^{\Delta_p}_{\Delta_s} 
	dE (E-\Delta_s^2/E) \sin(2E t)
	\notag\\
	&= -\frac{1}{4} \nu v_F  \Delta_p^{-2}   \left(
	\partial_t 
	+ 4\Delta_s^2 \int dt 
	\right)
	\frac{1}{t}\left(\sin(2\Delta_p t)-\sin(2\Delta_s t)  \right) 
	\label{eqn:chi_t}
\end{align}
is the response kernel which is time dependent due to the fact that $\Delta_s$,$\Delta_p$ changes with time. Note that $E=\sqrt{\xi_k^2+\Delta_s^2+\Delta_p^2 f_k^2}$ in the above integrals although we neglected its subscripts for notational simplicity. If one uses the values of  $\Delta_s,\,\Delta_p$  at $t^\prime$ and evaluate the polarization at $t\rightarrow \infty$ by $\int dt J(t)$, one recovers exactly the adiabatic current and the topological charge pumping. Therefore, the non-adiabatic correction is beyond the scope of \equa{eqn:chi_t}, but lies in the fact that the state at $t^\prime$ is not the ground state of the instantaneous mean field Hamiltonian assumed here. We will show that this physics can be addressed in terms of exact dynamics of pseudo-spins.

\section{Edge states\label{app:edgestates}} 
In this section we analyse the behavior of  edge states. For simplicity we foucs on the  weak coupling BCS limit of \equa{eqn:HS_action_SI} with open boundary condition. Linearizing the Hamiltonian near the two fermi points $\pm k_F$, we find that an edge state wave function may be written
\begin{equation}
	\psi(x)=\phi_1 e^{-ik_F x + k_0 x} + \phi_2 e^{ik_F x + k_0 x}
	\label{psi_edge}
\end{equation}
with energy $E^2= \Delta^2- v_F^2 k_0^2$ where $\Delta^2=\Delta_s^2 + \Delta_p^2 f_{k_F}^2=\Delta_s^2 + \Delta_p^2$ and we have made use of our convention $f_{k_F}=1$.
The spinor part of the wave function is 
\begin{equation}
	\phi_1 = (\Delta_s+ i\Delta_p,\, -iv_F k_0+E) \,, \quad
	\phi_2 = (\Delta_s- i\Delta_p,\, i v_F k_0+E)
	\,.
	\label{psi_edge_spinor}
\end{equation}
To satisfy the open boundary condition $\psi(0)=0$, one requires $\phi_1+\phi_2=0$ which yields
\begin{equation}
	\frac{\Delta_s+ i\Delta_p}{\Delta_s- i\Delta_p} = \frac{E-iv_F k_0}{E+ i v_F k_0}
	\,.
	\label{edge_condition}
\end{equation}
This and the relation $E^2= \Delta^2- v_F^2 k_0^2$  is satisfied by two solutions: $( k_0,\, E_{+})=(-\Delta_p/v_F,\, \Delta_s)$ and $(k_0,\, E_{-})=(\Delta_p/v_F,\, -\Delta_s)$. The corresponding wave functions are 
\begin{align}
	\psi_\pm = \frac{1}{C_\pm} \left(1, \pm1 \right) \sin (k_F x) e^{\mp x \Delta_p/v_F} 
	\,.
	\label{eqn:edge_state_SI}
\end{align}
Note that the subscript $\pm$ tracks each wave function smoothly as $\theta$ varies, but does not specify either the energy or the side where the state is localized at. They are determined by the signs of their energies and the exponential factors.

The relation between the two edge states follows from  symmetries. One may define two unitary operations, the `phase rotated inversion' $\hat{P}_1:(\psi_a(x),\psi_b(x))\rightarrow (\psi_a(-x),-\psi_b(-x))$ and the $\hat{P}_2: (\psi_a(x),\psi_b(x))\rightarrow (\psi_b(-x),-\psi_a(-x))$. Both operators inter converts the two edge states. $\hat{P}_1$ is a symmetry of the mean field Hamiltonian $H$ if the system is in a pure $p$-wave state while $\hat{P}_2$ always anti commutes with $H$. Therefore, $\psi_{\pm}$ have opposite energies and will be at zero energy in a pure $p$-wave state.

Note that in open 1D wires connecting two reservoirs in Fig.~3 of the main text, although the edge states seem to be responsible for the charge pumping, the actual carries are all electrons in the valence band moved by continuous deformation of their wave functions, which is a bulk property. Indeed, in macroscopically long wires, the expansion and shrinking of edges states happen only in a tiny vicinity of $\theta=0,\pi$, while the charge pumping is a continuous process as $\theta$ varies. For example in the $\theta=0_+$ state, although there is an occupied edge state localized on the right, the other electrons in the valence band form a density distribution that has a `hole' on the right, such that the total polarization is still nearly zero. In the $\theta \rightarrow \pi/2$ state, the background density distribution has a `half' hole on each edge. Together with the occupied edge state on the right, it look like there is  a half charge on the right edge and a half hole on the left, so that the polarization $P=1/2$.  

\section{The Ginzburg-Landau action \label{app:GL}}
In this section we present the derivation of the semiclassical action used in the main text.
The Ginzburg-Landau action for order parameter fields and the EM field is obtained by integrating out the fermions  ($e^{-S[\Delta_s,\Delta_p, A]}
\equiv \int  D[\bar{\psi},\psi]  e^{-S[\psi,\Delta_s,\Delta_p, A]}$) in \equa{eqn:HS_action_SI}, resulting in
\begin{equation}
	S[\Delta_s,\Delta_p, A]=\mathrm{Tr\, ln} \left[\partial_\tau +\xi_k\sigma_3+
	\Delta_s \sigma_1+\Delta_p f_k\sigma_2\right]
	+
	\int dr d\tau \left(\frac{\Delta_s^2}{g_s}+\frac{\Delta_p^2}{g_p} \right)
	\equiv \int dr d\tau L(\Delta_s,\Delta_p, A) \,.
	\label{eqn:grand_GL_si}
\end{equation}
The $\mathrm{Tr\, ln}$ means trace of logarithm of the infinite dimensional matrix where $k$ should be interpreted as the spatial derivative $-i\nabla$ acting on the fermion fields, i.e., the matrix is just the kernel $\partial_\tau +
H_m$ for all Fermion fields at all $(r, t)$ in \equa{eqn:HS_action_SI} (See Sec. 6.4 of Ref.~\cite{Altland.2010}). Performed in Fourier basis, it involves a summation over momenta $k$, the fermion Matsubara frequencies $\omega_n= (2n+1) \pi T$ ($n\in \mathbb{Z}$, $T$ is the temperature and we have set the Boltzmann constant to be $1$) and a trace of logarithm of the $2\times 2$ matrices. Note that we have removed the absolute value symbol from \equa{eqn:HS_action_SI} since $\Delta_s, \Delta_p$ are real numbers.
We interpret the action as the Lagrangian for the order parameter fields moving in the presence of  electric field $E=-\partial_t A$. Changing the argument $A$ to $E$, we write the Lagrangian
\begin{align}
	L(\Delta_s,\Delta_p;E) = F - K +L_{\text{drive}}
	\,
	\label{eqn:GL_lagrangian_SI}
\end{align}
as the sum of three terms: the static free energy landscape $F$, the `Kinetic energy' $K$ and drive terms. We consider each in turn. 

\subsection{Static free energy landscape}
By integrating out the fermions for static Hubbard-Stratonovich fields in \equa{eqn:grand_GL_si} we obtain the free energy density (see Ref.~\cite{Sun2020a} or Sec. 6.4, page 276 of Ref.~\cite{Altland.2010}):
\begin{align}
	F &= T  \int \frac{dk^d}{(2\pi)^d} 
	\left(\sum_{\omega_n} \mathrm{Tr ln} \left[i\omega_n+\xi_k\sigma_3 +\Delta_s\sigma_1+\Delta_pf_k\sigma_2\right] 
	+\xi_k
	\right)+
	\frac{\Delta_s^2}{g_s}+\frac{\Delta_p^2}{g_p}
	\notag\\
	&= T \int \frac{dk^d}{(2\pi)^d} 
	\left(\sum_{\omega_n} \ln \left( i\omega_n^2 -E_k^2 \right) +\xi_k \right) +\frac{\Delta_s^2}{g_s}+\frac{\Delta_p^2}{g_p}
	\notag\\
	&\xrightarrow{T \rightarrow 0}
	- \int \frac{dk^d}{(2\pi)^d} \left(E_k -\xi_k \right) +\frac{\Delta_s^2}{g_s}+\frac{\Delta_p^2}{g_p}
	\label{eqn:free_energy_SI}
\end{align} 
where $\mathrm{Tr ln}$ now means the trace of logarithm of the $2 \times 2$ matrix at momentum $k$. Note that the momentum integral has an ultraviolet (UV) cutoff $\Lambda$ which is at the order of fermi energy but also affected by the Thomas-Fermi screening length \cite{Kozlov1965}. 
Converting the integral variable to an energy $\xi$,  we find for (quasi) 1D systems in the BCS limit:
\begin{align}
	F &= - 2\nu \int_0^{\Lambda} d\xi \left( \sqrt{\xi^2+\Delta_s^2+\Delta_p^2} - \xi \right)  + 
	\frac{1}{g_s} \Delta_s^2 + 
	\frac{1}{g_p} \Delta_p^2
	\xrightarrow{\sqrt{\Delta_s^2 + \Delta_p^2} \ll \Lambda} -\nu \left(\Delta_s^2 + \Delta_p^2 \right) \ln \frac{2\Lambda}{\sqrt{\Delta_s^2 + \Delta_p^2}} + 
	\frac{1}{g_s} \Delta_s^2 + 
	\frac{1}{g_p} \Delta_p^2
	\,.
	\label{eqn:F_1D_SI}
\end{align} 
For a 2D isotropic Fermi surface, the first term is replaced by 
\begin{align}
	-\nu \int \frac{d\theta_k}{2\pi} \left(\Delta_s^2 + \Delta_p^2 \cos^2 \theta_k \right) \ln \frac{2\Lambda}{\sqrt{\Delta_s^2 + \Delta_p^2 \cos^2 \theta_k}}
	\label{eqn:F_2D}
\end{align}
where $\theta_k$ runs from $0$ to $2\pi$. In 3D, one just needs to make the replacement $\frac{d\theta_k}{2\pi} \rightarrow \frac{\sin \theta_k d\theta_k d\phi_k}{4\pi}$ where $\theta_k$ is the polar angle ranging from $0$ to $\pi$ and $\phi_k$ is the azimuthal angle ranging from $0$ to $2\pi$. In 2D, as long as $g_p< 2g_s$ \cite{sun2020BS}, the $s$-wave phase at $\Delta = 2\Lambda e^{-\frac{1}{g_s\nu}-\frac{1}{2}}$ is the ground state with energy $-\nu \Delta^2/2$ while the $p$-wave phase at $\Delta_{p0}=4\Lambda e^{-\frac{2}{g_p\nu} -1}$ is a saddle point that has energy $-\nu \Delta_{p0}^2/4$. For $g_p>2 g_s$, the ground state minimum shifts to an $s+ip$ one. Note that in d dimensions, the density of states is  $\nu=\frac{\Omega_d}{(2\pi)^d} k_F^{d-1}/v_F$ where $\Omega_d$ is the surface area of the $d$ dimensional sphere with radius $1$. For example, in 2D, $\Omega_d=2\pi$ and $\nu=\frac{1}{2\pi} k_F/v_F$.

\subsection{Kinetic energy}
The action for order parameter  fluctuations is obtained by expanding \equa{eqn:grand_GL_si} as
\begin{equation}
	S= F(\Delta_s, \Delta_p) +S_2(\delta\Delta_s, \delta\Delta_p) \,
	\label{eqn:grand_S_si}
\end{equation}
around the mean field configuration to second order in $\delta\Delta_s$, $\delta\Delta_p$, temporarily neglecting the EM field.
We assume spatially uniform, time-dependent order fluctuations $\hat{\Delta}_k(i\Omega_n)=\left(\Delta_s+\delta\Delta_s(i\Omega_n)\right)\sigma_1+\left(\Delta_p+\delta\Delta_p(i\Omega_n)\right)f_k\sigma_2=\Delta_s\sigma_1+\Delta_pf_k\sigma_2 + \delta\hat{\Delta}_k(i\Omega_n)$ and find the fluctuation term
\begin{equation}
	S_2=-\frac{1}{2}T\sum_{\omega_n}\sum_k \mathrm{Tr}\left[
	\frac{\delta\hat{\Delta}_k\left(i\omega_n+i\Omega_n-H^k_m\right)\delta\hat{\Delta}_k\left(i\omega-H^k_m\right)}
	{\left(\left(\omega_n+\Omega_n\right)^2+E_k^2\right)\left(\omega_n^2+E_k^2\right)}
	\right]+\frac{\left(\delta\Delta_s\right)^2}{g_s}+\frac{\left(\delta\Delta_p\right)^2}{g_p} \,.
\end{equation}
For convenience we will include the $f_k$ in the definition of $\Delta_p$ and $\delta \Delta_p$ in this subsection.
Evaluating the frequency integral at $T=0$, taking the trace explicitly, rearranging and keeping only the terms with $\Omega$ dependence gives
\begin{equation}
	S_2(\Omega)-S_2(\Omega=0)=-\sum_k\frac{ \frac{\Omega^2}{4}\left(\left(\delta\Delta_s\right)^2+\left(\delta\Delta_p\right)^2\right)+\left(\delta \Delta_s\Delta_s+\delta \Delta_p\Delta_p\right)^2}{2E_k\left(E_k^2+\frac{\Omega^2}{4}\right)} \,.
\end{equation}
Writing $\sum_k=\nu \int d\xi d\Omega_k$ with $\Omega_k$ the angular coordinates on the contours of constant energy, one obtains
\begin{equation}
	S_2(\Omega)-S_2(\Omega=0)=\nu \int d\Omega_k \int d\xi \frac{ \frac{\Omega^2}{4}\left(\left(\delta\Delta_s\right)^2+\left(\delta\Delta_p\right)^2\right)+\left(\delta \Delta_s\Delta_s+\delta \Delta_p\Delta_p\right)^2}{2\sqrt{\xi^2+\Delta^2}\left(\xi^2+\Delta^2+\frac{\Omega^2}{4}\right)}
\end{equation}
Defining $\xi=\Delta \tan x$ we find for the energy integral
\begin{equation}
	\frac{1}{2} \int dx\frac{\cos x}{1+\frac{\Omega^2}{4\Delta^2}\cos^2 x}= \int_0^1d(\sin x)\frac{1}{1+\frac{\Omega^2}{4\Delta^2}-\frac{\Omega^2}{4\Delta^2}\sin^2 x}=
	\frac{1}{2}\frac{\frac{2\Delta }{|\Omega|}}{\sqrt{1+\frac{\Omega^2}{4\Delta^2}}} \ln\frac{\sqrt{1+\frac{\Omega^2}{4\Delta^2}}+\frac{|\Omega|}{2\Delta}}{\sqrt{1+\frac{\Omega^2}{4\Delta^2}}-\frac{|\Omega|}{2\Delta}} \,.
\end{equation}
In the adiabatic limit (lowest order in $\Omega$ expansion), the kinetic energy is thus
\begin{equation}
	K= \nu \int d\Omega_k \frac{1}{12\Delta^4}\left(3\Delta^2\left(\left(\partial_t\Delta_s\right)^2+\left(\partial_t\Delta_p\right)^2\right)-2\left(\Delta_s\partial_t\Delta_s+\Delta_p\partial_t\Delta_p\right)^2\right) \,
	\label{Kfinal}
\end{equation}
where $\Delta^2 = \Delta_s^2+\Delta_p^2$.
In 1D, writing $\Delta_s+i\Delta_p =R e^{i\theta}$,
the kinetic term becomes 
\begin{equation}
	K=\frac{\nu}{12 R^2}\left((\partial_t R)^2+3R^2(\partial_t\theta)^2\right) \,.
	\label{Kfinal1}
\end{equation}
If $\Delta_s$ is very small and the system has a closed Fermi surface in $d=2$ or $d=3$ then the adiabatic expansion breaks down  in the regions where the gap vanishes. In this case the operator $K$ becomes nonlocal in time, and the physics is most efficiently treated directly from the action \equa{eqn:HS_action_SI}.

\subsubsection{Dissipative terms}
For convenience, we first introduce the general form of correlation functions at zero temperature:
\begin{align}
	\chi_{\sigma_i,\, \sigma_j}(\omega, q) 
	= \frac{1}{2}
	\sum_{k} \frac{1}{\omega^2 - (E_k+E_{k^\prime})^2}
	\Bigg\{& 
	(E_k+E_{k^\prime})
	\mathrm{Tr}\left[
	\sigma_i \sigma_j 
	-
	\frac{H_m^k \sigma_i H_m^{k^\prime} \sigma_j}{E_k E_{k^\prime}}
	\right]
	+
	\omega \mathrm{Tr}\left[
	\frac{\sigma_i H_m^{k^\prime} \sigma_j}{E_{k^\prime}}
	-
	\frac{H_m^{k} \sigma_i \sigma_j}{E_k}
	\right]
	\Bigg\}
	\label{eqn:chi_zero_t}
\end{align}
where $H_m^k=\xi_k \sigma_3 + \Delta_s \sigma_1+ \Delta_p f_k \sigma_2$.
We re-derive the kinetic terms by expanding the order parameter correlation functions in frequency:
\begin{align}
	S_2=\sum_\omega
	\left(
	\begin{array}{cc}
		\Delta_s(-\omega) &
		\Delta_p(-\omega)
	\end{array}
	\right)
	\left(
	\begin{array}{cc}
		\frac{1}{g_s} +\chi_{\Delta_s,\Delta_s}(\omega) & \chi_{\Delta_s,\Delta_p}(\omega)\\
		\chi_{\Delta_p,\Delta_s}(\omega) &  \frac{1}{g_p} +\chi_{\Delta_p,\Delta_p}(\omega)
	\end{array}
	\right)
	\left(
	\begin{array}{c}
		\Delta_s(\omega) \\
		\Delta_p(\omega)
	\end{array}
	\right)
	\,
	\label{eqn:s_kinetic}
\end{align}
where the subscripts of the correlation functions correspond to their channels in $H_m^k$. For example, $\chi_{\Delta_s,\Delta_p} \equiv \chi_{\sigma_1,f_k\sigma_2}$ which is \equa{eqn:chi_zero_t} with $\sigma_i,\, \sigma_j$ replaced by $\sigma_1,f_k\sigma_2$ everywhere.
The $\Delta_s^2$ term in $S_2$ is
\begin{align}
	\chi_{\Delta_s,\Delta_s}(\omega,0) &= 4 \sum_k \frac{\xi_k^2+\Delta_p^2f^2_k}{(\omega^2-4E_k^2)E_k} =-\sum_k \frac{1}{E_k} - \int \frac{d\Omega_k}{\Omega_d} (\omega^2-4\Delta_s^2) F_0(\Delta_{\Omega_k},\omega)
	= \chi_{\Delta_s,\Delta_s}(0,0) -\omega^2 \nu 
	\left\{
	\begin{array}{lc}
		\frac{1}{2\Delta^2} -\frac{\Delta_s^2}{3\Delta^4}
		&  
		\, d=1
		\\
		\frac{\Delta_s^2+2\Delta_p^2}{6|\Delta_s|\Delta^{3}}
		&  \, d=2
	\end{array}
	\right. 
	+O(\omega^4)
	\label{eqn:chi_deltas}
\end{align}
where $\Delta_{\Omega_k}^2=\Delta_s^2 + \Delta_p^2 f^2_{k({\Omega_k})}$, $\Delta^2=\Delta_s^2 + \Delta_p^2$,
$F_0(\Delta,\omega)=
\frac{\nu}{4\Delta^2} \frac{2\Delta}{\omega} \frac{\mathrm{sin}^{-1}\left(\frac{\omega}{2\Delta}\right)}{\sqrt{1-\left(\frac{\omega}{2\Delta}\right)^2}}$  and $\Omega_k$ is the angular variable in $d$-dimension.
The $\Delta_p^2$ term is
\begin{align}
	\chi_{\Delta_p,\Delta_p}(\omega,0) &= 4 \sum_k  \frac{f^2_k\left({\xi_k}^2+\Delta_s^2 \right)}{{E_k}(\omega^2-4{E_k}^2)}
	=-\sum_k \frac{f^2_k}{E_k} - \int \frac{d\Omega_k}{\Omega_D} (\omega^2-4\Delta_p^2 f^2_{k({\Omega_k})}) F_0(\Delta_{\Omega_k},\omega)
	\notag\\
	&= \chi_{\Delta_p,\Delta_p}(0,0) -\omega^2 \nu 
	\left\{
	\begin{array}{lc}
		\frac{1}{2\Delta^2} -\frac{\Delta_p^2}{3\Delta^4}
		&  
		\, d=1
		\\
		\frac{1}{6\Delta_p^2} \left[ 1-\frac{\Delta_s^3}{\Delta^3}
		\right]
		&  \, d=2
	\end{array}
	+O(\omega^4)
	\right. 
	\,.
	\label{eqn:chi_deltap}
\end{align}
The $\Delta_p \Delta_s$ term is
\begin{align}
	\chi_{\Delta_s,\Delta_p}(\omega,0) &= 4\sum_k f^2_k\frac{-\Delta_s \Delta_p}{{E_k}(\omega^2-4{E_k}^2)}
	=4\Delta_s \Delta_p \int \frac{d{\Omega_k}}{\Omega_D}  f^2_{k({\Omega_k})} F_0(\Delta_{\Omega_k},\omega)
	\notag\\
	&= \chi_{\Delta_s,\Delta_p}(0,0) +\omega^2 \nu 
	\left\{
	\begin{array}{lc}
		\frac{\Delta_s\Delta_p}{3\Delta^4}
		&  
		\, d=1
		\\
		\frac{\Delta_s}{3\Delta_p^3} \left[ 1-\frac{\Delta_s(2\Delta_s^2+3\Delta_p^2)}{2\Delta^3}
		\right]
		&  \, d=2
	\end{array}
	\right. 
	+O(\omega^4) \,.
	\label{eqn:chi_deltasp}
\end{align}

The above expansions in $\omega$ fails as $\omega \sim \Delta_s$, the minimal gap around the fermi surface, especially when $\Delta_s=0$ such that there are nodes at $k=(0,\pm k_F)$ in 2D. We next evaluate the kernels in the pure $p$-wave case $\Delta_s=0$ to gain a rough idea of the crossover of dynamical behavior.  
The dissipative part of  $\Delta_s$ kernel is 
\begin{align}
	\mathrm{Im}\left[\chi_{\Delta_s,\Delta_s}(\omega,0)\right] &=\mathrm{Im}\left[ 4 \sum_k \frac{{E_k}}{(\omega+i\eta)^2-4{E_k}^2} \right]
	=- \pi \sum_k \left( \delta(\omega-2{E_k}) - \delta(\omega+2{E_k}) \right)
	\xrightarrow{\omega \ll \Delta_p,\, d=2} - \frac{1}{2} \nu \frac{\omega}{\Delta_p}
	\label{eqn:chi_deltas_dis} 
\end{align}
where $\eta$ is an infinitesimal positive number and  we have made use of the quasi-particle density of states due to the nodes: $g(E)=\frac{1}{2\pi}k_F E/(v_F \Delta_p) $. The linear in frequency dissipation continues with a cutoff of about $\Delta_p$ beyond which it scales as a constant. Kramers-Kronig relation implies that 
\begin{align}
	\chi_{\Delta_s,\Delta_s}(\omega,0) \approx - \frac{1}{2} \nu \left( i\frac{\omega}{\Delta_p} + \frac{\omega^2}{\Delta_p^2} \right)
	\,.
	\label{eqn:chi_deltas_dis} 
\end{align}
The dissipative part of  $\Delta_p$ kernel is 
\begin{align}
	\mathrm{Im}[\chi_{\Delta_p,\Delta_p}(\omega,0)] 
	&= \mathrm{Im}\left[4 \sum_k \frac{f^2_k {\xi_k}^2}{{E_k}^2} \frac{ {E_k}}{(\omega+i\eta)^2-4{E_k}^2} \right]
	=- \pi \sum_k \frac{f^2_k {\xi_k}^2}{{E_k}^2} \left( \delta(\omega-2{E_k}) - \delta(\omega+2{E_k}) \right)
	\xrightarrow{\omega \ll \Delta_p,\, d=2} - \frac{\pi}{2^7} \nu \frac{\omega^3}{\Delta_p^3}
	\,
	\label{eqn:chi_deltas_dis} 
\end{align}
and the cubic behavior has the cutoff $\Delta_p$. This together with the $\Delta_s=0$ limit of  \equa{eqn:chi_deltap} gives 
\begin{align}
	\chi_{\Delta_p,\Delta_p}(\omega,0) \approx - \frac{\pi}{2^7} \nu \left( i\frac{\omega^3}{\Delta_p^3} + \frac{\omega^2}{6\Delta_p^2} \right)
	\,.
	\label{eqn:chi_deltap_dis} 
\end{align}

\subsubsection{In time domain}
With the adiabatic approximation so at time $t_0+t$ we can write $\Delta(t_0+t)=\Delta(t_0)+\delta\Delta(t_0+t)$, the action from \equa{eqn:grand_S_si} reads
\begin{equation}
	S=\int dt F\left[\Delta\right]+\frac{1}{2}\int dt dt^\prime \frac{\partial \delta\Delta}{\partial t}\ M^R(t-t^\prime)\frac{\partial \delta\Delta}{\partial  t^\prime}
	\label{Sadiabatic}
\end{equation}
where $M^R$ is the retarded kernel in time domain as a $2\times 2 $ matrix and $\Delta \equiv(\Delta_s, \Delta_p)$ in this subsection.
The  instantaneous (force) term in the Euler-Lagrange equations comes from the equal time correlator (potential) and the dynamics comes from expanding in derivatives, in other words
\begin{equation}
	\frac{\delta F}{\delta  \Delta}=\partial_t\int^t dt^\prime M^R(t-t^\prime)\partial_{t^\prime}\delta\Delta(t^\prime)
\end{equation}
Noting that $M^R(0)=0$ we have
\begin{equation}
	\frac{\delta F}{\delta  \Delta}=\int^t dt^\prime \partial_tM^R(t-t^\prime)\partial_{t^\prime}\delta\Delta(t^\prime)
\end{equation}
The adiabatic approximation is reasonable if the change in $\Delta$ over a time corresponding to the range of $M$ is small ($\partial_t\Delta/|\Delta|\ll 1$) so that we can evaluate $M$ at fixed $\Delta$. If we have a fully gapped configuration (open Fermi surface or $\Delta_s$ not small), $M$ decays on times larger than $|\Delta|^{-1}=1/\sqrt{\Delta_s^2+\Delta_p^2}$ so we can shift the derivative to the $t^\prime$ and integrate by parts to get 
\begin{equation}
	\frac{\delta V}{\delta  \Delta}=\int^t dt^\prime M^R(t-t^\prime)\partial^2_{t^\prime}\delta\Delta(t^\prime)\rightarrow M\partial_t^2\Delta
\end{equation}
with $M=\int^t dt^\prime M^R(t-t^\prime)$.
However, for closed Fermi surfaces, the vanishing of $\Delta_p f_k$ at some Fermi surface points means that when $\Delta_s$ is small $M$ has a part that decays slowly, actually on the time-scale of $1/\Delta_s$ and a more careful analysis is needed.
In the isotropic 2D case, we have
\begin{equation}
	S_2=\frac{1}{2}\int dt_1dt_2\left(\begin{array}{cc}\partial_t\delta \Delta_s(t_1) & \partial_t\delta \Delta_p(t_1)\end{array}\right)\mathbf{M}_R(t_1-t_2)\left(\begin{array}{c}\partial_t\delta\Delta_s(t_2) \\\partial_t\delta\Delta_p(t_2)\end{array}\right)
\end{equation}
and the (retarded) correlator is given by
\begin{equation}
	\mathbf{M}_R(t)={\Theta}(t)\sum_k\frac{\sin2E_kt}{4E_k^4}\left(\begin{array}{cc}\xi_k^2+\Delta_p^2 &- \Delta_s\Delta_p f_k \\-\Delta_s\Delta_p f_k & (\xi_k^2+\Delta_s^2)f^2_k\end{array}\right)
	\,
\end{equation}
where $\Theta$ is the Heaviside step function.
Performing the integral over momentum, one obtains the low energy kernel
\begin{eqnarray}
	\partial_t M_R^{11}(t)&\approx&\Theta(t)\frac{\nu}{2\Delta_p}\int_{\Delta_s}^{\Delta_p} 2dv\left(1-\frac{\Delta_s^2}{v^2}\right)\cos2vt + \text{high energy contribution}
	\\ \notag
	&=&\Theta(t)\frac{\nu}{2\Delta_p}\left[\frac{\sin2\Delta_pt-\sin2\Delta_st}{t} +\Delta_s\left[2\Delta_st\left(\frac{\pi}{2}-Si[2\Delta_st]\right)-\cos2\Delta_s t \right]\right]
	+ \frac{\nu}{6(\Delta_s^2 +\Delta_p^2)} \partial_t \delta(t)
	\\ \notag
	&\approx&\Theta(t)\frac{\nu}{2\Delta_p}\frac{\sin2\Delta_pt-\sin2\Delta_st}{t} 
	+ \frac{\nu}{6(\Delta_s^2 +\Delta_p^2)} \partial_t \delta(t)
	\,,
	\\ \notag
	\partial_t M_R^{22}(t)
	&\approx&\frac{\nu}{6(\Delta_s^2 +\Delta_p^2)} \partial_t \delta(t)
	\label{eqn:k_kernel}
\end{eqnarray}
The off diagonal terms don't affect the qualitative dynamics which we neglect. At small $\Delta_s$ we can neglect the second term of $\partial_t M_R^{11}$.
Therefore, in 2D, a smooth crossover between non dissipative and dissipative behaviors during the swiping across $\theta=\pi/2$ can be described by the retarded Kinetic kernel
\begin{align}
	S_{\text{dis}} &=\frac{1}{2}\int dt dt^\prime \dot{\Delta}_s(t)\ M_R(t-t^\prime)
	\dot{\Delta}_s(t^\prime)
	\,, \quad
	M_R(t) \approx \frac{\nu}{2|\Delta_p|} \int_0^{t} dt^\prime \frac{\sin2\Delta_pt^\prime-\sin2\Delta_st^\prime}{t^\prime} 
	\label{eqn:k_kernel_simple}
	\,.
\end{align}
\equa{eqn:k_kernel} implies the  equation of motion
\begin{equation}
	\frac{\delta F}{\delta  \Delta_i}=\frac{\nu}{6(\Delta_s^2 +\Delta_p^2)} \partial_t^2 \Delta_i +
	\frac{\nu \delta{i,s}}{2\Delta_p} \int^t_{-\infty} dt^\prime \frac{\sin\left[2\Delta_p(t-t^\prime)\right]-
		\sin\left[2\Delta_s(t-t^\prime)\right]}{t-t^\prime}  \Delta_i(t^\prime)
	\label{eqn:retarded_dynamics}
\end{equation}
which describes the dissipationless-dissipative crossover behavior when $\Delta_s$ crosses zero during the dynamics.

\subsection{The drive term}
\label{SI:E2}
In the drive term
$
L_{\text{drive}}= -P(\theta) E -s(\Delta_s,\Delta_p) E^2 + O(E^3)
$,
the linear coupling of electric field to the polarization is obvious. We derive the second term in this section.
The kernel of the $O(A^2)$ term is \cite{sun2020BS}
\begin{align}
	K_{ij}(\omega)= \left(\frac{n}{m} + \chi_{J_i,J_j}(\omega) \right) \delta_{ij}
\end{align}
where $J$ is the current operator in \equa{eqn:current} and $m$ is the electron mass in our model. Since the second term in the current in \equa{eqn:current} is suppressed by the factor $\Delta_p/\varepsilon_F$ in the BCS limit, its contribution can be neglected. In 1D, the current correlation function is thus  
\begin{align}
	\chi_{J,J}(\omega) = \chi_{\sigma_3 v_k,\sigma_3 v_k}(\omega) = - 4 v_F^2 \Delta^2 F_0(\Delta,\omega)
	=-v_F^2 \nu \left(1+ \frac{2}{3} \left(\frac{\omega}{2\Delta}\right)^2 + O\left(\left(\frac{\omega}{2\Delta}\right)^4 \right) \right)
\end{align}
where $\Delta^2=\Delta_s^2+\Delta_p^2$ and $
F_0(\omega) = \sum_{k} \frac{1}{E_k(4E_k^2 - \omega^2)}
=
\frac{\nu}{4\Delta^2} \frac{2\Delta}{\omega} \frac{\mathrm{sin}^{-1}\left(\frac{\omega}{2\Delta}\right)}{\sqrt{1-\left(\frac{\omega}{2\Delta}\right)^2}} = \frac{\nu}{4\Delta^2} \left(1+ \frac{2}{3} \left(\frac{\omega}{2\Delta}\right)^2 + O\left(\left(\frac{\omega}{2\Delta}\right)^4 \right) \right)
\,
$. The constant term cancels the diamagnetic contribution $n/m$ and what remains in the kernel is the $O(\omega^2)$ term that corresponds to the static polarizability from `scattering states' of the electron hole pair.
In 2D, the current correlator up to $O(\omega^2)$ is 
\begin{align}
	\chi_{J_i,J_j}(\omega)
	=- \delta_{ij} \frac{1}{d}v_F^2 \nu \int \frac{d\theta_k}{2\pi}\left(1+ \frac{2\cos^2\theta_k}{3} \frac{\omega^2}{4(\Delta_s^2+\Delta_p^2 \cos^2\theta_k)}\right) 
	= -\delta_{ij} \frac{1}{d}v_F^2 \nu  \left(1+ \frac{1}{6} \frac{\omega^2}{\Delta_s^2+\Delta_p^2+|\Delta_s|\sqrt{\Delta_s^2+\Delta_p^2}}  \right) 
	\,.
\end{align}
Therefore, the $O(E^2)$ term in the action reads
\begin{align}
	L_2
	= \frac{1}{\omega^2}K_{ij} E_iE_j= 
	-\frac{1}{6}\nu \Delta^2 \left(\frac{E}{E_0} \right)^2
	\left\{
	\begin{array}{lc}
		\frac{\Delta^2}{\Delta_s^2+\Delta_p^2} 
		&  
		\,d=1
		\\
		\frac{\Delta^2}{\Delta_s^2+\Delta_p^2+|\Delta_s|\sqrt{\Delta_s^2+\Delta_p^2}} 
		&  d=2
	\end{array}
	\right. \,
	\,
\end{align}
where the coefficient can be interpreted as $s=\lim_{\omega\rightarrow 0} \sigma(\omega)/(2i\omega)$.
The higher order terms is $E$ are in higher powers of $\left(\frac{E}{E_0} \right)^2 \frac{\Delta^2}{\Delta_s^2+\Delta_p^2}$ where $E_0$ is defined in the main text.

\section{The adiabatic transport scheme}
\label{SI:adiabatic_transport}
\subsection{Description}
If the sin pulse is wide enough in time, it is possible to make the dynamics perfectly adiabatic since the system simply follows the instantaneous minimum on the free energy landscape. As the field increases, the minimum shifts away from $(\Delta,0)$ counter clockwisely while the maximum at $(0,\Delta_{p0})$ shifts clockwisely. The maximum field needed is simply that making the instantaneous minimum and maximum coincide. In 1D, this field can be computed analytically: 
\begin{align}
	E_m(g_p)=2t \sqrt{1-x^2} e^{-1/2-t+\sqrt{t^2+1/4}}
	\label{eqn:adiabatic_threshold}
\end{align}
where $x=(-1/t+\sqrt{1/t^2 +4})/2$ and $t=1/(\nu g_p)-1/(\nu g_s)$. After reaching the maximum value (a little higher than that), the field starts to decrease, shifting back the two extrema. The order parameter is moved to the immediate left of the maximum, which gradually shifts back to $(0,\Delta_{p0})$ as the field decreases to zero. The second half of the sin pulse would therefore transport the order parameter to the minimum at $(-\Delta,0)$, completing a half cycle. However, if the decreasing field phase of the pulse is too slow, unstable fluctuations of order parameter tend to grow exponentially\cite{Sun2019metastable} and get comparable to its mean field value within the `spinodal time' $\frac{1}{|\Delta|}\ln \frac{1}{G_0}$ where $G_0\sim \frac{|\Delta|}{\varepsilon_F} \ll 1$ is the Ginzburg parameter of the landau theory. Therefore, the time scale of the pulse has to be smaller than the spinodal time. 

If $g_p$ is too small, the requires maximum field is so large that the $O(E^2)$ term $L_2=-\frac{1}{6} \nu  \frac{E^2}{E_0^2} \frac{\Delta^4}{\Delta_s^2+\Delta_p^2}$ would pull the order parameter to the origin and destroy the above adiabatic trajectory. This imposes an lower bound for the $p$-wave pairing strength $g_{pc}=g_s/(1+\sqrt{3/8} \nu g_s)$. For the adiabatic transport scheme to work, $g_p$ has to be larger than $g_{pc}$. These conclusions apply qualitatively to higher dimensions.

If the adiabatic scheme is realized, experimental measurement of the threshold electric field  gives the estimation of $g_p$ through \equa{eqn:adiabatic_threshold}.
In the fast scheme described in the main text, if the full frequency spectrum of the current can be measured, it is possible to reconstruct the angular dynamics through, e.g., Eq.~(6) of the main text.
\subsection{Derivation}
\begin{figure}
	\includegraphics[width=\linewidth]{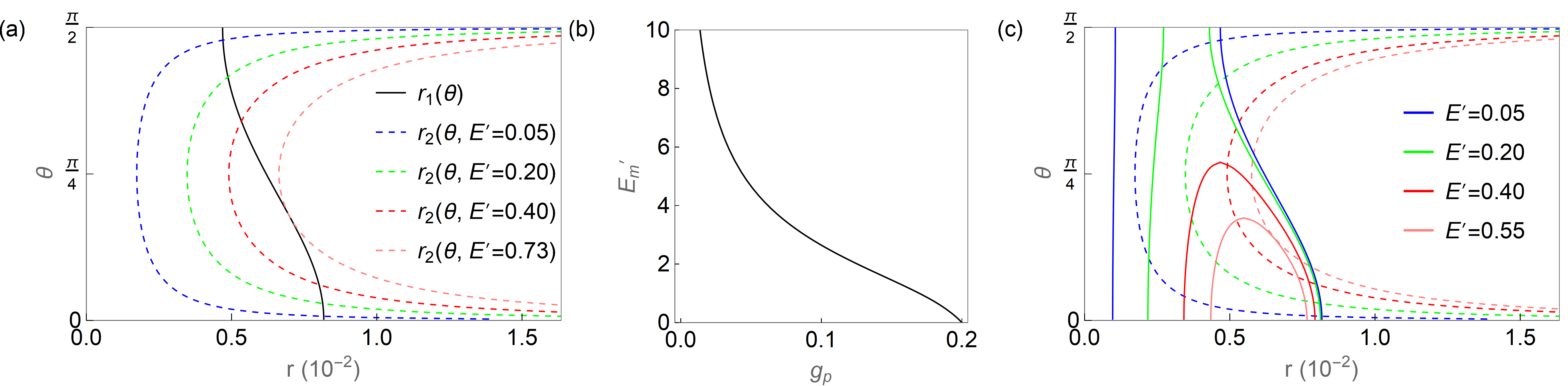}
	\caption{(a) The curves $r_1$ and $r_2$ on the $(r,\theta)$ plane at various values of electric field $E^\prime$, neglecting $O(E^2)$ terms in the free energy. Solid curves are $r_1$ and dashed curves are $r_2$. The intersections between the solid black curve and the dashed curves are the saddle points. The parameters are $g_s=0.2$ and $g_p=0.18$. (b) The maximum field $E_m$ as a function of $g_p$ for $g_s=0.2$. (c) Same as (a) but with the $O(E^2)$ terms taken into account. The $r_2$ curves are not affected by the $O(E^2)$ terms while $r_1$ curves are deformed. Each $r_1$ curve can be separated into two branches: the left branch has $\partial_r^2 f <0$ (maxima) while the right branch has $\partial_r^2 f >0$ (minima).}
	\label{fig:saddle}
\end{figure}
In 1D, incorporating the effect of a static electric field up to $O(E^2)$, the free energy is 
\begin{align}
	f(\Delta_s,\Delta_p) = \nu \left(
	-r^2 \ln \frac{2\Lambda}{r} + \frac{1}{\nu g_s} r^2 + \left(\frac{1}{\nu g_p} -\frac{1}{\nu g_s}\right) r^2\sin^2\theta -\frac{1}{2}\Delta_0^2 E^\prime\theta -\frac{1}{6} E^{\prime 2} \frac{\Delta_0^4}{r^2}
	\right)
	\label{eqn:f_1D}
\end{align}
where the `polar' coordinate is defined as $(\Delta_s,\Delta_p)=r(\cos\theta,\sin\theta)$, the dimensionless electric field is $E^\prime=E/E_0$, $E_0=\Delta_0^2/v_F$ and $\Delta_0=\Delta$. Note that $r$ is $R$ in the main text. We look for saddle points on the free energy landscape within the domain $\theta \in [0,\pi/2]$. There are two curves defined by $\partial_r f=0$ and $\partial_\theta f=0$ respectively, whose solutions read 
\begin{align}
	r_1(\theta)=\Lambda e^{- \frac{1}{g_s \nu} -t\sin^2\theta-\frac{1}{2}} ,\quad
	r_2(\theta)=\sqrt{\frac{1}{2t} \frac{\Delta_0^2 E^\prime}{\sin2\theta}}
\end{align}
where $t=\frac{1}{\nu g_p}-\frac{1}{\nu g_s}$, as shown in Fig.~\ref{fig:saddle}(a). Note that we temporarily neglected the $O(E^2)$ terms in the free energy. The intersections of the two curves are the saddle points. At zero field, the two saddle point are just the two minima at $(\theta,r)=(0,\Delta_0),\,(\pi/2,\Delta_{p0})$. For weak field, the two saddle points shift towards each other in angular direction. As the field further increases to the critical value $E_m$, the two saddle point meet which means the two lines are tangent to each other: $r_1=r_2,\, \partial_\theta r_1 = \partial_\theta r_2$ is satisfied at the intersection. This condition gives the angle at intersection as $\cos(2\theta_m)=\frac{1}{2}\left(-\frac{1}{t} + \sqrt{\frac{1}{t^2}+4} \right)$ and critical field
\begin{align}
	E_m^\prime=2t \sin(2\theta_m) e^{-1/2-t+\sqrt{t^2+1/4}}
	\label{eqn:E_m}
	\,.
\end{align}
It increases from zero as $g_p$ decreases from $g_s$, and diverges as $1/\sqrt{g_p}$ as $g_p \rightarrow 0$, as shown in Fig.~\ref{fig:saddle}(b).

The $O(E^2)$ term in \equa{eqn:f_1D} lowers the energy dramatically close to $r=0$, and therefore tends to pull the system to the zero order state. As a result, the free energy has a maximum in the $r$ direction, followed by the minimum as $r$ increases. Thus the $r_1$ curve has two branches: the left one has $\partial_r^2 f <0$ (maxima) while the right one has $\partial_r^2 f >0$ (minima), as shown by the solid curves in Fig.~\ref{fig:saddle}(c) for weak fields. For strong enough field $E$, it can happen that the two branches meet each other at certain $\theta(E)$ such that there will be no saddle points along $r$ if $\theta > \theta(E)$, as shown by the solid curves in Fig.~\ref{fig:saddle}(c) for stronger fields. The summits of those curves satisfy $(\partial_r, \partial_r^2)f=(0,0)$ which yields 
\begin{align}
	E^\prime= \frac{3}{2} \frac{r^4}{\Delta_0^4} ,\quad
	r= 2\Lambda e^{-\left( \frac{1}{\nu g_s} +t \sin^2\theta \right)-\frac{3}{4}}
	\,.
\end{align}
Making the summit at $\theta=\pi/2$, the pure $p$-wave order line, one obtains the minimal field  $E^\prime_c=\sqrt{3/2}e^{-2t-1/2}$ for the $r_1$ curves to be closed, i.e., for the minima in $r$ direction to disappear at certain angles. 

As the field increases, the intersections $A,B$ between the  $r_2$ curve and the right branch of $r_1$ curve moves towards each other. If they successfully meet each other at certain field $E_m$, the order parameter is handed by $B$ to $A$ and the subsequent decreasing field phase pushes $A$ back to the $p$-wave order, i.e., adiabatic transport works. However, if $g_p$ is too weak, it can happen that $A$ annihilates with another intersection on the left branch of $r_1$. In this situation, the order parameter will be transported to zero order instead of to the $p$-wave state. The critical $p$-wave pairing strength can be estimated roughly in this way: the summit of $r_1$ collides with the left most point of $r_2$ as field increases. This condition leads to the equality $r^2=\sqrt{2/3}\Delta_0^2 E^\prime=\frac{1}{2t}\Delta_0^2 E^\prime$ which renders $g_{pc}=g_s/(1+\sqrt{3/8} \nu g_s)$.

\section{Exact mean field dynamics \label{app:exactmeanfield}}

\subsection{Pseudo spin representation}
\label{app:exactmeanfield_pseudo_spin}
The degree of freedom at each momentum $k$ in Eq.~(2) of the main text can be mapped to an Anderson pseudo spin $\mathbf{s}_k$. 
In the second quantized language, each momentum $k$ labels two single particle states from the two bands whose annihilation operators are $\psi_{c,k}$ and $\psi_{v,k}$, giving a 4-dimensional Hilbert space. However, the mean field dynamics here implies that the total occupation number $n_k=\psi^\dagger_{c,k}\psi_{c,k}+\psi^\dagger_{v,k}\psi_{v,k}=\psi^\dagger_{k} \sigma_0 \psi_{k}$ at $k$ is always one ($n_k$ commutes with the Hamiltonian) where $\psi^\dagger_{k} = (\psi^\dagger_{c,k},\, \psi^\dagger_{v,k})$ and $\sigma_0$ is the identity matrix. Therefore, it is enough to consider a 2-dimensional subspace of single occupancy which can be mapped to a pseudo spin-$1/2$ defined as $\hat{\mathbf{s}}_k \equiv \psi^\dagger_{k} \boldsymbol{\sigma} \psi_{k}$ where $\boldsymbol{\sigma}$ are the three Pauli matrices.

The general mean field Hamiltonian at momentum $k$ can be written as 
\begin{align}
	\psi^\dagger_{k} H_m^k \psi_{k}=\psi^\dagger_{k}
	\begin{pmatrix}
		\xi_k & \Delta_s - i\Delta_p f_k\\
		\Delta_s^\ast + i \Delta_p^\ast f_k & -\xi_k
	\end{pmatrix}
	\psi_{k}
	=\mathbf{b}_k \cdot \hat{\mathbf{s}}_k
	\label{eqn:pseudofield_si}
\end{align}
and the dynamics implied by Eq.~(1) of the main text is mapped to the procession of the Anderson pseudo spins $\mathbf{s}_k=\langle \hat{\mathbf{s}}_k \rangle$ in the time dependent self consistent mean field $\mathbf{b}_k$ \cite{Barankov.2004}:
\begin{align}
	\dot{\mathbf{s}}_k &=(\mathbf{b}_k-\gamma \mathbf{b}_k \times \mathbf{s}_k ) \times \mathbf{s}_k \,, 
	\notag\\
	\mathbf{b}_k &=
	\left(\mathrm{Re}[\Delta_s] + \mathrm{Im}[\Delta_p] f_k,\, 
	-\mathrm{Im}[\Delta_s] + \mathrm{Re}[\Delta_p] f_k,\, \xi_k \right)
	=\left(\frac{g_s}{2}\sum_{k^{\prime}} s_{1 k^{\prime}} +\frac{g_p}{2} f_k\sum_{k^{\prime}} f_{k^{\prime}} s_{1 k^{\prime}},\,\,\,\,
	\frac{g_s}{2} \sum_{k^{\prime}} s_{2 k^{\prime}} +
	\frac{g_p}{2}f_k\sum_{k^{\prime}} f_{k^{\prime}} s_{2 k^{\prime}},\,\,\,\,
	\xi_k \right)
	\label{eqn:pseudospin_dynamics}
\end{align}
where the last equality is known as the gap equation. The EM vector potential  $A(t)$ enters by $k\rightarrow k-A(t)$ and we use a phenomenological damping $\gamma$ to account for the effect of energy loss due to, e.g., the phonon bath. Starting with the ground state with a real $\Delta_s$, we simulated the dynamics induced by a pulse and the current 
$
j_i=\sum_k \mathbf{s}_k \cdot \partial_{k_i} \mathbf{b}_k
$ is integrated over time to obtain the pumped charge.
Some numerical solutions to \equa{eqn:pseudospin_dynamics} are shown in Figs.~\ref{fig:charge_field} and \ref{fig:charge_field_2D}.

Note that Eqs.~\eqref{eqn:pseudofield_si} and \eqref{eqn:pseudospin_dynamics} are general which allows for appearance of imaginary components of $\Delta_s$ and $\Delta_p$.  However, as a result of an emergent `particle-hole' symmetry  in the BCS weak coupling case that maps $(\Delta_s, \Delta_p)$ to $(\Delta_s^\ast, \Delta_p^\ast)$,
the order parameter dynamics is restricted to the $s+ip$ plane where $\Delta_s(t), \Delta_p(t)$ are always real numbers which we prove in the next subsection. Away from  BCS weak coupling, the order parameter trajectory departs  from the $s+ip$ plane in a continuous way: e.g., $\Delta_p$ can temporarily develop an imaginary component that couples to $\sigma_1$ and $\Delta_s$ can have an imaginary component that couples to $\sigma_2$. 
However, since the initial and final states are the same ones, and the charge pumping is still quantized for 1D systems because it is a topological effect.

\subsubsection{Proof of why the dynamics is confined within the $s+ip$ plane in the BCS weak coupling case}
\begin{figure}
	\includegraphics[width=0.6 \linewidth]{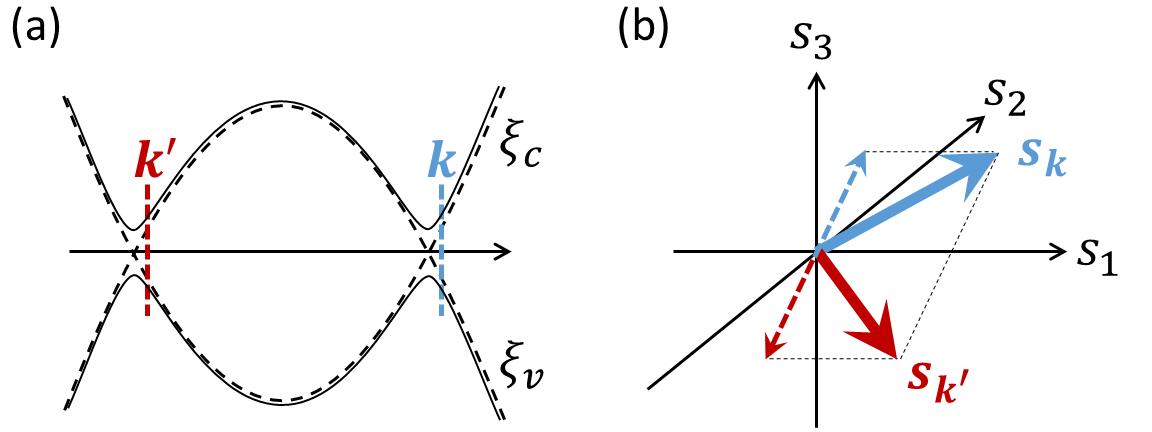}
	\caption{(a) Illustration of the momentum $\mathbf{k}$ and its `particle hole' image $\mathbf{k}^\prime$ on the diagram of quasi particle energy dispersion as a function of momentum. (b) Solid arrows are the pseudo spins $\mathbf{s_k}$ and $\mathbf{s_{k^\prime}}$ in the 1-2-3 space. Dashed arrows are their projections on the 2-3 plane. }
	\label{fig:particle_hole}
\end{figure}

The physical argument is that in the BCS weak coupling case, there will be no force pushing the order parameter away from the $s+ip$ plane during the dynamics. Following we provide a more mathematical proof.
Define the `particle-hole' operation 
\begin{align}
	\hat{P}_h: \quad
	\left(\psi_{c, \mathbf{k}}\,,\, \psi_{v, \mathbf{k}} \right) \rightarrow  
	\left(\psi_{v, \mathbf{k}^\prime}\,,\, \psi_{c, \mathbf{k}^\prime} \right)\,,\quad
	\Delta_{\mathbf{k}} \rightarrow \Delta^\ast_{\mathbf{k}^\prime}
	\,,\quad \mathbf{k}^\prime = \mathbf{k}-2k_F \hat{\mathbf{k}}
	\label{eqn:Ph_si}
\end{align}
where $k_F$ is the magnitude of Fermi momentum, $\hat{\mathbf{k}}$ is the unit vector along the direction of $\mathbf{k}$ and we have used bold fonts for the momentum to emphasize its vector nature (although every $k$ in the main text and supplemental material already means a vector). The operation $\hat{P}_h$ changes $\mathbf{k}$ to its `particle-hole' image  $\mathbf{k}^\prime$, as illustrated in \fig{fig:particle_hole}. In the BCS weak coupling case ($|\Delta_s|, |\Delta_p| \ll G$), what is relevant are the low energy states close to the Fermi surface (band crossing point) where $\hat{P}_h$ becomes a symmetry of the action in Eq.~(1) of the main text (note that $\xi_{\mathbf{k}}=-\xi_{\mathbf{k}^\prime}$). Considering that $f_{\mathbf{k}^\prime}=-f_{\mathbf{k}}$ in the BCS weak coupling case, $(\Delta_s, \Delta_p)$ is changed to its complex conjugate $(\Delta_s^\ast, \Delta_p^\ast)$ under operation of $\hat{P}_h$ according to \equa{eqn:Ph_si} and \equa{eqn:pseudofield_si}.  Therefore, given a solution $(\Delta_s(t), \Delta_p(t))$ to the order parameter dynamics, its `image' $(\Delta_s^\ast(t), \Delta_p^\ast(t))$ must also be a solution. Given the initial condition of $(\Delta_s(0), \Delta_p(0))$=$(|\Delta|, 0)$, since the solution is unique, the order parameter trajectory must lie within the plane of $\Delta_s, \Delta_p \in \mathbb{R}$, i.e., the $s+ip$ plane studied in the main text. 

An alternative way to prove this is using the pseudo-spin language. According to \equa{eqn:pseudospin_dynamics}, to prove that $(\Delta_s(t), \Delta_p(t))$ is always real, we just need to prove $\mathbf{s}_{\mathbf{k}}$ and  $\mathbf{s}_{\mathbf{k}^\prime}$ are always related to each other by the `mirror' operation $\hat{M}$ with respect to `$2-3$' plane: $(s_1,s_2,s_3)_{\mathbf{k}}=(s_1,-s_2,-s_3)_{\mathbf{k}^\prime}$, as shown in \fig{fig:particle_hole}(b). Note that the pseudo spins are defined as axial vectors such that the mirror operation $\hat{M}=\hat{s}_1$ transforms the spins as $(s_1,s_2,s_3)\rightarrow (s_1,-s_2,-s_3)$. For the initial ground state, this is obviously true  and the `magnetic field' $\mathbf{b}_{\mathbf{k}/\mathbf{k}^\prime}$ on the two pseudo-spins are also mirror images of each other under $M$, so are $(\mathbf{b} \times \mathbf{s})_{\mathbf{k}/\mathbf{k}^\prime}$. Therefore, this `mirror' relation between the pseudo spins at $\mathbf{k}/\mathbf{k}^\prime$  is sustainable during the dynamics according to \equa{eqn:pseudospin_dynamics}, and guarantees that the order parameter stays on the $s+ip$ plane.

\begin{figure}
	\includegraphics[width=0.3\linewidth]{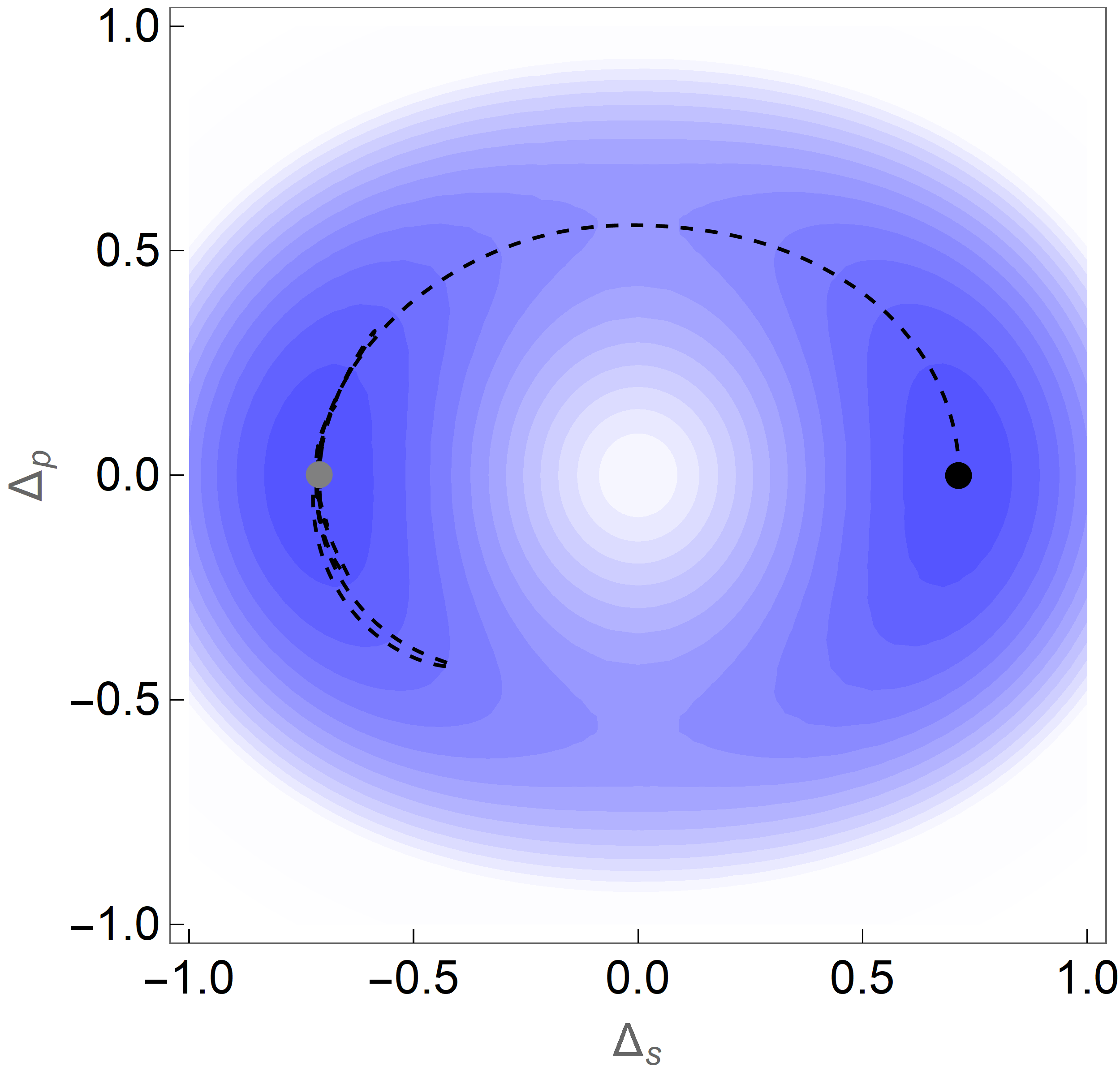}
	\includegraphics[width=0.3\linewidth]{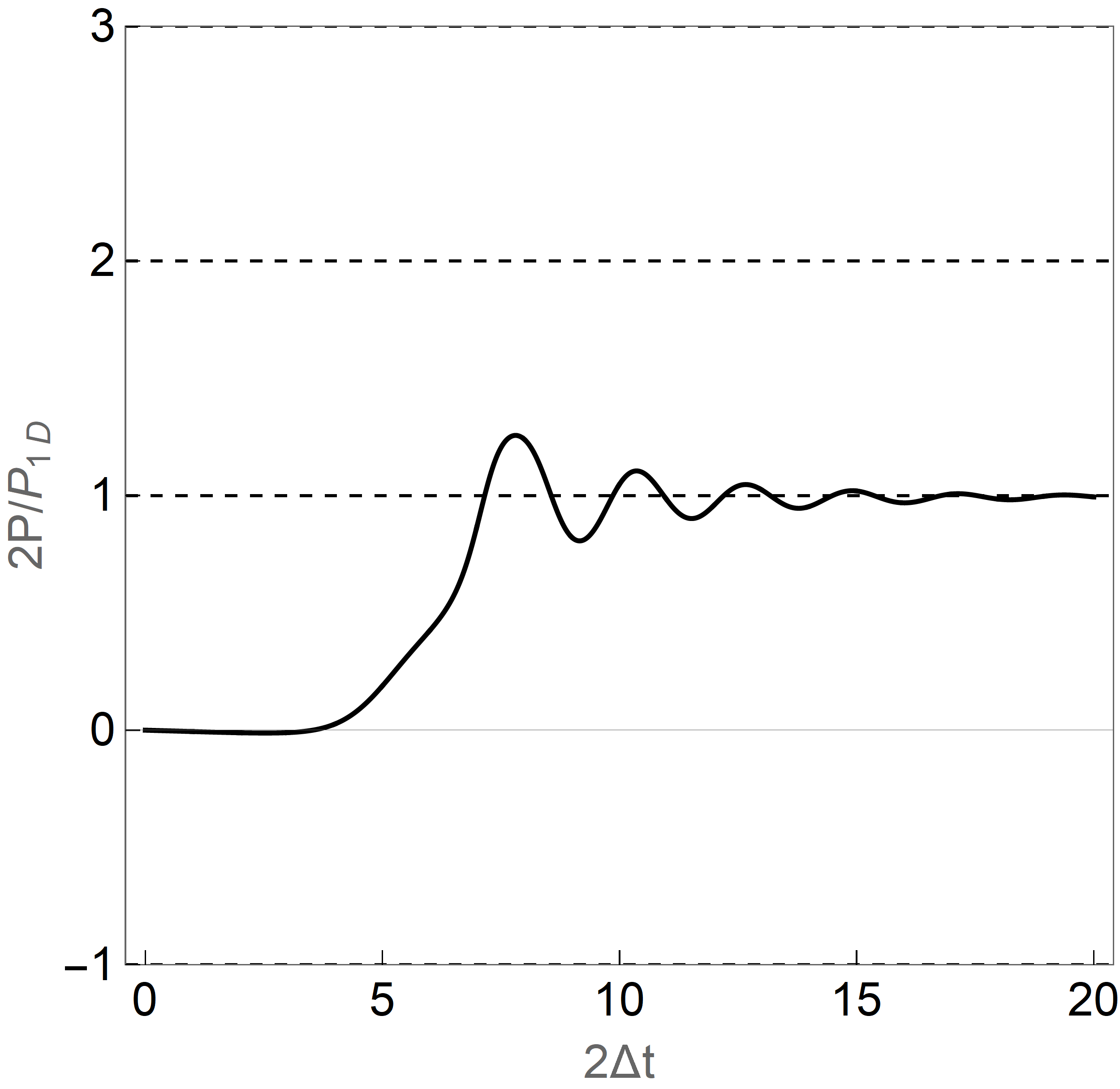}
	\includegraphics[width=0.3\linewidth]{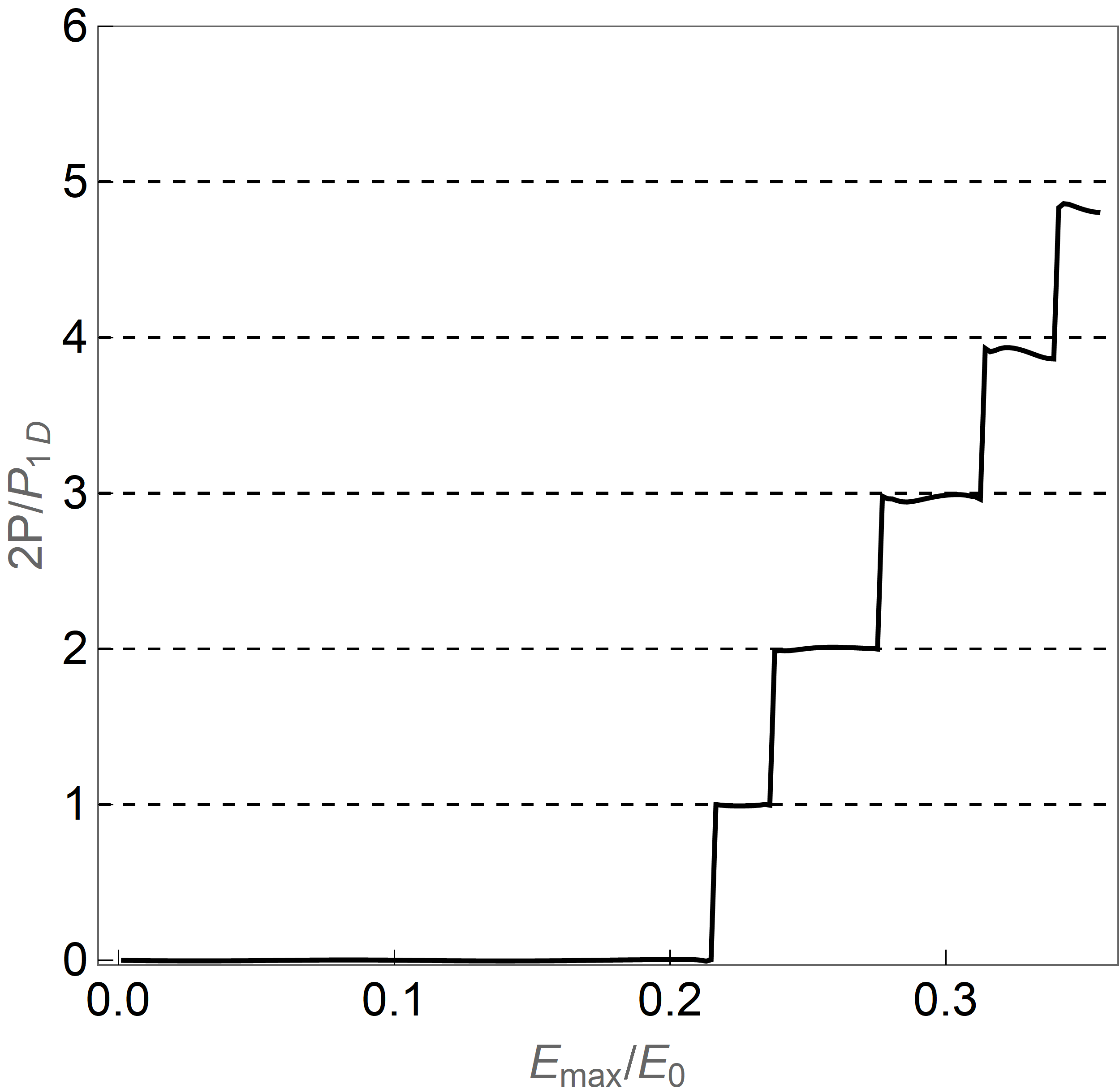}
	\caption{Order parameter dynamics of an 1D excitonic insulator subject to a pump pulse described by the vector potential $A(t)=-E_{\text{max}}w \left(\tanh(\frac{t-t_0}{w})+1\right)$. Left panel is the trajectory on the free energy landscape plotted on the $s+ip$ plane for $E_{\text{max}}=0.22 E_0$. Middle panel is the polarization as a function of time. Right panel is the pumped charge as a function of $E_{\text{max}}$. The parameters are $w=1/(2\Delta)$, $g_s \nu=0.3$, $g_p \nu=0.28$, $\Delta=2\Lambda e^{-1/(g_s \nu)}=0.071 \Lambda$, $\gamma=0.07 \Delta$, $E_{\text{max}}=0.22 E_0$. The grid in time direction is $10^4$.}
	\label{fig:charge_field}
\end{figure}

\begin{figure}
	\includegraphics[width=0.3\linewidth]{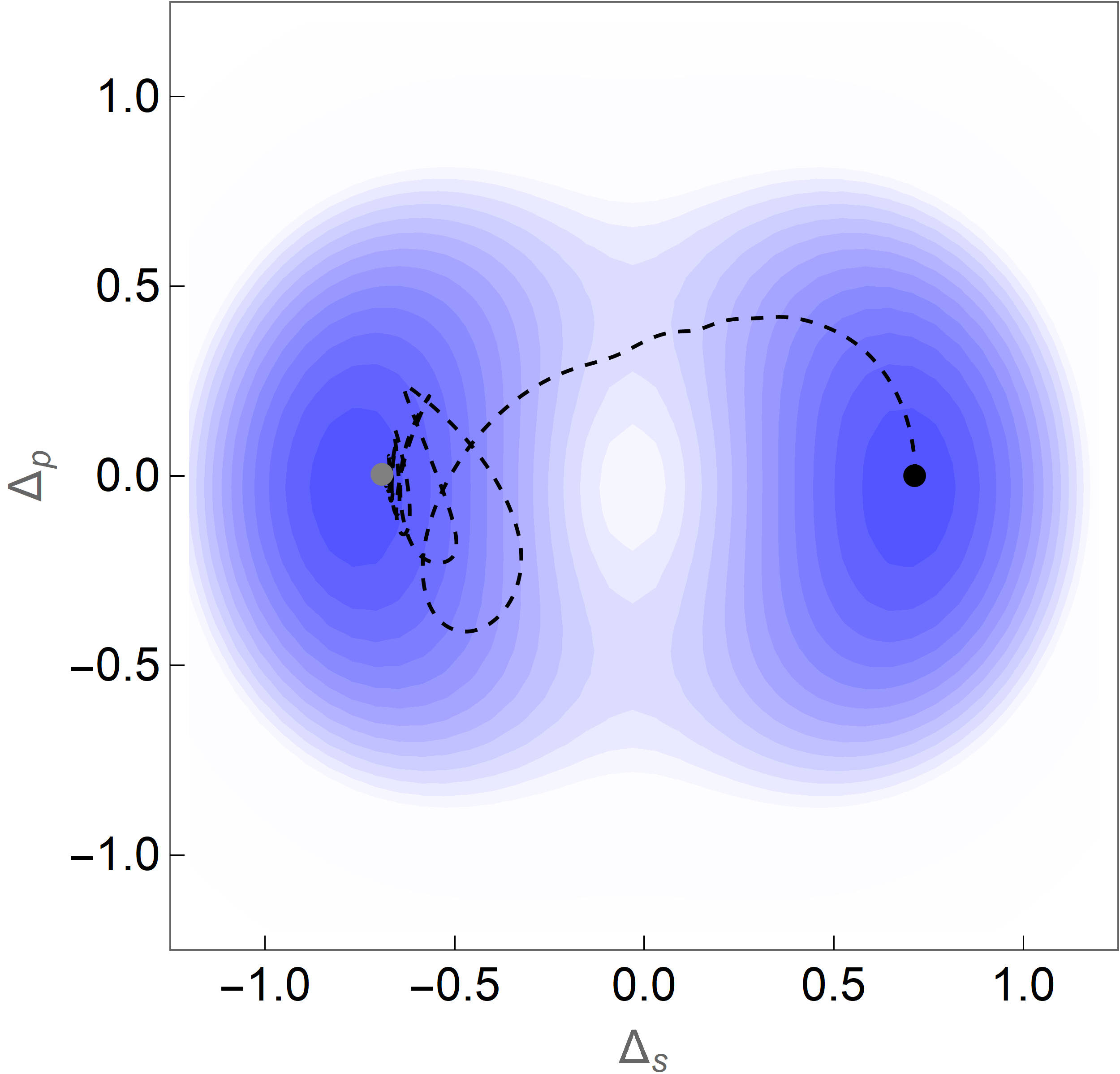}
	\includegraphics[width=0.3\linewidth]{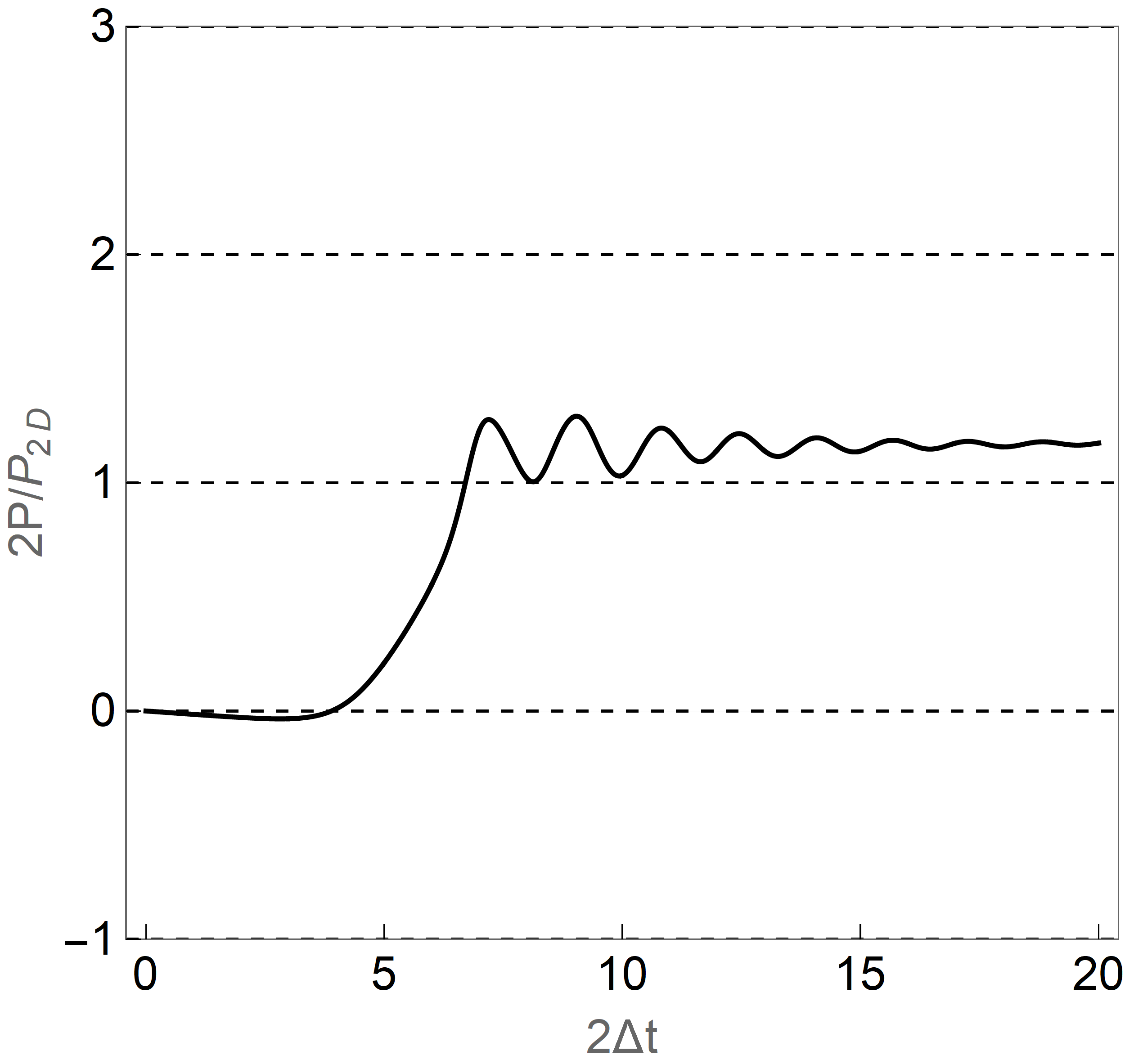}
	\includegraphics[width=0.3\linewidth]{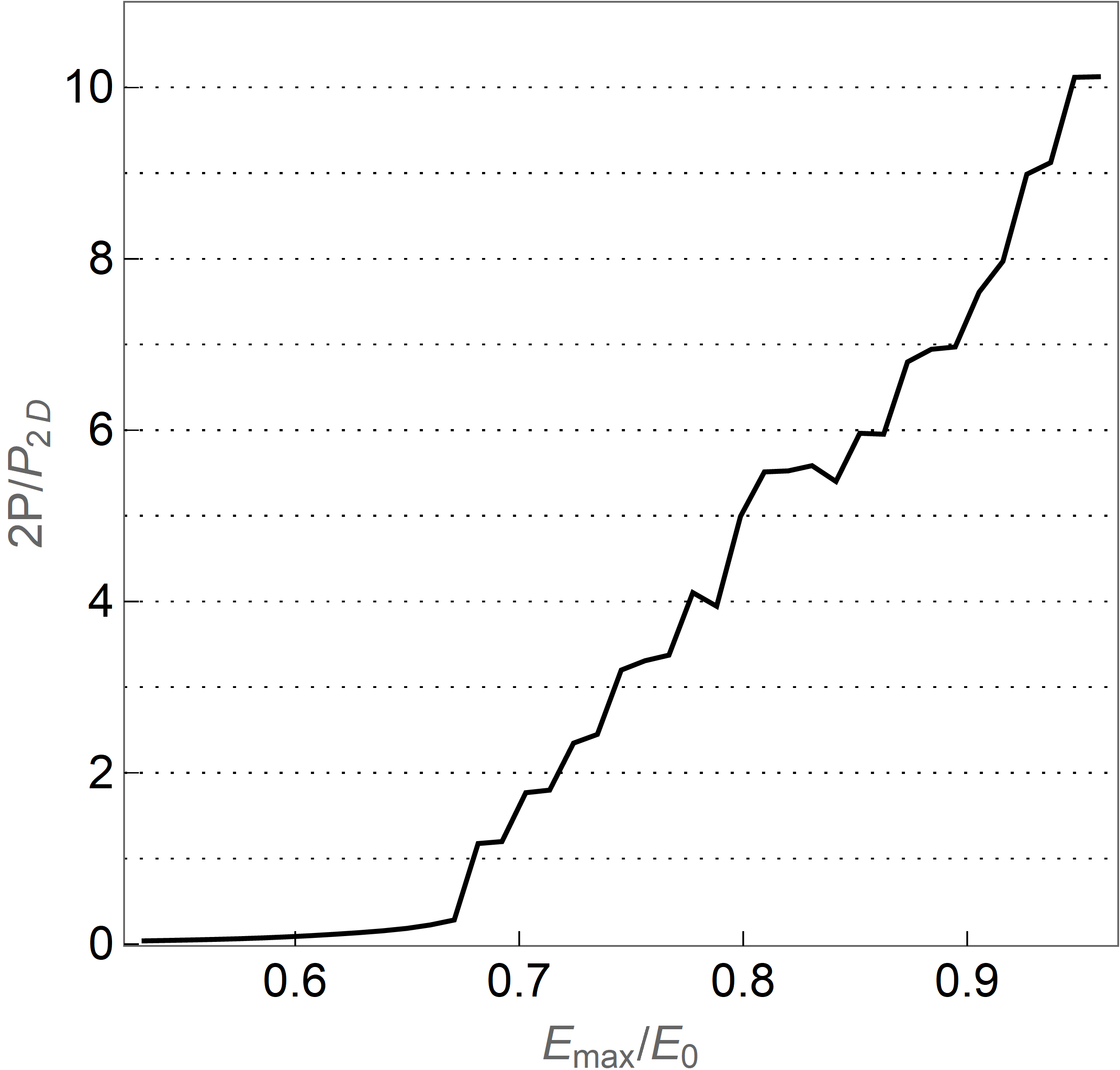}
	\caption{Order parameter dynamics of a 2D excitonic insulator subject to a pump pulse described by the vector potential $A(t)=-E_{\text{max}}w \left(\tanh(\frac{t-t_0}{w})+1\right)$. Left panel is the trajectory on the free energy landscape plotted on the $s+ip$ plane for $E_{\text{max}}=0.686 E_0$. Middle panel is the polarization as a function of time. Right panel is the pumped charge as a function of $E_{\text{max}}$. The parameters are $w=1/(2\Delta)$, $g_s \nu=0.3$, $g_p \nu=0.5$, $\Delta=2\Lambda e^{-1/(g_s \nu)}=0.071 \Lambda$, $\gamma=0.07 \Delta$. The grid in time direction is $10^4$. Here the $p$-wave pairing interaction $g_p$ is weaker than Fig.~4 of the main text, so that stronger field is needed to induce the dynamics and the nonadiabatic correction to polarization is larger.}
	\label{fig:charge_field_2D}
\end{figure}

\subsection{`Super-current' in 1D systems}
\label{app:exactmeanfield_1D}
The solution is trivial in the degenerate case $g_s=g_p$ in the BCS limit where the effect of the pairing function $f_k$ is captured by $f_{\pm k_F}=\pm 1$ on the right/left fermi point. We start from a ground state $(\Delta_s, \Delta_p)=(\Delta,0)$ where all spins are pointing in $1-3$ plane: $\mathbf{s}_k=(\Delta,0,\xi_k)/E_k$. Consider the electric field pulse at $t=0$  applied through $A=A_0\Theta(t)$. The leading driving term due to electric field is $\mathbf{b}_k=(0,0, v_k A)$ where $v_k=\pm v_F$ around the right/left fermi point. The diamagnetic term $\sim A^2$ is subleading in driving the spinor dynamics but contributes a diamagnetic current we will discuss in the end. After the kick, the spinors start to rotate around the `$3$' axis with angular frequency $\omega=2 v_F A_0$. The mean field rotates at the same speed: $(\Delta_s, \Delta_p)=\Delta(\cos(\omega t),\sin(\omega t))$ such that $\mathbf{b}_k-(0,0, v_k A)$ is always parallel to each spinor, not affecting the spin rotation. Thus the solution is that each spin synchronize and keeps rotating around `$3$' with angular frequency $v_F A_0$. Now we evaluate the current $J=J_P+J_D$. The paramagnetic current $J_P=\sum_k \langle v_k \sigma_3 \rangle$ vanishes in this state. The diamagnetic current is $j_D=\frac{2}{\pi} v_F A_0=2\omega/(2\pi)$. Therefore, the system behaves like a `superconductor' with the superfluid density $n$.

\subsection{Dynamics of the node in 2D: Landau-Zener formula}
\label{app:node}
The node contribution to the polarization is captured by a Dirac Hamiltonian with time dependent gap:
\begin{align}
	H_k(t)= \frac{\Delta_p}{k_F v_F} v_F k_x \sigma_2 + v_F k_y \sigma_3+ \Delta_s(t) \sigma_1
\end{align}
which is an approximation to Eq.~(2) of the main text around $k_0=(0,k_F)$, and is valid for $k_x,k_y \ll k_F$.
Taking into account the higher order term $\frac{k_x^2}{2m} \sigma_3$ in the Hamiltonian, the current in $x$ direction is $J_x=\frac{v_F}{k_F} k_x\sigma_3+\frac{\Delta_p}{k_F } \sigma_2$.
In the BCS limit we are concerned here, during the dynamics, the spinor at $(k_x,k_y)$ is always the mirror image (under $\hat{M}$ defined in Sec.~\ref{app:exactmeanfield}) of that at $(-k_x,-k_y)$ with respect to the $2-3$ plane. Therefore, the $\sigma_2$ contributions to the current will always sum to zero, and it is enough to consider $J_x=\frac{v_F}{k_F} k_x\sigma_3$. 
Define the energy variables $k_x^\prime=\frac{\Delta_p}{k_F }  k_x$, $k_y^\prime=v_F  k_y$, the Hamiltonian becomes
\begin{align}
	H_k(t)=  k_x^\prime \sigma_2 + k_y^\prime \sigma_3+ \Delta_s(t) \sigma_1
	\label{eqn:node_simple}
\end{align}
and the current reads $J_x=\frac{v_F}{\Delta_p} k_x^\prime \sigma_3$. We now use \equa{eqn:node_simple} to study the spinor dynamics and the current generated.

As the order parameter passes the $(0,\Delta_p)$ point with nearly constant velocity, the nodal gap $\Delta_s$ changes sign.
For the spinor at certain $k$, as $\Delta_s$ swipes from the positive value $\Delta$ at time $-t_0$ to negative value $-\Delta$ at time $t_0$, the energy splitting starts from $\sqrt{\delta_k^2 +\Delta^2}$, passes through the minimal splitting $|\delta_k|= |k^\prime|$, and ends up with $\sqrt{\delta_k^2 +\Delta^2}$. If the initial state is the low energy state, the probability of finally tunneling into the high energy state is given by the Landau-Zener formula \cite{Wittig.2005}: 
\begin{align}
	P_k= e^{-2\pi \frac{\delta_k^2}{|\partial_t 2 \Delta_s|}} 
	\label{eqn:landau_zener}
\end{align}
which is exact if $\Delta \gg \delta_k$. Therefore, the tunneling probability is unity at the node and decays to zero away from the node within a range of $\sim \sqrt{|\partial_t \Delta_s|}$.  Considering there are two nodes, the total number of quasiparticles excited is thus 
\begin{align}
	N=2 \sum_k P_k=\frac{2}{4\pi^2} \frac{k_F }{v_F \Delta_p}\int dk_x^\prime dk_y^\prime 
	e^{-\pi \frac{k^{\prime 2}}{|\partial_t \Delta_s|}}=
	\frac{2}{4\pi^2} \frac{k_F}{v_F \Delta_p}
	\pi \frac{|\partial_t \Delta_s|}{\pi}
	=
	\frac{1}{2\pi^2} \frac{k_F}{v_F}
	\frac{|\partial_t \Delta_s|}{\Delta_p }
	=
	\frac{k_F^2}{2\pi^2} \frac{1}{k_F v_F}
	\frac{|\partial_t \Delta_s|}{\Delta_p }
	\label{eqn:quasiparticle_number}
	\,.
\end{align} 
Since we have assumed Dirac dispersion in the integral, \equa{eqn:quasiparticle_number} is accurate if $\sqrt{|\partial_t \Delta_s|} \ll \Delta_p$.

\subsubsection{The pumped charge around the node}
We now compute the pumped charge, which reads
\begin{align}
	P= 2 \sum_k \int dt \langle J_x(k) \rangle_t =\frac{2}{4\pi^2} \frac{k_F }{ \Delta_p^2}\int dk_x^\prime dk_y^\prime dt
	k_x^\prime  \langle \sigma_3 \rangle_{k,t}
	=P_0+ P_{dis}
	\label{eqn:charge_landau_zener}
	\,.
\end{align} 
The integral is completely determined by the dynamics governed by \equa{eqn:node_simple}, the evalution of which requires more detailed analysis of the time evolution of each spinor. Before that, we can guess the result simply from dimensional analysis. The nonadiabatic correction $P_{dis}$ comes from spinors with $\delta_k \ll \Delta$, and the contribution arises during the anti crossing time regime when $\Delta_s(t)$ is not much larger than $\delta_k$. Therefore, neither the momentum cutoff nor the maximum value of $\Delta_s$ should enter the result. The only remaining energy scale in \equa{eqn:node_simple} is provide by $\partial_t \Delta_s$ which has the unit of energy squared. Since the integral in $P_{dis}$ has the unit of energy squared, one obtains 
$
P_{dis}=\kappa \frac{k_F}{2\pi^2} \frac{|\partial_t \Delta_s|}{ \Delta_p^2} 
$ 
where $\kappa$ is a universal $O(1)$ constant. 

Now we compute $P_{dis}$ exactly. It is more convenient to perform a permutation of the Pauli matrices: $(\sigma_2, \sigma_3, \sigma_1) \rightarrow (\sigma_1, \sigma_2, \sigma_3)$ such that the node Hamiltonian reads 
\begin{align}
	H_k(t)=  k_x^\prime \sigma_1 + k_y^\prime \sigma_2+ \Delta_s(t) \sigma_3
	\label{eqn:node_simple_new}
\end{align}
and the current becomes $J_x=\frac{v_F}{\Delta_p} k_x^\prime \sigma_2$. The dynamics of the pseudo spin at $k^\prime$ is a Landau-Zener problem \cite{Wittig.2005}. At time $-t_0$, we have $\Delta_s=\Delta \gg k^\prime$ and the spin is in the ground state:  $\psi=(0,1)^T$. The time evolution can be written as $\psi=(A(t) e^{-i \phi(t)},\, B(t) e^{i \phi(t)})^T$ where $\phi(t)=\int dt \Delta_s(t)$ should not be confused with that from Sec.~\ref{app:polarization}. The Schrodinger equation for the amplitudes reads 
\begin{align}
	\partial_t A= -i (k_x^\prime-i k_y^\prime) B e^{i\phi} \,, \quad \partial_t B= -i (k_x^\prime+i k_y^\prime) A e^{-i\phi} \,
	\label{eqn:AB}
\end{align} 
which leads to
\begin{align}
	\partial_t^2 A - i 2 \Delta_s(t) \partial_t A + k^{\prime 2} A=0 \,, \quad
	\partial_t^2 B + i 2 \Delta_s(t) \partial_t B + k^{\prime 2} B=0 
	\,.
	\label{eqn:AB2}
\end{align} 
The current involves the expectation value of $\sigma_2$:
\begin{align}
	\langle \sigma_2 \rangle=i \left( B^\ast A e^{i\theta} - c.c. \right)
	=-\left( \frac{1}{k_x^\prime-i k_y^\prime} B \partial_t B^\ast + c.c. \right)
	\,
	\label{eqn:sigma2}
\end{align} 
whose time integral gives the charge:
\begin{align}
	\int dt \langle \sigma_2 \rangle=
	-\text{Re}\left[ \frac{1}{k_x^\prime-i k_y^\prime} \right]  \left( |B(t_0)|^2- |B(-t_0)|^2\right) 
	-i\text{Im}\left[ \frac{1}{k_x^\prime-i k_y^\prime} \right] \int dt |B(t)|^2 \partial_t \ln \left(\frac{B^\ast}{B} \right)
	\,.
	\label{eqn:int_sigma2}
\end{align} 
It can be seen from \equa{eqn:AB2} that the time dependent wave function is the same between the spins at $(k_x^\prime, k_y^\prime)$ and $(-k_x^\prime, k_y^\prime)$.
Since the current is $j_x=\frac{v_F}{\Delta_p} k_x^\prime \sigma_2$, the second term in \equa{eqn:int_sigma2} will be canceled out by the two spins. The first term just needs the initial and final state information:
\begin{align}
	\int dt \langle \sigma_2 \rangle=
	-\text{Re}\left[ \frac{1}{k_x^\prime-i k_y^\prime} \right]  \left( |B(t_0)|^2- |B(-t_0)|^2\right)
	= \text{Re}\left[ \frac{1}{k_x^\prime-i k_y^\prime} \right]  \left(1- P_k\right)
	\,.
	\label{eqn:int_sigma2_simple}
\end{align}
which is provide by the Landau-Zener formula. Summing over all the spins, the pumped charge reads
\begin{align}
	P &=\frac{2}{4\pi^2} \frac{k_F }{ \Delta_p^2}\int dk_x^\prime dk_y^\prime k_x^\prime 
	\text{Re}\left[\frac{1}{k_x-ik_y}\right] (1-P_k)
	=P_0+ P_{dis}
\end{align} 
where the nonadiabatic correction is identified as 
\begin{align}
	P_{dis}&=-\frac{2}{4\pi^2} \frac{k_F }{ \Delta_p^2}\int dk_x^\prime dk_y^\prime k_x^\prime 
	\text{Re}\left[\frac{1}{k_x-ik_y}\right] P_k 
	\notag\\
	&=-\frac{2}{4\pi^2} \frac{k_F }{ \Delta_p^2}\int dk_x^\prime dk_y^\prime 
	\frac{k_x^2}{k^2} 
	e^{-\pi \frac{k^{\prime 2}}{|\partial_t \Delta_s|}}
	=-\frac{2}{4\pi^2} \frac{k_F }{ \Delta_p^2}\int dk^\prime d\theta k^\prime
	\cos^2\theta
	e^{-\pi \frac{k^{\prime 2}}{|\partial_t \Delta_s|}}
	=-\frac{k_F}{8\pi^3} 
	\frac{|\partial_t \Delta_s|}{\Delta_p^2}
	\label{eqn:charge_landau_zener}
	\,.
\end{align} 
Due to the negative relative sign of the non-adiabatic  correction to the adiabatic one, we conclude that 
\begin{align}
	P_{dis}=-\frac{k_F}{8\pi^3} 
	\frac{|\partial_t \Delta_s|}{\Delta_p^2} = -P_0 \frac{1}{8\pi^2} 
	\frac{|\partial_t \Delta_s|}{\Delta_p^2}
	\label{eqn_P_dis_final}
	\,.
\end{align} 
Therefore, \equa{eqn_P_dis_final} gives the non adiabatic correction during each half cycle of order parameter rotation, which is valid if $\sqrt{|\partial_t \Delta_s|} \ll \Delta_p$. This formula is nonperturbative in the swiping speed in the sense that, it can not be obtained by integrating over instantaneous linear or nonlinear current response functions perturbatively over the time evolution.

\end{document}